\documentclass[aps,pra,preprint,showpacs,groupedaddress]{revtex4}
\usepackage[dvips]{graphicx}
\usepackage{epsfig}
\usepackage{graphicx}
\usepackage{subfigure}
\def\be{\begin{equation}}
\def\ee{\end{equation}}
\def\ba{\begin{eqnarray}}
\def\ea{\end{eqnarray}}

\begin{document}
\title{Entanglement in a time-dependent coupled XY spin chain in an external magnetic field}
\author{Gehad Sadiek,$^{1,2,3}$ \footnote{
Corresponding author: gehad@ksu.edu.sa} Bedoor Alkurtass,$^{1}$ Omar Aldossary$^{1}$}
\affiliation{
${}^1$ Department of Physics, King Saud University, Riyadh 11451, Saudi Arabia\\
${}^2$ Department of Physics, Ain Shams University, Cairo 11566, Egypt\\
${}^3$ Department of Physics, Purdue University, West Lafayette, Indiana 47907, USA\\
}
\begin{abstract}
We consider an infinite one-dimensional anisotropic $XY$ spin chain with a nearest-neighbor time-dependent Heisenberg coupling $J(t)$ between the spins in presence of a time-dependent magnetic field $h(t)$. We discuss a general solution for the system and present an exact solution for particular choice of $J$ and $h$ of practical interest. We investigate the dynamics of entanglement for different degrees of anisotropy of the system and at both zero and finite temperatures. We find that the time evolution of entanglement in the system shows non-ergodic and critical behavior at zero and finite temperatures and different degrees of anisotropy. The asymptotic behavior of entanglement at the infinite time limit at zero temperature and constant $J$ and $h$ depends only the parameter $\lambda = J/h$ rather than the individual values of $J$ and $h$ for all degrees of anisotropy but changes for nonzero temperature. Furthermore, the asymptotic behavior is very sensitive to the initial values of $J$ and $h$ and for particular choices we may create finite asymptotic entanglement regardless of the final values of $J$ and $h$. The persistence of quantum effects in the system as it evolves and as the temperature is raised is studied by monitoring the entanglement. We find that the quantum effects dominate within certain regions of the $kT$-$\lambda$ space that vary significantly depending on the degree of the anisotropy of the system. Particularly, the quantum effects in the Ising model case persist in the vicinity of both its critical phase transition point and zero temperature as it evolves in time. Moreover, the interplay between the different system parameters to tune and control the entanglement evolution is explored.
\end{abstract}
\maketitle
\section{Introduction}
\label{sec:introduction}
Quantum entanglement represents one of the corner stones of the quantum mechanics theory and is of fundamental interest in modern physics \cite{Peres}. In the early days of the theory, the notion of entanglement was first noted and introduced by Einstein, Podolsky, and Rosen as a paradox in the formalism of the quantum theory \cite{EPR}. Nowadays entanglement is treated as a well-established concept and experimentally verified phenomenon in modern physics. Quantum entanglement is a nonlocal correlation between two (or more) quantum systems such that the description of their states has to be done with reference to each other even if they are spatially well separated. Understanding and quantifying entanglement may provide an answer for many questions regarding the behavior of complex quantum systems. Particularly, entanglement is considered as the physical property responsible for the long-range quantum correlations accompanying a quantum phase transition in many-body systems at zero temperature \cite{RMP,Osborne-QPT,Zhang-criticality}.
Particular fields where entanglement plays a crucial role are quantum teleportation, quantum cryptography, and quantum computing, where it is considered as the physical basis for manipulating linear superpositions of quantum states to implement the different proposed quantum computing algorithms \cite{Nielsen, Boumeester}.

Different physical systems have been proposed as reliable candidates for the underlying technology of quantum computing and quantum information processing \cite{Barenco,Vandersypen,Chuang,Jones,Cirac,Monroe,Turchette,Averin,Shnirman}. The basic idea in each one of these systems is to define certain quantum degree of freedom to serve as a qubit, such as the charge, orbital, or spin angular momentum. This is usually followed by finding a controllable mechanism to form an entanglement between a two-qubit system in such a way to produce a fundamental quantum computing gate such as an exclusive Boolean $XOR$. In addition, we have to be able to coherently manipulate such an entangled state to provide an efficient computational process. Such coherent manipulation of entangled states has been observed in different systems such as isolated trapped ions \cite{Chiaverini} and superconducting junctions \cite{Vion}. The coherent control of a two-electron spin state in a coupled quantum dot was achieved experimentally, in which the coupling mechanism is the Heisenberg exchange interaction between the electron spins \cite{Johnson,Koppens,Petta}.

The obvious demand for a controllable mechanism led to one of the most interesting proposals in that regard which is to introduce a time-dependent exchange interaction between the two valence spins on a doubled quantum dot system as the coupling mechanism \cite{Spin_QD1,Spin_QD2}. The coupling can be pulsed over definite intervals resulting a swap gate which can be achieved by raising and lowering the potential barrier between the two dots through controllable gate voltage. The ground state of the two-coupled electrons is a spin singlet, which is a highly entangled spin state. 

The interacting Heisenberg spin chain model represents a very reliable model for constructing quantum computing schemes in different solid state systems and a very rich model for studying the novel physics of localized spin systems. This spin chain can be experimentally realized, for instance, as a one-dimensional chain of coupled nano quantum dots.

There has been many studies focusing on the entanglement at zero and finite temperature for isotropic and anisotropic Heisenberg spin chains in presence and absence of an external magnetic field \cite{Wang1,Wang3,Asoudeh,Zhang,Asoudeh2,Rossignoli,Abdalla,Physica_B_404,Hannu_09,Sodano_10}. Particularly, the dynamics of thermal entanglement has been studied in an $XY$ spin chain considering a constant nearest-neighbor exchange interaction, in the presence of a time-varying magnetic field represented by a step, exponential, and sinusoidal functions of time \cite{HuangQInfo,HuangPhysRev}. 

Recently, the dynamics of entanglement in a one-dimensional Ising spin chain at zero temperature was investigated numerically where the number of spins was seven at most \cite{Dyn_Ising}. The generation and transportation of the entanglement through the chain under the effect of an external magnetic field and irradiated by a weak resonant field were studied. It was shown that the remote entanglement between the spins is generated and transported although only nearest-neighbor coupling was considered. Later, the anisotropic $XY$ model for a small number of spins, with a time-dependent nearest-neighbor coupling at zero temperature was studied, too \cite{Driven_xy_model}. The time-dependent spin-spin coupling was represented by a dc part and a sinusoidal ac part. It was found that there is an entanglement resonance through the chain whenever the ac coupling frequency matches the Zeeman splitting.

In this work, we investigate the evolution of quantum entanglement in an infinite one-dimensional $XY$ spin chain system coupled through nearest-neighbor interaction under the effect of a time-varying magnetic field $h(t)$ at zero and finite temperature. We consider a time-dependent nearest-neighbor Heisenberg coupling $J(t)$ between the spins on the chain. We discuss a general solution for the problem for any time dependence form of the coupling and magnetic field and present an exact solution for a particular case of practical interest, namely a step function form for both the coupling and the magnetic field. We focus on the dynamics of entanglement between any two spins in the chain and its asymptotic behavior under the interplay of the time-dependent coupling and magnetic field. Moreover, we investigate the persistence of quantum effects in the system as it evolves in time and as its temperature increases. We show that the time evolution and asymptotic behavior of entanglement for static coupling and magnetic field at zero temperature depends only the ratio of the coupling to the magnetic field rather than their individual values but not at finite temperatures. The entanglement was found to be very sensitive to the initial values of the coupling and the magnetic field and in particular cases they may dictate the asymptotic entanglement regardless of the final values of the parameters. The quantum effects were shown to dominate within certain regions of the temperature, coupling, and magnetic field space which depend significantly on the degree of anisotropy of the coupling which are manifested by the asymptotic behavior of entanglement.

This article is organized as follows. In Sec. II we present our model and discuss a general solution for the the $XY$ spin chain for a general form of the coupling and magnetic field and focus on a particular case where the system is exactly solvable. In Sec. III we evaluate the magnetization and the spin-spin correlation functions of the system and use them to evaluate the entanglement. In secs. IV, V, and VI we study the entanglement dynamics in the completely anisotropic, partially anisotropic, and isotropic cases of the system respectively. We conclude in Sec. VII and discuss future directions.    
\section{THE TIME DEPENDENT XY MODEL}
\label{sec:themodel}
In this section, we present an exact solution for the $XY$ model of a one-dimensional lattice with $N$ sites in a time-dependent external magnetic field $h(t)$. We consider a time-dependent coupling $J(t)$ between the nearest-neighbor spins on the chain. The Hamiltonian for such a system is given by

\begin{equation}
H=-\frac{J(t)}{2} (1+\gamma) \sum_{i=1}^{N} \sigma_{i}^{x} \sigma_{i+1}^{x}-\frac{J(t)}{2}(1-\gamma)\sum_{i=1}^{N} \sigma_{i}^{y} \sigma_{i+1}^{y}- \sum_{i=1}^{N} h(t) \sigma_{i}^{z}\, ,
\label{eq:H}
\end{equation}
where $\sigma_{i}$'s are the Pauli matrices and $\gamma$ is the anisotropy parameter. For simplicity, we will consider $\hbar=1$ throughout this article.
Introducing the raising and lowering operators $a^{\dagger}_{i}$, $a_{i}$
\begin{equation}
a^{\dagger}_{i} = \frac{1}{2} (\sigma_{i}^{x}+ i \sigma_{i}^{y}), \;\;\; a_{i} = \frac{1}{2} (\sigma_{i}^{x}- i \sigma_{i}^{y})\, .
\label{eq:raisinglowering}\end{equation}
Hence, Pauli matrices can be written as follows:
\begin{equation}
	\sigma^{x}_{i}=a^{\dagger}_{i}+a_{i},\;\;\;\sigma^{y}_{i}=\frac{a^{\dagger}_{i}-a_{i}}{i},\;\;\;\sigma^{z}_{i}=2 a^{\dagger}_{i} a_{i}-I\, .
	\label{eq:pauli}
\end{equation}
Following the standard procedure to treat the Hamiltonian (\ref{eq:H}), we transform the Pauli spin operators into fermionic creation and annihilation operators $b^{\dagger}_{i}$, $b_{i}$ \cite{LSM}
\begin{equation}
a_{i}^{\dagger}=b_{i}^{\dagger} \exp(i \pi \sum_{j=1}^{i-1}b_{j}^{\dagger}b_{j}), \;\;\; a_{i}= \exp(-i \pi \sum_{j=1}^{i-1}b_{j}^{\dagger}b_{j})b_{i}\, ,
\label{eq:fraisinglowering}\end{equation}
then, applying a Fourier transformation, we obtain
\begin{equation}
b^{\dagger}_{i} = \frac{1}{\sqrt{N}} \sum_{p=-N/2}^{N/2} e^{i j \phi_{p}} c^{\dagger}_{p}, \;\;\; b_{i} = \frac{1}{\sqrt{N}} \sum_{p=-N/2}^{N/2} e^{-i j \phi_{p}} c_{p}\, .
\label{eq:fourierraisinglowering}\end{equation}
where $\phi_{p}=\frac{2 \pi p}{N}$. As a result, the Hamiltonian can be written as
\begin{equation}
H=\sum_{p=1}^{N/2} \tilde{H}_{p}\, ,
\label{eq:Hsum}\end{equation}
with $\tilde{H}_{p}$ given by
\begin{equation}
\tilde{H}_{p}=\alpha_{p}(t) [c_{p}^{\dagger} c_{p}+c_{-p}^{\dagger} c_{-p}]+i J(t) \delta_{p} [c_{p}^{\dagger} c_{-p}^{\dagger}+c_{p} c_{-p}]+2 h(t)\, ,
\label{eq:Hp}\end{equation}
where $\alpha_{p}(t)=-2 J(t) \cos \phi_{p} - 2 h(t)$ and  $\delta_{p}=2 \gamma \sin \phi_{p}$.

The decomposition of the Hamiltonian was only possible because $[\tilde{H}_{l},\tilde{H}_{m}]=0$, for $l,m=0,1,2,\dots,N/2$. Thus the Hamiltonian in the $2^N$-dimensional Hilbert space has been split into $N/2$ noncommuting sub-Hamiltonians, each in a four-dimensional independent subspace.
Writing the matrix representation of $\tilde{H}_{p}$ in the basis $\{ \left|0\right\rangle, c_{p}^{\dagger}c_{-p}^{\dagger}\left|0\right\rangle, c_{p}^{\dagger}\left|0\right\rangle, c_{-p}^{\dagger}\left|0\right\rangle \}$ we obtain
\begin{equation}
\tilde{H}_{p}=\left(\begin{array} {cccc}
2 h(t) & -i J(t)\delta_{p} & 0 & 0\\
i J(t) \delta_{p} & -4 J(t)\cos \phi_{p}-2 h(t) & 0 & 0\\
0 & 0 & -2 J(t)\cos \phi_{p} & 0\\
0 & 0 & 0 & -2 J(t)\cos \phi_{p}\\
\end{array}\right)\, .
\label{eq:Hmatrix}\end{equation}

Initially the system is assumed to be in a thermal equilibrium state and therefore its initial density matrix is given by
\begin{equation}
\rho_{p}(0)=e^{-\beta \tilde{H}_{p}(0)}\, ,
\label{eq:rho0}\end{equation}
where $\beta=1/k T$, $k$ is Boltzmann constant, and $T$ is the temperature. Using Eq.(\ref{eq:Hp}) the matrix representation of $\rho_{p}(0)$ reads

\begin{equation}
\rho_{p}(0)=e^{2\beta (\cos \phi_{p}+\Gamma[h(0),J(0)])}\left(\begin{array} {cccc}
\zeta_{11}^{p} & \zeta_{12}^{p} & 0 & 0\\
\zeta_{21}^{p} & \zeta_{22}^{p} & 0 & 0\\
0 & 0 & \zeta_{33}^{p} & 0\\
0 & 0 & 0 & \zeta_{44}^{p}\\
\end{array}\right)\, ,
\label{eq:rho0matrix}\end{equation}
where
\begin{eqnarray}
\nonumber \zeta_{11}^{p}=\frac{1}{2\Gamma[h(0),J(0)]}\Bigl[\left\{\Gamma[h(0),J(0)]+J(0)\cos \phi_{p}+h(0)\right\} e^{-4\beta\Gamma[h(0),J(0)]}\\
+\left\{\Gamma[h(0),J(0)]-J(0)\cos \phi_{p}-h(0)\right\} \Bigr] \, ,\quad\quad\quad\quad \label{eq:k11}\end{eqnarray}
\begin{equation}\zeta_{12}^{p}=\frac{i\delta_{p}J_{0}\left\{1-e^{-4\beta\Gamma[h(0),J(0)]}\right\}}{4\Gamma[h(0),J(0)]}\, ,\label{eq:k12}\end{equation}
\begin{equation}\zeta_{21}^{p}=\frac{-i\delta_{p}J_{0}\left\{1-e^{-4\beta\Gamma[h(0),J(0)]}\right\}}{4\Gamma[h(0),J(0)]}\, ,\label{eq:k21}\end{equation}
\begin{eqnarray} \nonumber \zeta_{22}^{p}=\frac{1}{2\Gamma[h(0),J(0)]}\Bigl[\left\{\Gamma[h(0),J(0)]-J(0)\cos \phi_{p}-h(0)\right\} e^{-4\beta\Gamma[h(0),J(0)]}\\+\left\{\Gamma[h(0),J(0)]+J(0)\cos \phi_{p}+h(0)\right\}\Bigr]\, ,\quad\quad\quad\quad\label{eq:k22}\end{eqnarray}
\begin{equation}\zeta_{33}^{p}=\zeta_{44}^{p}=e^{-2\beta\Gamma[h(0),J(0)]}\, ,\label{eq:k33}\end{equation}
and
\begin{equation}
\Gamma[h(t),J(t)]=\left\{[J(t)\cos \phi_{p} + h(t)]^{2}+\gamma^2 J^2(t) \sin^2\phi_{p}\right\}^{\frac{1}{2}}\, .
\label{eq:Gamma}\end{equation}
Since the Hamiltonian is decomposable we can find the density matrix at any time $t$, $\rho_{p}(t)$, for the $p$th subspace by solving the Liouville equation given by
\begin{equation}
i \dot{\rho}_{p}(t)=[H_p(t),\rho_{p}(t)] \, ,
\label{eq:Liouville}
\end{equation}
which gives
\begin{equation}
\rho_{p}(t)=U_{p}(t) \rho_{p}(0) U_{p}^{\dagger}(t) \, ,
\label{eq:UrhoU}
\end{equation}
where $U_{p}(t)$ is the time evolution matrix which can be obtained by solving the equation
\begin{equation}
i \dot{U}_{p}(t)=U_{p}(t) \tilde{H}_{p}(t) \, .
\label{eq:Udot}
\end{equation}
Since $\tilde{H}_{p}$ is block diagonal $U_{p}$ should take the form

\begin{equation}
U_{p}(t)=\left(\begin{array}{cccc}
U_{11}^{p} & U_{12}^{p} & 0 & 0\\
U_{21}^{p} & U_{22}^{p} & 0 & 0\\
0 & 0 & U_{33}^{p} & 0\\
0 & 0 & 0 & U_{44}^{p}\\
\end{array}\right)\, .
\label{eq:U}
\end{equation}
Fortunately, Eq. (\ref{eq:Udot}) may have an exact solution for a time-dependent step function form for both exchange coupling and the magnetic field which we adopt in this work. Other time-dependent function forms will be considered in a future work where other techniques can be applied. The coupling and magnetic field are represented respectively by 

\begin{equation}
J(t) = J_0 + (J_1 - J_0) \theta(t) \, ,
\label{eq:stepJ}\end{equation}

\begin{equation}
h(t) = h_0 + (h_1 - h_0) \theta(t) \, ,
\label{eq:steph}\end{equation}
where $\theta(t)$ is the usual mathematical step function.
With this set up, the matrix elements of $U_p$ were evaluated to be
\begin{equation}
U_{11}^{p}=e^{2 i t J_{1} \cos\phi_{p}} \Bigl\{\frac{-i[J_{1}\cos\phi_{p}+h_{1}]\sin[2t \Gamma(h_{1},J_{1})]}{\Gamma(h_{1},J_{1})}+\cos[2t \Gamma(h_{1},J_{1})]\Bigr\}\, ,
\label{eq:U11}
\end{equation}
\begin{equation}
U_{12}^{p}=e^{2 i t J_{1} \cos\phi_{p}} \Bigl\{\frac{- J_{1} \delta_{p} \sin[2t\Gamma(h_{1},J_{1})]}{2\Gamma(h_{1},J_{1})}\Bigr\}\, ,
\label{eq:U12}\end{equation}
\begin{equation}
U_{21}^{p}=e^{2 i t J_{1} \cos\phi_{p}} \Bigl\{\frac{J_{1} \delta_{p} \sin[2t \Gamma(h_{1},J_{1})]}{2\Gamma(h_{1},J_{1})}\Bigr\}\, ,
\label{eq:U21}\end{equation}
\begin{equation}
U_{22}^{p}=e^{2 i t J_{1} \cos\phi_{p}} \Bigl\{\frac{i[J_{1}\cos\phi_{p}+h_{1}]\sin[2t \Gamma(h_{1},J_{1})]}{\Gamma(h_{1},J_{1})}+\cos[2t \Gamma(h_{1},J_{1})]\Bigr\}\, ,
\label{eq:U22}\end{equation}
\begin{equation}
U_{33}^{p}=U_{44}^{p}=e^{2 i t J_{1} \cos\phi_{p}} \, .
\label{eq:U33}\end{equation}
Consequently, the density matrix takes the form
\begin{equation}
\rho_{p}(t)=e^{2\beta J_{0}\cos\phi_{p}+2\beta\Gamma(h_{0},J_{0})}\left(\begin{array}{cccc}
\rho_{11}^{p} & \rho_{12}^{p} & 0 & 0\\
\rho_{21}^{p} & \rho_{22}^{p} & 0 & 0\\
0 & 0 & \rho_{33}^{p} & 0\\
0 & 0 & 0 & \rho_{44}^{p}\\
\end{array}\right)\, ,
\label{eq:rhot}\end{equation}
where
\begin{eqnarray}
\nonumber \rho_{11}^{p}&=& \frac{1}{4\Gamma(h_{0},J_{0})\Gamma^2(h_{1},J_{1})} \biggl\{\bigl\{J_{1} [J_{0} h_{1}-J_{1} h_{0}] \delta_{p}^2 \sin^2[2t \Gamma(h_{1},J_{1})]\\
\nonumber &+&2\Gamma^2(h_{1},J_{1})[\Gamma(h_{0},J_{0})+J_{0}\cos\phi_{p}+h_{0}]\bigr\} e^{-4\beta\Gamma(h_{0},J_{0})}\\
\nonumber &+& J_{1} [J_{1} h_{0} -J_{0} h_{1}] \delta_{p}^2 \sin^2[2t \Gamma(h_{1},J_{1})] \\
&+&2\Gamma^2(h_{1},J_{1})[\Gamma(h_{0},J_{0})-J_{0}\cos\phi_{p}-h_{0}]\biggr\}\, ,
\label{eq:rho11}\end{eqnarray}
\begin{eqnarray}
\nonumber \rho_{12}^{p}= \frac{\delta_{p}(1-e^{-4\beta\Gamma(h_{0},J_{0})})}{4\Gamma(h_{0},J_{0})\Gamma^2(h_{1},J_{1})}\biggl\{\Gamma(h_{1},J_{1})(J_{0} h_{1} - J_{1} h_{0})\sin[4t \Gamma(h_{1},J_{1})]\quad \quad \quad \\
+i \left\{J_{0}\Gamma^2(h_{1},J_{1})+2(J_{1} h_{0} -J_{0} h_{1}) (J_{1}\cos\phi_{p}+h_{1}) \sin^2[2t \Gamma(h_{1},J_{1})] \right\}\biggr\}\, ,\quad
\label{eq:rho12}
\end{eqnarray}
\begin{equation}
\rho_{21}^{p}=(\rho_{12}^{p})^{*}\, ,
\label{eq:rho21}\end{equation}
\begin{eqnarray}
\nonumber \rho_{22}^{p}&=&\frac{1}{4\Gamma(h_{0},J_{0})\Gamma^2(h_{1},J_{1})}\biggl\{\bigl\{J_{1} [J_{1} h_{0}-J_{0} h_{1}] \delta_{p}^2 \sin^2[2t \Gamma(h_{1},J_{1})]\\
\nonumber &+&2\Gamma^2(h_{1},J_{1})[\Gamma(h_{0},J_{0})-J_{0}\cos\phi_{p}-h_{0}]\bigr\}e^{-4\beta\Gamma(h_{0},J_{0})}\\
\nonumber &+& J_{1} [J_{0} h_{1} -J_{1} h_{0}] \delta_{p}^2 \sin^2[2t \Gamma(h_{1},J_{1})]\\
\nonumber &+&2\Gamma^2(h_{1},J_{1})[\Gamma(h_{0},J_{0})+J_{0}\cos\phi_{p}+h_{0}]\biggr\}\, ,
\label{eq:rho22}\end{eqnarray}
\begin{equation}
\rho_{33}^{p}=\rho_{44}^{p}=e^{-2\beta \Gamma(h_{0},J_{0})}\, .
\label{eq:rho33}
\end{equation}
\section{Spin Correlation Functions and Entanglement Evaluation}
\label{sec:QuantifyingQuantumEntanglement}
In this section we evaluate different magnetic functions of the $XY$ model which we utilize afterward to evaluate the spin-spin entanglement in the chain. The first function is the magnetization in the $z$ direction which is defined as
\begin{equation}
M=\frac{1}{N}\sum_{j=1}^{N}(S_{j}^{z})=\frac{1}{N}\sum_{p=1}^{1/N}M_p \:,
\label{eq:Mdef}
\end{equation}
where $M_p=c_{p}^{\dagger} c_{p}+c_{-p}^{\dagger} c_{-p}-1$. In terms of the density matrix, it is given by
\begin{equation}
\left\langle M_{z}\right\rangle=\frac{Tr[M\rho(t)]}{Tr[\rho(t)]} = \frac{1}{N}\sum_{p=1}^{1/N}\frac{Tr[M_{p}\rho_{p}(t)]}{Tr[\rho_{p}(t)]} \,,
\label{eq:Mexp}
\end{equation}
which yields
\begin{eqnarray}
\nonumber M_{z}= \frac{1}{4N}\sum_{p=1}^{N/2}\frac{\tanh[\beta \Gamma(h_{0},J_{0})]}{\Gamma^2(h_{1},J_{1})\Gamma(h_{0},J_{0})}\quad \quad \quad \quad \quad \quad \quad \quad \quad \quad \quad \quad \quad \quad \quad \quad \quad \\
\nonumber \biggl\{2 J_{1}(J_{0} h_{1} -J_{1} h_{0}) \delta_{p}^2 \sin^2[2t \Gamma(h_{1},J_{1})]+4\Gamma^2(h_{1},J_{1})(J_{0}\cos\phi_{p}+h_{0})\biggr\}\, .
\label{eq:M}\end{eqnarray}
The other functions needed are the spin correlation functions defined by
\begin{equation}
S^{x}_{l,m}=\left\langle S^{x}_{l} S^{x}_{m} \right\rangle, \;\;\;S^{y}_{l,m}=\left\langle S^{y}_{l} S^{y}_{m} \right\rangle, \;\;\; S^{z}_{l,m}=\left\langle S^{z}_{l} S^{z}_{m} \right\rangle\, ,
\label{eq:Sdef}\end{equation}
which can be written in terms of the fermionic operators as follows \cite{LSM}:

\begin{equation}
S_{l,m}^{x}=\frac{1}{4}\left\langle B_{l} A_{l+1} B_{l+1}\ldots A_{m-1} B_{m-1} A_{m}\right\rangle\, ,
\label{eq:Sxdef}\end{equation}

\begin{equation}
S_{l,m}^{y}=\frac{\left(-1\right)^{l-m}}{4}\left\langle A_{l} B_{l+1} A_{l+1}\ldots B_{m-1} A_{m-1} B_{m}\right\rangle\, ,
\label{eq:Sydef}\end{equation}

\begin{equation}
S_{l,m}^{z}=\frac{1}{4}\left\langle A_{l} B_{l} A_{m} B_{m}\right\rangle\, ,
\label{eq:Szdef}\end{equation}
where
\begin{equation}
A_{i}=b_{i}^{\dagger}+b_{i}, \;\;\;B_{i}=b_{i}^{\dagger}-b_{i}\, .
\label{eq:AandB}\end{equation}
Using Wick theorem \cite{Wick}, the expressions (\ref{eq:Sxdef})-(\ref{eq:Szdef}) can be evaluated as Pfaffians of the form
\begin{equation}
S_{l,m}^{x}=\frac{1}{4} pf \left(\begin{array}{cccccc}
0 &
 F_{l, l+1} & G_{l, l+1} & \cdots & G_{l, m-1} & F_{l, m}\\
 & 0 & P_{l+1, l+1} & \cdots & P_{l+1, m-1} & Q_{l+1, m}\\
 &  &  & \cdots & . & .\\
 &  &  & & P_{m-1, m-1} & Q_{m-1, m}\\
 &  &  & & 0 & F_{m-1, m}\\
 &  &  & &  & 0 \end{array}\right) \, ,
\label{eq:Sxpf}
\end{equation}
\begin{equation}
S_{l,m}^{y}=\frac{\left(-1\right)^{l-m}}{4} pf \left(\begin{array}{cccccc}
0 & P_{l, l+1} & Q_{l, l+1} & \cdots & Q_{l, m-1} & P_{l, m}\\
 & 0 & F_{l+1, l+1} & \cdots & F_{l+1, m-1} & G_{l+1, m}\\
 &  &  & \cdots & . & .\\
 &  &  & & F_{m-1, m-1} & G_{m-1, m}\\
 &  &  &  & 0 & P_{m-1, m}\\
 &  &  &  &  & 0 \end{array}\right) \, ,
\label{eq:Sybf}
\end{equation}
\begin{equation}
S_{l,m}^{z}=\frac{1}{4} pf \left(\begin{array}{cccc}
0 &  P_{l, l} &  Q_{l, m} &  P_{l, m}\\
 & 0 &  F_{l, m} &  G_{l, m}\\
 &  & 0 &  P_{m, m}\\
 &  &  & 0
\end{array}\right) \, ,
\label{eq:Szbf}\end{equation}
where
\begin{equation}
F_{l,m}=\left\langle B_{l} A_{m}\right\rangle, \;\;\; P_{l,m}=\left\langle A_{l} B_{m}\right\rangle,\;\;\; Q_{l,m}=\left\langle A_{l} A_{m}\right\rangle,\;\;\; G_{l,m}=\left\langle B_{l} B_{m}\right\rangle,
\end{equation}
and
\begin{eqnarray}
\nonumber Q_{l, m}= \frac{1}{N} \sum_{p=1}^{N/2} \biggl\{2\cos[(m-l)\phi_{p}]\quad \quad \quad \quad \quad \quad \quad \quad \quad \quad \quad \quad \quad \quad \quad \quad \quad \\
+\frac{i(J_{1}h_{0}- J_{0}h_{1}) \delta_{p} \sin[(m-l)\phi_{p}]\sin[4t\Gamma(h_{1},J_{1})]\tanh[\beta \Gamma(h_{0},J_{0})]}{\Gamma(h_{1},J_{1})\Gamma(h_{0},J_{0})}\biggr\}\, ,\quad
\label{eq:AA}\end{eqnarray}
\begin{eqnarray}
\nonumber G_{l, m}= \frac{1}{N} \sum_{p=1}^{N/2} \biggl\{-2\cos[(m-l)\phi_{p}]\quad \quad \quad \quad \quad \quad \quad \quad \quad \quad \quad \quad \quad \quad \quad \quad \quad \\
+\frac{i(J_{1}h_{0}- J_{0}h_{1}) \delta_{p} \sin[(m-l)\phi_{p}]\sin[4t\Gamma(h_{1},J_{1})]\tanh[\beta \Gamma(h_{0},J_{0})]}{\Gamma(h_{1},J_{1})\Gamma(h_{0},J_{0})}\biggr\}\, ,\quad
\label{eq:BB}\end{eqnarray}
\begin{eqnarray}
\nonumber F_{l, m}= \frac{1}{N} \sum_{p=1}^{N/2} \frac{\tanh[\beta \Gamma(h_{0},J_{0})]}{\Gamma^2(h_{1},J_{1})\Gamma(h_{0},J_{0})} \Biggl\{\cos[(m-l)\phi_{p}] \quad \quad \quad \quad \quad \quad \quad \quad \quad\\
\nonumber \biggl\{J_{1} [J_{0} h_{1} - J_{1} h_{0}] \delta^2_{p} \sin^2[2t\Gamma(h_{1},J_{1})] + 2\Gamma^2(h_{1},J_{1})(J_{0}\cos\phi_{p}+h_{0})\biggr\}\\
\nonumber + \delta_{p} \sin[(m-l)\phi_{p}] \quad \quad \quad \quad \quad \quad \quad \quad \quad \quad \quad \quad \quad \quad \quad \quad \quad \quad \quad \quad \quad\\
\nonumber \biggl\{J_{0}\Gamma^2(h_{1},J_{1}) +2 (J_{1} h_{0} - J_{0} h_{1}) (J_{1}\cos\phi_{p}+h_{1}) \sin^2[2t\Gamma(h_{1},J_{1})]\biggr\}\Biggr\}\, ,
\label{eq:BA}\end{eqnarray}
\begin{equation}
P_{l, m}=-F_{l, m}\, .
\label{eq:AB}
\end{equation}

The amount of entanglement between two quantum systems, bipartite entanglement, is a monotonic function of what is called the concurrence \cite{Wootters}. The concurrence varies from a minimum value of zero to a maximum of one coinciding with the entanglement function range and behavior. Therefore, the concurrence itself is considered as a measure of entanglement. The concurrence $C(t)$ is defined as
\begin{equation}
C(\rho)=\max(0,\lambda_{a}-\lambda_{b}-\lambda_{c}-\lambda_{d})\, ,
\label{eq:C}
\end{equation}
where the $\lambda_{i}$'s are the positive square root of the eigenvalues, in a descending order, of the matrix $R$ defined by
\begin{equation}
R=\sqrt{\sqrt{\rho}\tilde{\rho}\sqrt{\rho}}\, ,
\label{eq:R}
\end{equation}
and $\tilde{\rho}$ is the spin-flipped density matrix given by
\begin{equation}
\tilde{\rho}=(\sigma_{y}\otimes\sigma_{y})\rho^{*}(\sigma_{y}\otimes\sigma_{y})\, .
\label{eq:tilderho}
\end{equation}
Knowing that $\rho$ is symmetrical and real due to the symmetries of the Hamiltonian and particularly the global phase flip symmetry, there will be only six nonzero distinguished matrix elements of $\rho$ which takes the form \cite{H_rho_symmetry}
\begin{equation}
\rho=\left(\begin{array}{cccc}
\rho_{1,1} & 0 & 0 & \rho_{1,4}\\
0 & \rho_{2,2} & \rho{2,3} & 0\\
0 & \rho_{2,3} & \rho{3,3} & 0\\
\rho_{1,4} & 0 & 0 & \rho_{4,4}\\
\end{array}\right)\, .
\label{eq:rhomatrix}
\end{equation}
As a result, the roots of the matrix $R$ come out to be $\lambda_{a}=\sqrt{\rho_{1,1}\rho_{4,4}}+\left|\rho_{1,4}\right|$, $\lambda_{b}=\sqrt{\rho_{2,2}\rho_{3,3}}+\left|\rho_{2,3}\right|$, $\lambda_{c}=\left|\sqrt{\rho_{1,1}\rho_{4,4}}-\left|\rho_{1,4}\right|\right|$ , and $\lambda_{d}=\left|\sqrt{\rho_{2,2}\rho_{3,3}}-\left|\rho_{2,3}\right|\right|$.

To find the nonzero matrix elements of $\rho$, one can utilize the formula of the expectation value of an operator in terms of density matrix $\left\langle \hat{G} \right\rangle=Tr(\rho \: \hat{G})/ \: Tr(\rho)$ along with the magnetization Eq.(\ref{eq:Mexp}) and the spin correlation functions Eq.(\ref{eq:Sxdef})-(\ref{eq:Szdef}) which give
\begin{equation}\rho_{1,1}=\frac{1}{2} M^z_{l}+\frac{1}{2} M^z_{m}+S^z_{l,m}+\frac{1}{4}\, ,\label{eq:frho11}\end{equation}
\begin{equation}\rho_{2,2}=\frac{1}{2} M^z_{l}-\frac{1}{2} M^z_{m}-S^z_{l,m}+\frac{1}{4}\, ,\label{eq:frho22}\end{equation}
\begin{equation}\rho_{3,3}=\frac{1}{2} M^z_{m}-\frac{1}{2} M^z_{m}-S^z_{l,m}+\frac{1}{4}\, ,\label{eq:frho33}\end{equation}
\begin{equation}\rho_{4,4}=-\frac{1}{2} M^z_{l}-\frac{1}{2} M^z_{m}+S^z_{l,m}+\frac{1}{4}\, ,\label{eq:frho44}\end{equation}
\begin{equation}\rho_{2,3}=S^x_{l,m}+S^y_{l,m}\, ,\label{eq:frho23}\end{equation}
\begin{equation}\rho_{1,4}=S^x_{l,m}-S^y_{l,m}\, .\label{eq:frho14}\end{equation}
\section{Transverse Ising Model}

Considering a completely anisotropic $XY$ model by setting $\gamma=1$ in the Hamiltonian (\ref{eq:H}), we obtain the transverse Ising model Hamiltonian
\begin{equation}
H=-J(t) \sum_{i=1}^{N} \sigma_{i}^{x} \sigma_{i+1}^{x} - \sum_{i=1}^{N} h(t) \sigma_{i}^{z} \, .
\label{eq:H-Ising}
\end{equation}
Defining a dimensionless coupling parameter $\lambda=J/h$, the ground state of the Ising model is characterized by a quantum phase transition that takes place at $\lambda$ close to the critical value $\lambda_c = 1$ \cite{Osborne-QPT}. The order parameter is the magnetization $\langle \sigma^x \rangle$ which differs from zero for $\lambda \geq \lambda_c$ and zero otherwise. The ground state of the system is paramagnetic when $\lambda \rightarrow 0$ where the spins get aligned in the magnetic field direction, the $z$ direction. For the other extreme case when $\lambda \rightarrow \infty$ the ground state is ferromagnetic and the spins are all aligned in the $x$ direction. 
The ground state of the Ising model as $\lambda \rightarrow 0$ is a product of individual spin states pointing in the $z$ direction, while for $\lambda \rightarrow \infty$ is product of spin states pointing in the $x$ direction. This means that in both cases the state is minimally entangled. Quantum phase transition takes place at zero temperature as the thermal fluctuations destroy the quantum correlations in the ground state of the system. The effect of the temperature on entanglement near the critical point in the Ising model has been studied in Ref. \cite{Osborne-QPT}, where it has been shown how the entanglement decays abruptly as the temperature raises; nevertheless, it is sustained in the vicinity of the critical point close to $kT=0$. In this section we study the dynamics of entanglement in the Ising model under the effect of nearest-neighbor coupling and external magnetic field where both are considered time-dependent. The number of spins $N$ in the system is set to $1000$ throughout this study, where testing larger values of $N$ showed no effect on the results.
\begin{figure}[htbp]
\begin{minipage}[c]{\textwidth}
 \centering 
   \subfigure{\label{fig:1a}\includegraphics[width=6cm]{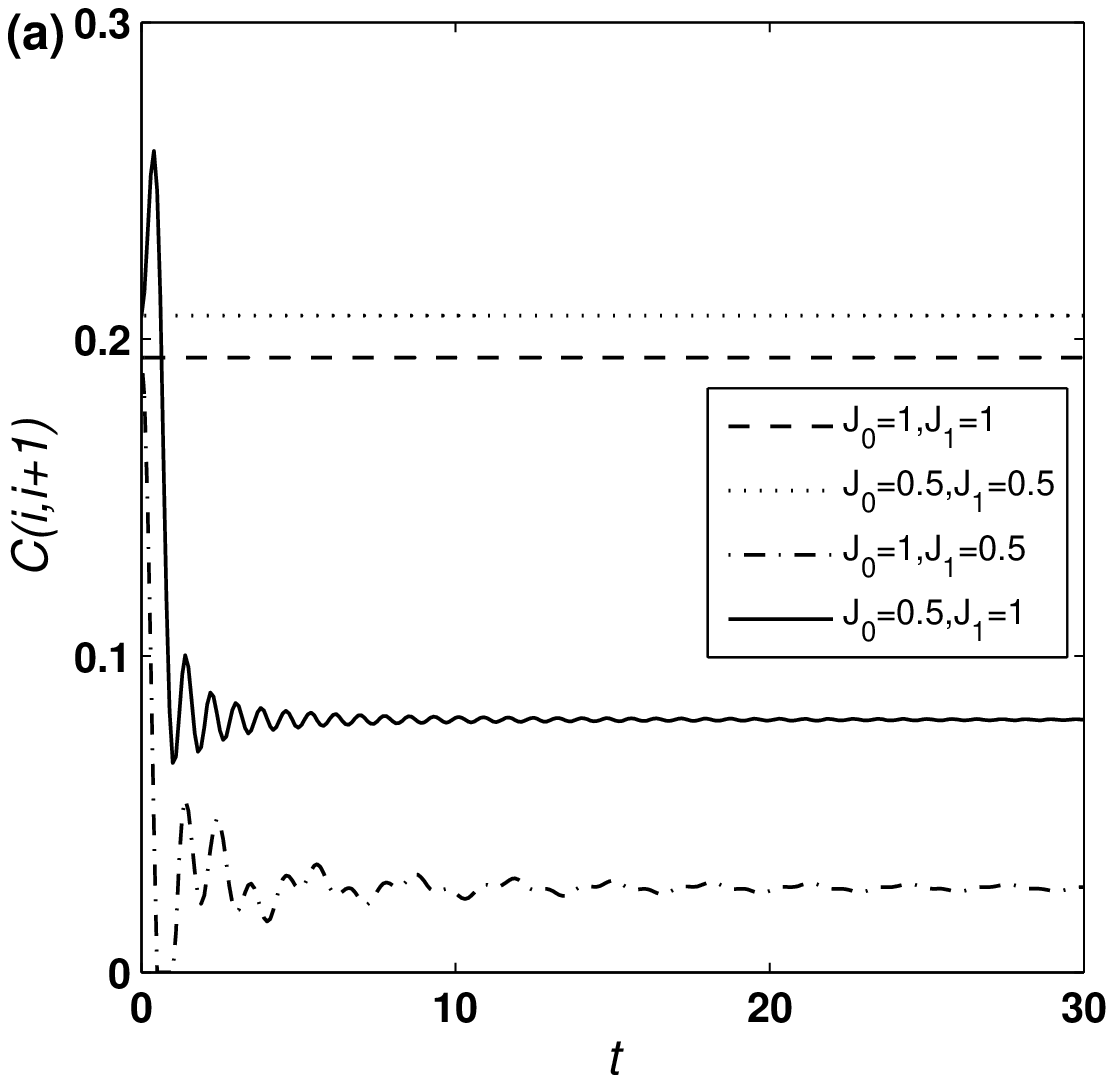}}\quad
   \subfigure{\label{fig:1b}\includegraphics[width=6cm]{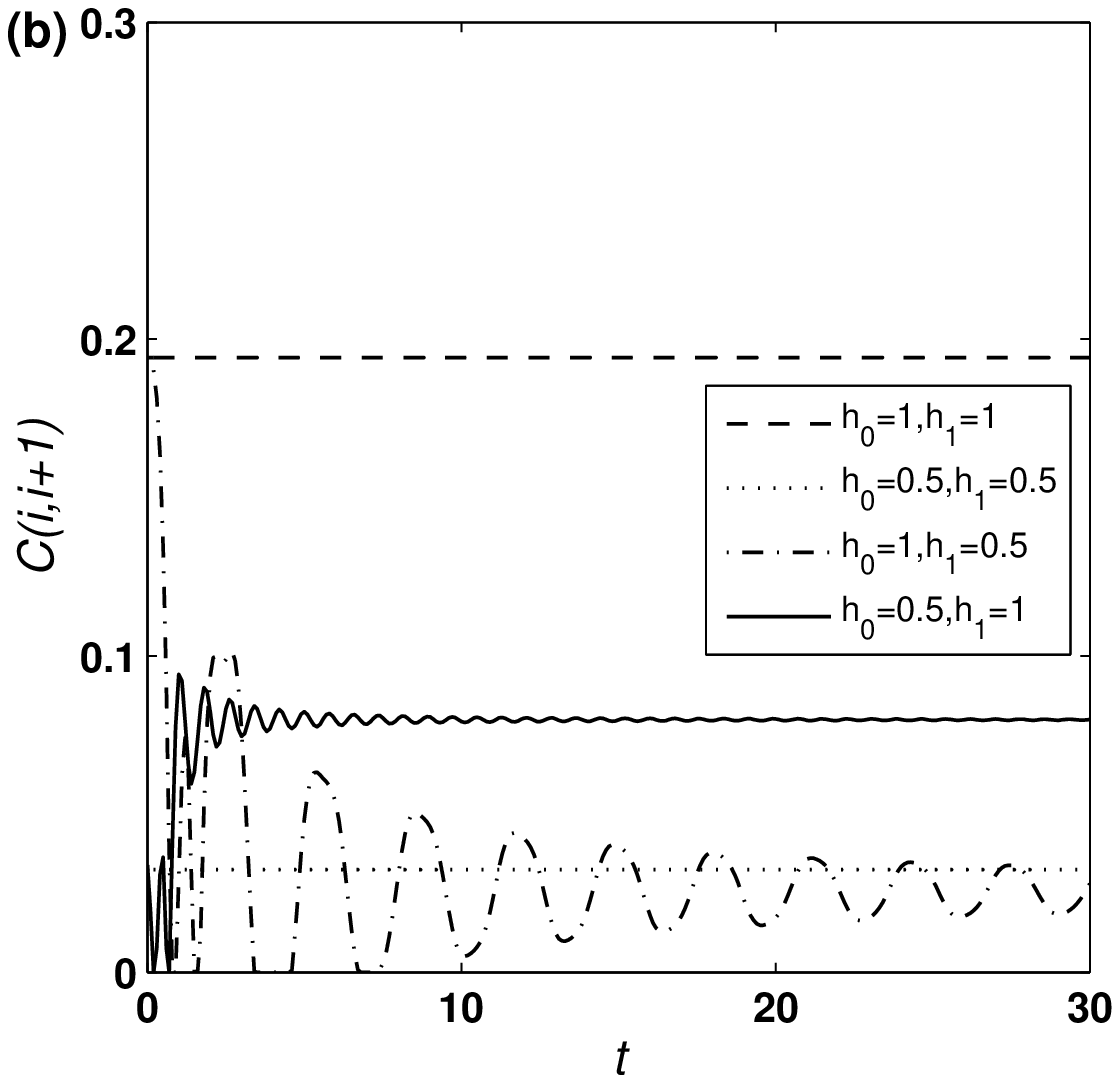}}\\
	 \subfigure{\label{fig:1c}\includegraphics[width=6cm]{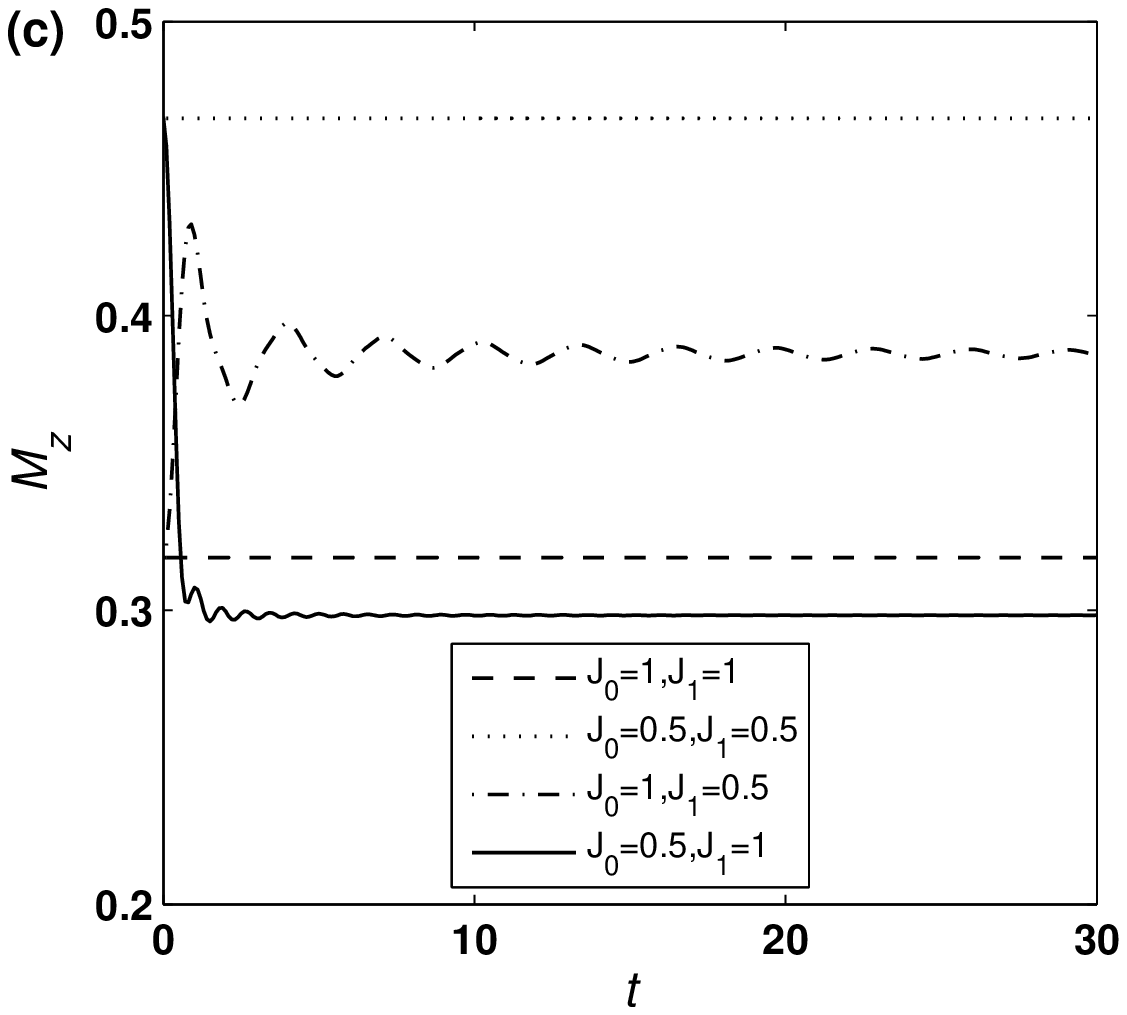}}\quad
   \subfigure{\label{fig:1d}\includegraphics[width=6cm]{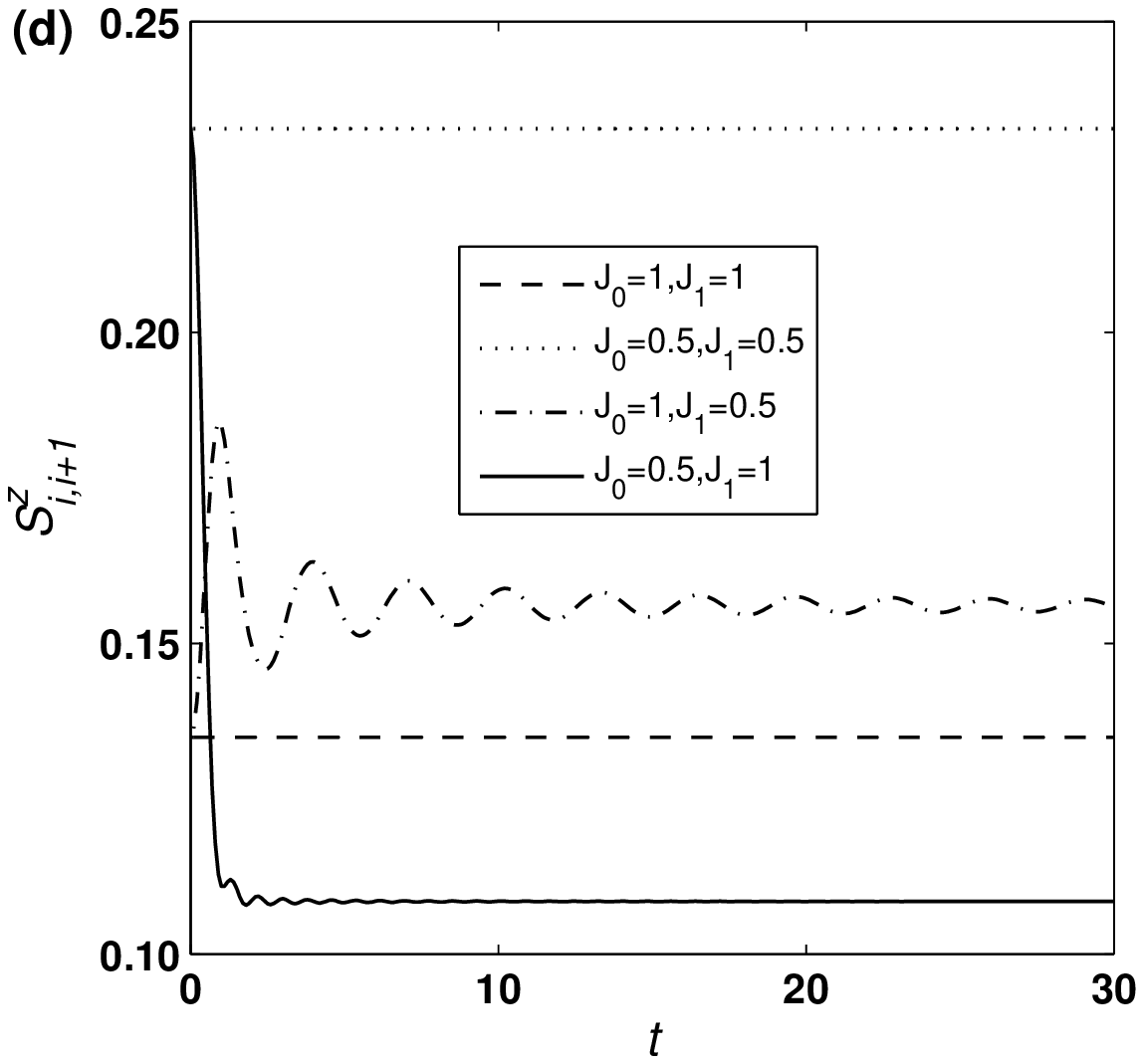}}
   \caption{{\protect\footnotesize Dynamics of the nearest-neighbor concurrence $C(i,i+1)$, where $t$ is in units of $J_{1}^{-1}$, with $\gamma=1$ and (a) $h_{0}=h_{1}=1$ for various values of $J_{0}$ and $J_{1}$ at $kT=0$; (b) $J_{0}=J_{1}=1$ for various values of $h_{0}$ and $h_{1}$ at $kT=0$. (c) Dynamics of the magnetization per spin; (d) dynamics of the spin-spin correlation function in the $z$-direction for fixed $h=h_{0}=h_{1}=1$ for various values of $J_{0}$ and $J_{1}$ at $kT=0$ with $\gamma=1$.}}
 \label{fig:1}
 \end{minipage}
\end{figure}

In Fig.~\ref{fig:1}, we explore the dynamics of the nearest-neighbor concurrence $C(i,i+1)$ at zero temperature. In Fig.~\ref{fig:1a} we choose the magnetic field to have a constant value of 1 while the coupling parameter takes the value 1 or 0.5 or a step function changing between 0.5 and 1 (or 1 and 0.5). In Fig.~\ref{fig:1b} we set the coupling parameter to be constant this time with a value 1 while the magnetic field can take the values 1 or 0.5 or a step function changing between 0.5 and 1 (or 1 and  0.5). As one can see, when the coupling parameter (the magnetic field) is a step function, the concurrence reaches a value that is neither its value when $J=J_{0}$ ($h=h_{0}$) nor $J_{1}$ ($h=h_{1}$), i.e., concurrence $C(i,i+1)$ shows a nonergodic behavior. This behavior follows from the nonergodic properties of the magnetization and the spin-spin correlation functions as reported by previous studies \cite{HuangQInfo, nonergodic_Mazur, nonergodic_Barouch}. The nonergodic behavior of the magnetization and the spin-spin correlation function in the $z$ direction are shown in Figs.~\ref{fig:1c} and \ref{fig:1d}. The spin-spin correlation functions in the $x$ and $y$-directions show similar behavior. At higher temperatures the nonergodic behavior of the system sustains but with reduced magnitude of the asymptotic concurrence (as $t \rightarrow \infty$).
\begin{figure}[htbp]
\begin{minipage}[c]{\textwidth}
 \centering 
   \subfigure{\label{fig:2a}\includegraphics[width=6cm]{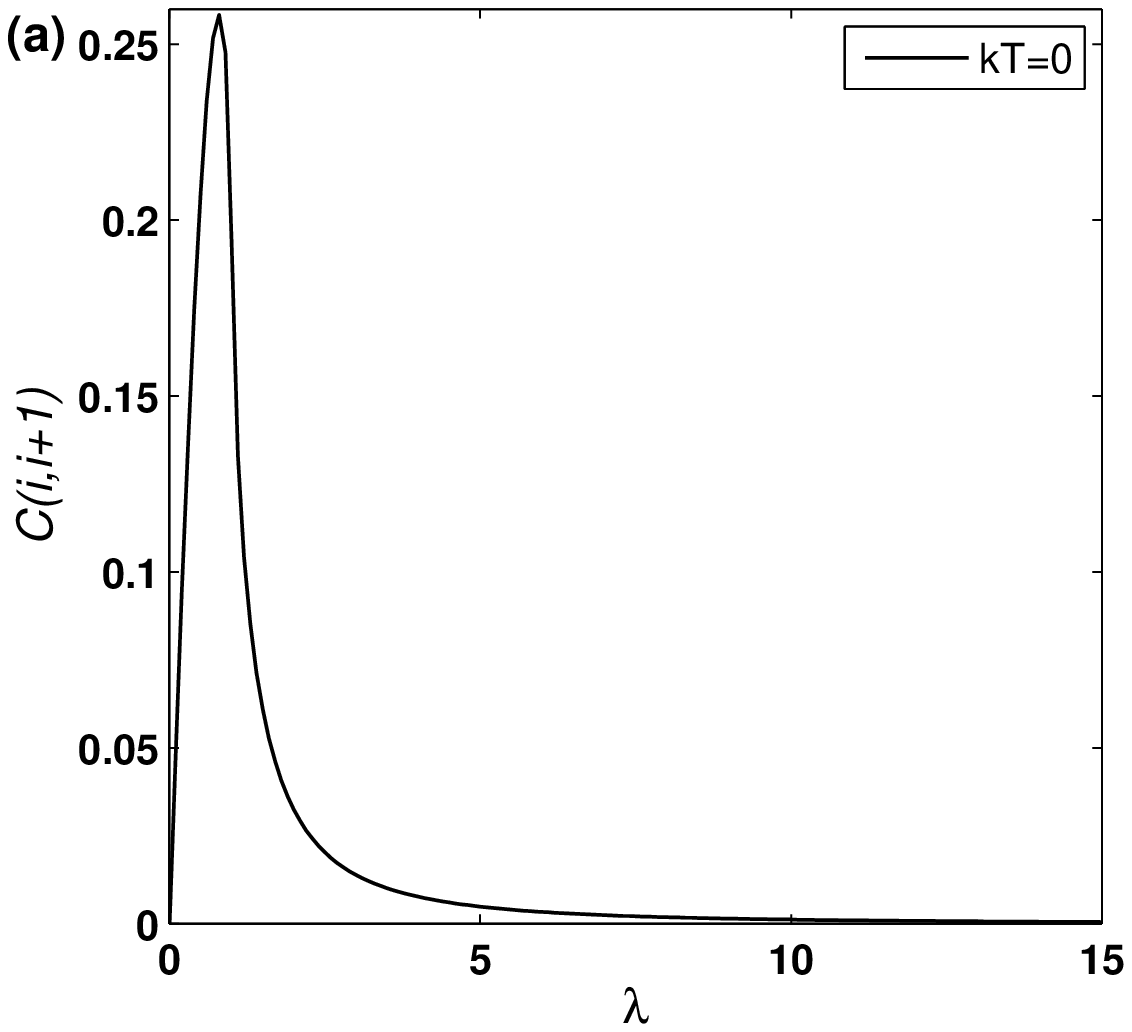}}\quad
   \subfigure{\label{fig:2b}\includegraphics[width=6cm]{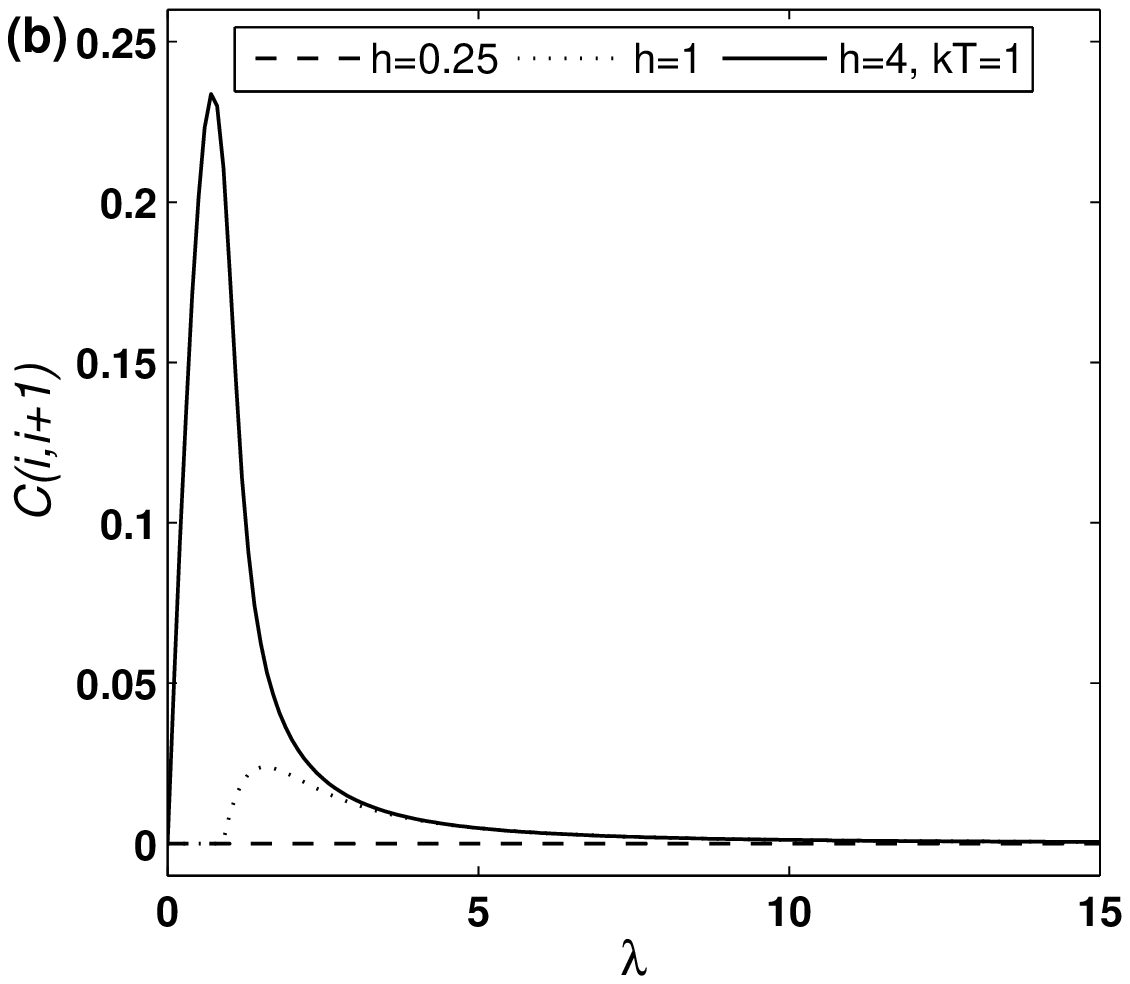}}\\
   \subfigure{\label{fig:2c}\includegraphics[width=6cm]{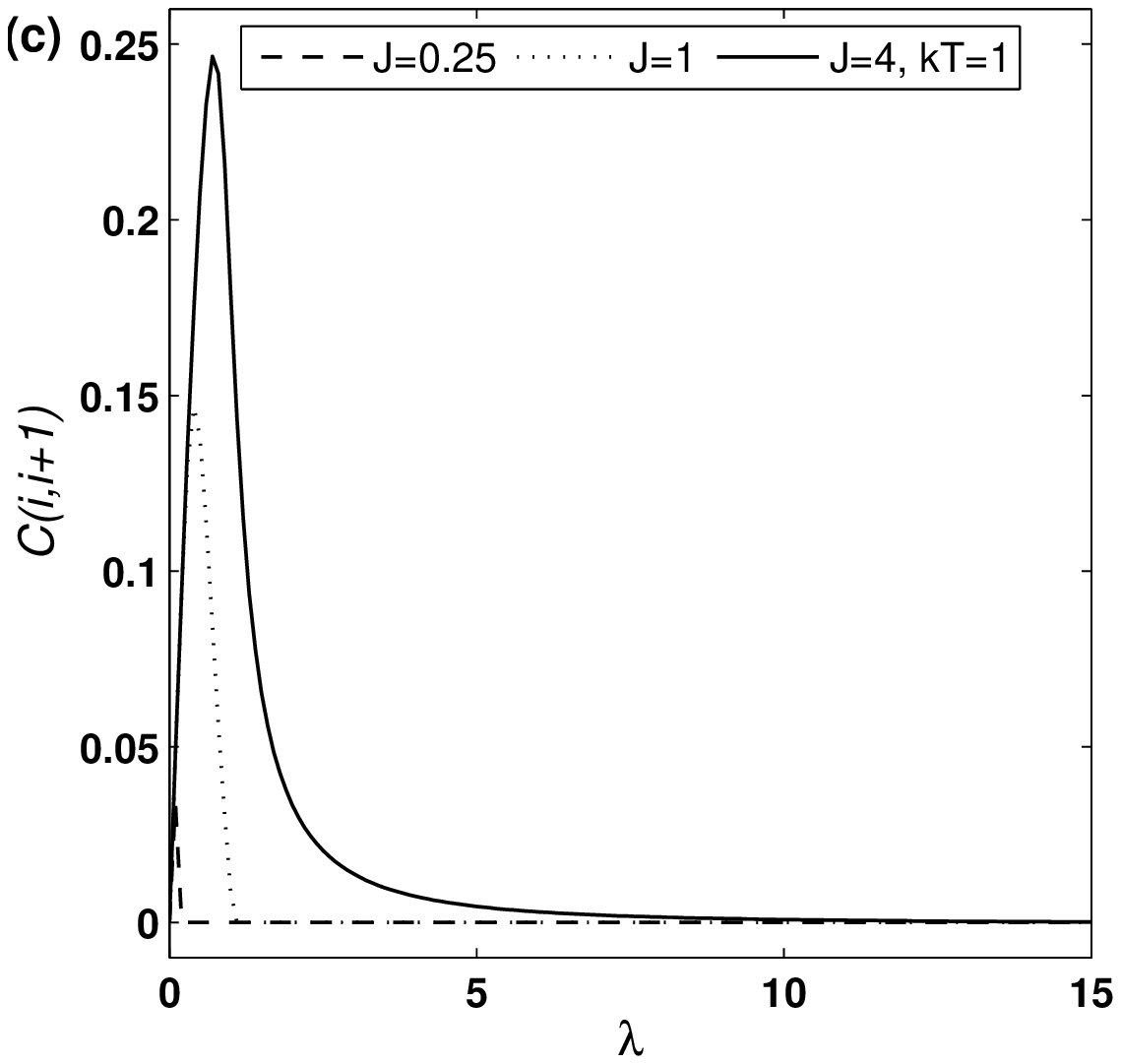}}\quad
   \subfigure{\label{fig:2d}\includegraphics[width=6cm]{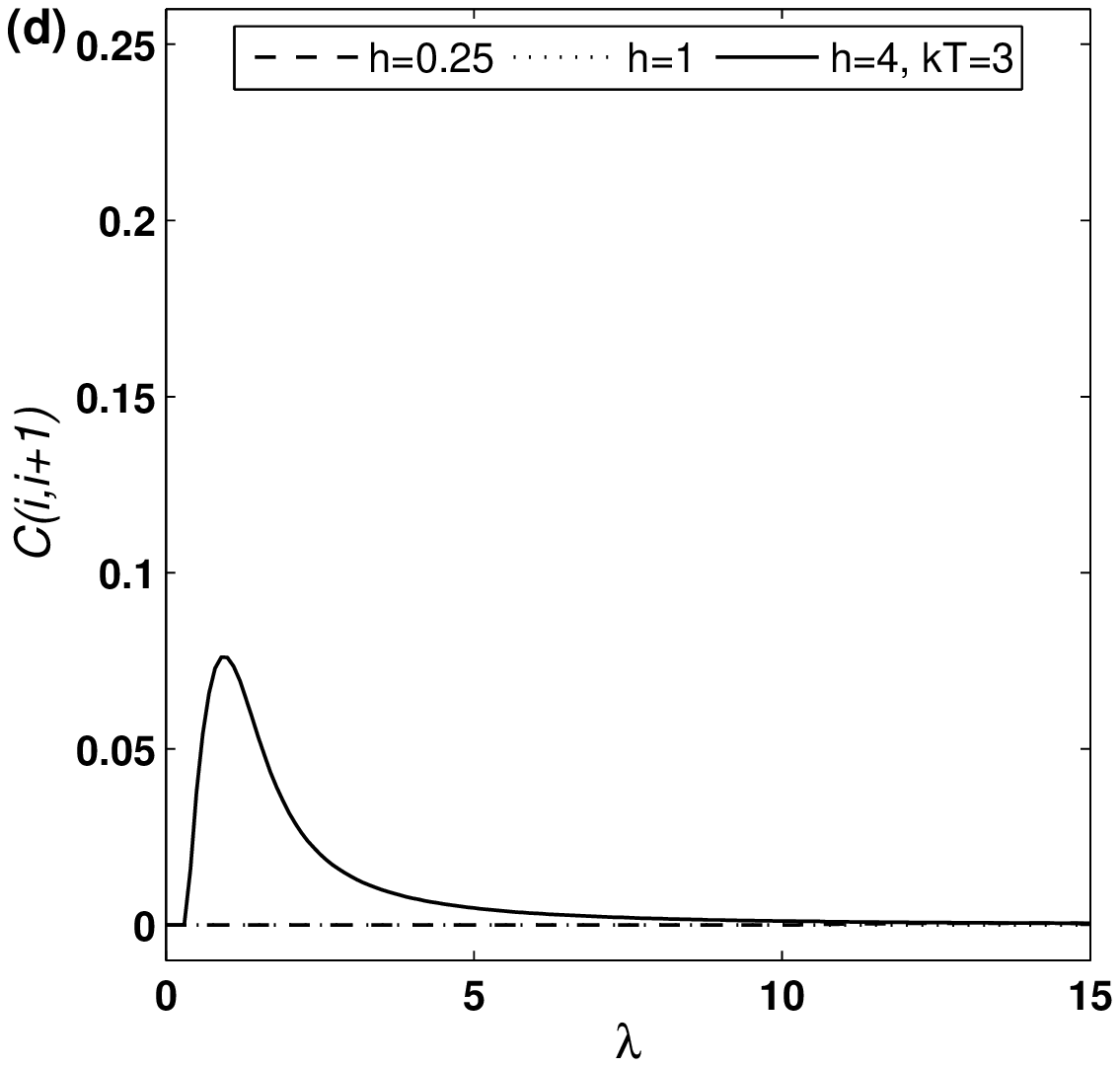}}
   \caption{{\protect\footnotesize $C(i,i+1)$ as a function of $\lambda$ for $h=h_{0}=h_{1}$ and $J=J_{0}=J_{1}$ at (a) $kT=0$ with any combination of $J$ and $h$; (b) $kT=1$ with $h_{0}=h_{1}=0.25, 1, 4$; (c) $kT=1$ with $J_{0}=J_{1}=0.25, 1, 4$; (d) $kT=3$ with $h_{0}=h_{1}=0.25, 1, 4$.}}
 \label{fig:2}
 \end{minipage}
\end{figure}

In Fig.~\ref{fig:2} we study the behavior of the nearest-neighbor concurrence $C(i,i+1)$ as a function of $\lambda$ for different values of $J$ and $h$ at different temperatures. In this figure we set $J=J_0=J_1$ and $h=h_0=h_1$. In Fig.~\ref{fig:2a}, we study the zero temperature case, where we fix $h$ (or $J$) and vary $J$ (or $h$). As one can see, the behavior of $C(i,i+1)$ depends only on the ratio $J/h$ (i.e., $\lambda$) rather than on their individual values. As expected $C(i,i+1)$ starts at zero, reaches a maximum value at $\lambda\approx\lambda_c=1$, and then vanishes for larger values of $\lambda$. Studying entanglement at nonzero temperatures shows that the maximum value of $C(i,i+1)$ decreases as the temperature increases. Furthermore, $C(i,i+1)$ shows a dependence on the individual values of $J$ and $h$, not only their ratio. In Fig.~\ref{fig:2b} one can see that increasing $h$ at $kT=1$ causes the maximum concurrence to increase and to shift toward smaller values of $\lambda$. In Fig.~\ref{fig:2c}, increasing $J$ at $kT=1$ causes the maximum concurrence to decrease and to move toward larger values of $\lambda$. Figure~\ref{fig:2d} shows the significant reduction in the entanglement as the temperature increases further ($kT=3$).
 
\begin{figure}[htbp]
\begin{minipage}[c]{\textwidth}
 \centering 
   \subfigure{\label{fig:3a}\includegraphics[width=6cm]{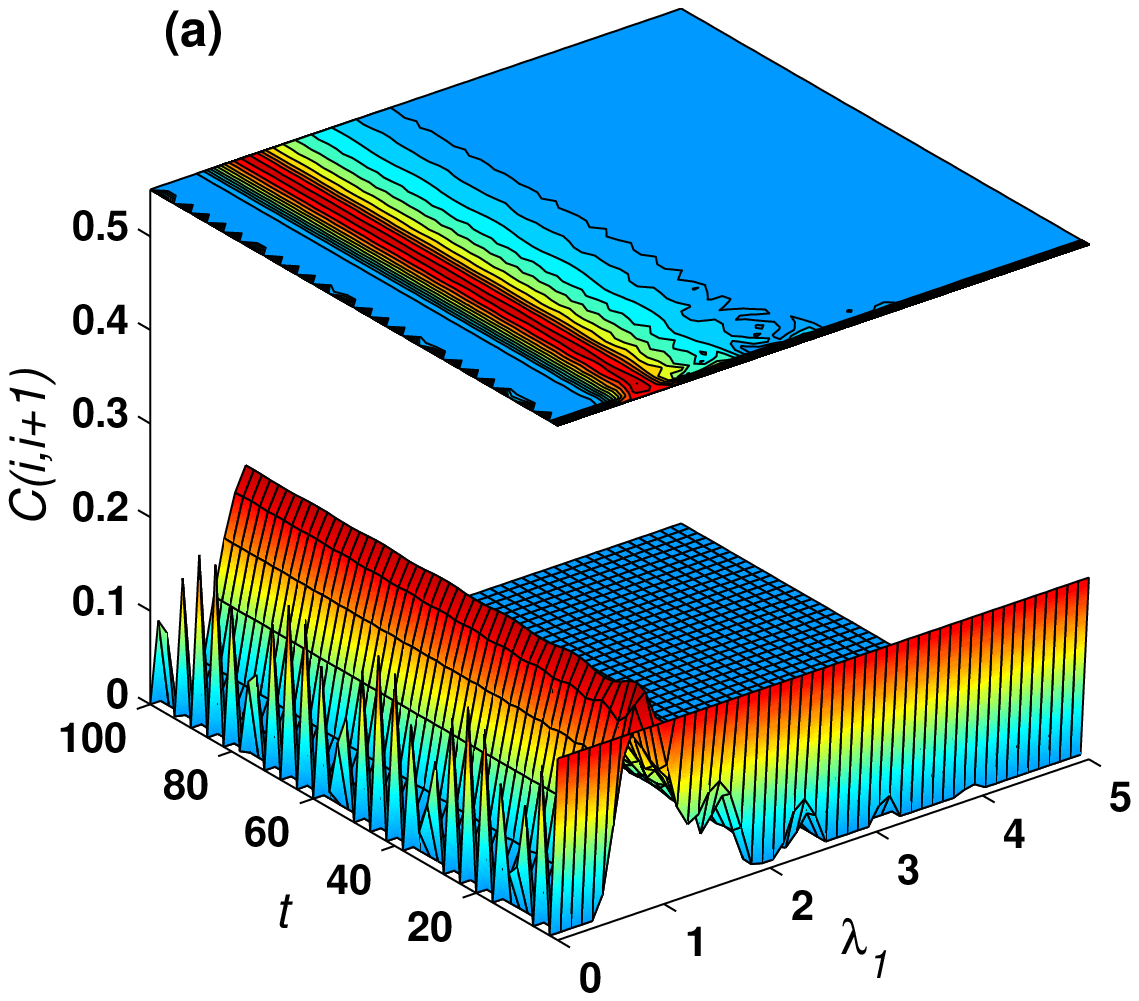}}\quad
   \subfigure{\label{fig:3b}\includegraphics[width=6cm]{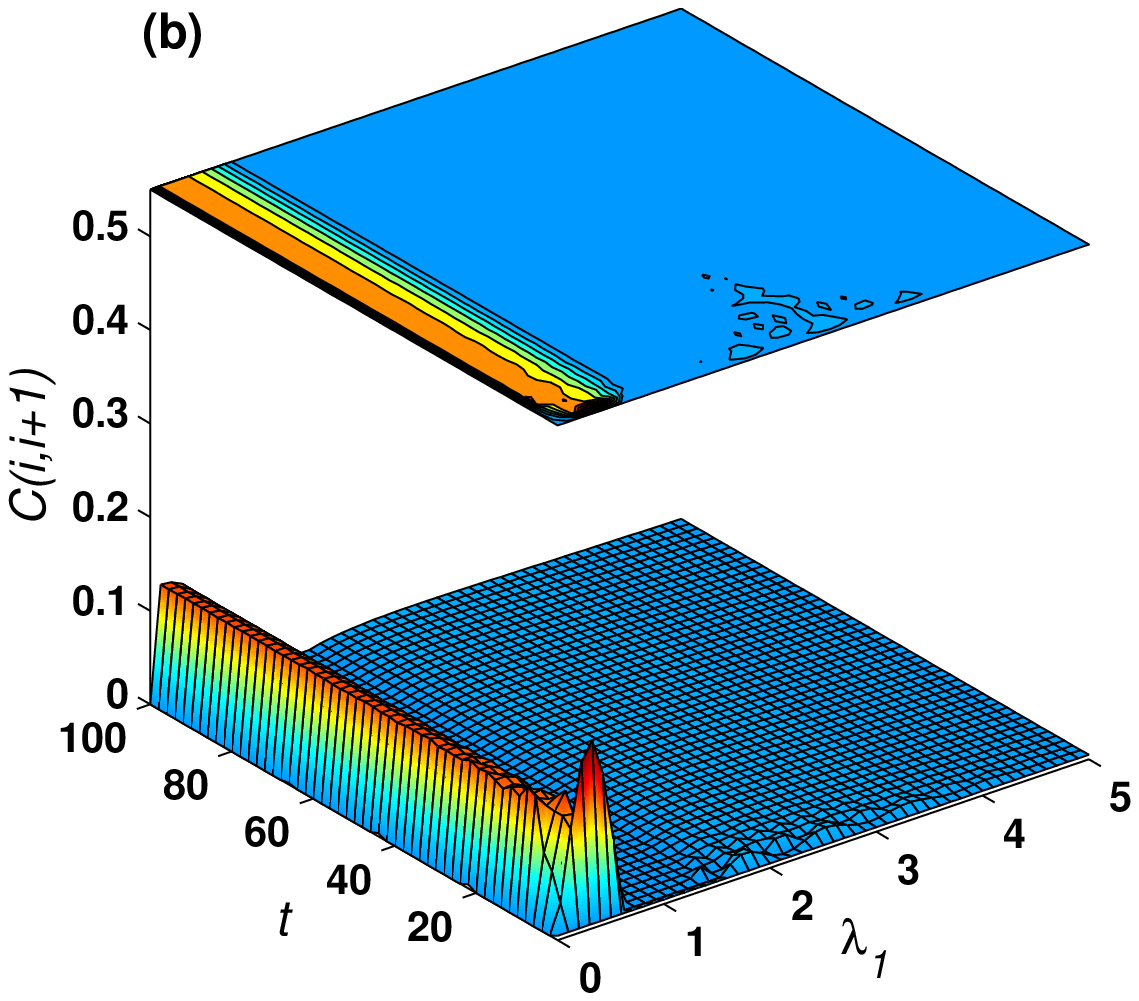}}\\
   \subfigure{\label{fig:3c}\includegraphics[width=6cm]{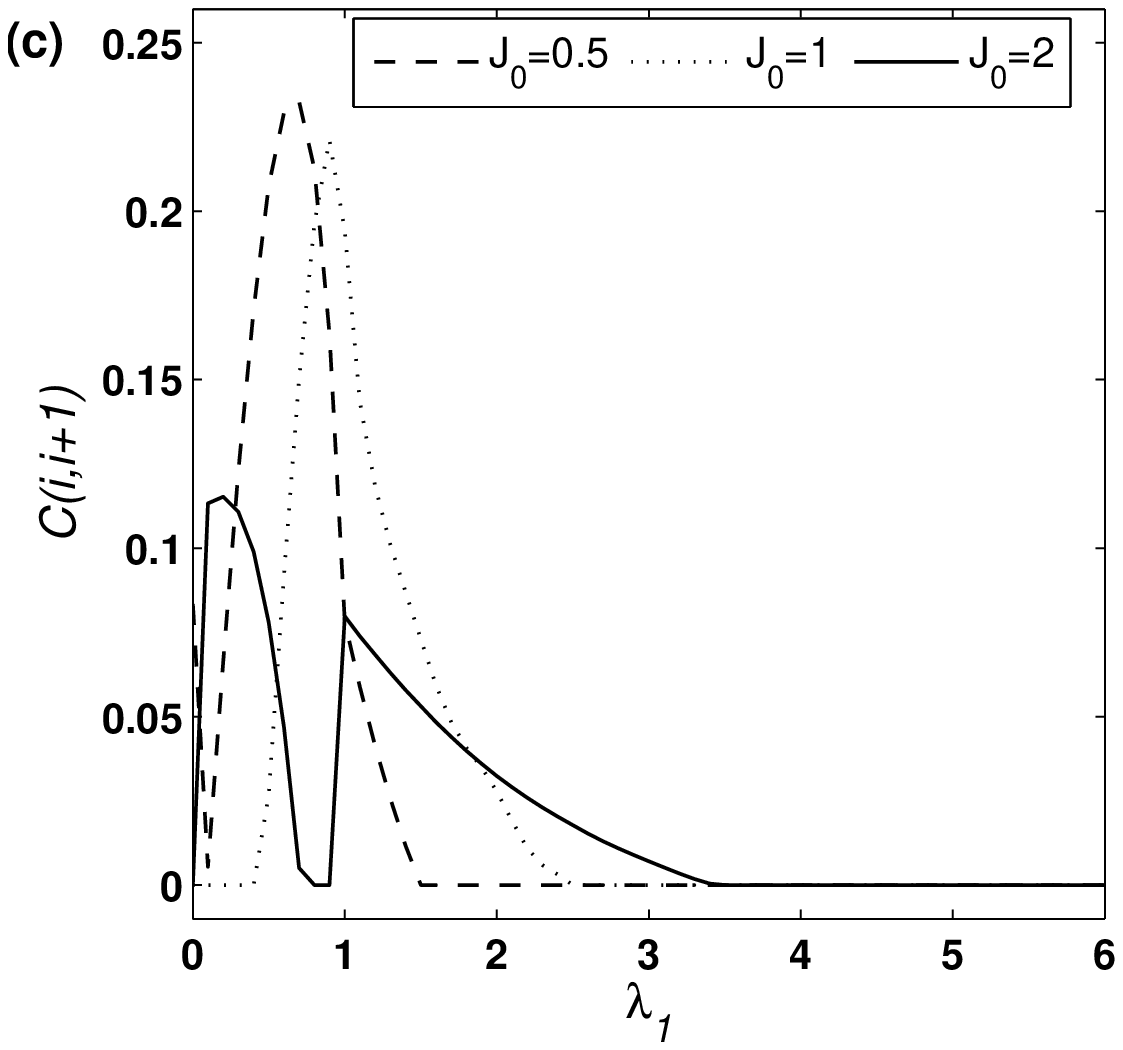}}\quad
   \subfigure{\label{fig:3d}\includegraphics[width=6cm]{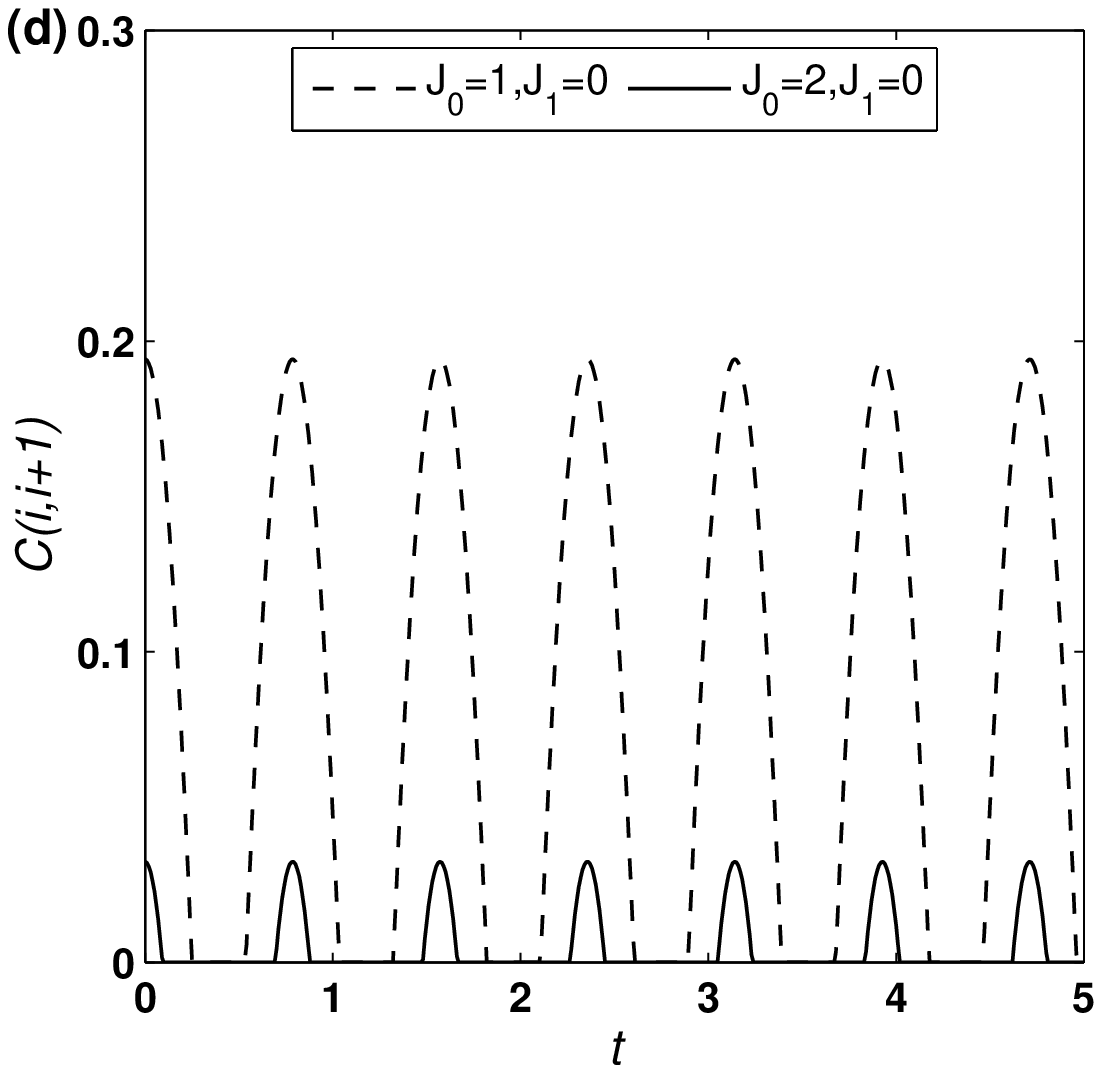}}
   \caption{{\protect\footnotesize (Color online) $C(i,i+1)$ as a function of $\lambda_{1}$ and $t$, in units of $J_{1}^{-1}$, at $kT=0$ with $\gamma=1$, $h_{0}=h_{1}=1$, and (a) $J_{0}=1$; (b) $J_{0}=5$. (c) The asymptotic behavior of $C(i,i+1)$ as a function of $\lambda_{1}$ with $\gamma=1$, $h_{0}=h_{1}=1$, and $J_{0}=0.5, 1, 2$ at $kT=0$; (d) dynamics of $C(i,i+1)$ with $\gamma=1$ , $h_{0}=h_{1}=1$, $J_{0}=1,2$, and $J_{1}=0$ at $kT=0$.}}
 \label{fig:3}
 \end{minipage}
\end{figure}

In Fig.~\ref{fig:3} we investigate the dynamics of entanglement as a function of the coupling parameter $J(t)$. We plot $C(i,i+1)$ as a function of time and $\lambda_1 (\equiv J_1/h_1)$ where we set $h=h_0=h_1$. In Fig.~\ref{fig:3a} we set the parameter values as $h=1$, $J_{0}=1$, and $kT=0$, i.e., the system is initially prepared in a state of maximum entanglement ($\lambda_0=\lambda_c$). As one can see, for zero $\lambda_1$, the entanglement shows an oscillatory behavior in time, where the spins precess about the magnetic field in the $z$ direction, and its magnitude increases as we increase $\lambda_1$ until it reaches its maximum value close to $\lambda_c$. As $\lambda_1$ exceeds $\lambda_c$ the entanglement decreases and eventually vanishes for $\lambda_1 > 2.5$ where in this case $J$ is dominating over $h$ and the spins are completely aligned in the $x$ direction. On the other hand, when the system is prepared in an initial state with $\lambda_0$ different from $\lambda_c$, the maximum entanglement it can reach is much lower than the previous case and appears at lower value of $\lambda_1 \approx 0.5$ as shown in Fig.~\ref{fig:3b} where $h=1$ and $J_{0}=5$. In Fig.~\ref{fig:3c} we exploit the asymptotic concurrence (as $t \rightarrow \infty$) versus $\lambda_1$ for different values of $J_0$ while the magnetic field is set as $h_0=h_1=1$. Clearly, the asymptotic value of the concurrence varies significantly depending on the initial values of the parameters. For $\lambda_0 < \lambda_c$ the asymptotic concurrence peaks at $\lambda_1 \approx 0.5$, and for $\lambda_0 = \lambda_c$ the peak takes place at $\lambda_1 \approx 1$. For higher values of the parameter $\lambda_0 > \lambda_c$ two peaks show up with a smaller second peak that decreases and shifts toward larger values of $\lambda_{1}$ as $\lambda_{0}$ increases. For instance, for $\lambda_0=2$ the $\lambda_1$ values at the peaks are 0.2 and 1 with much lower maximum values of concurrence than the previous cases. In Fig.~\ref{fig:3d} we examine the dynamical behavior of the concurrence as the coupling parameter is switched off, i.e., $J_{0}$ has a finite value while $J_{1}=0$. As one can see, the oscillation sustains as time elapses. Testing different values of $h$ and $J_{0}$ we find that the oscillation amplitude is largest when $0.5 \leq \lambda_{0} \leq 1$ while it almost vanishes and loses uniformity when $\lambda_{0} > 2$. The period of the oscillation decreases as $h$ increases and is independent of $J_{0}$. We observe a similar behavior for the concurrence when the magnetic field is switched off while setting $J_{0}=J_{1}$.
\begin{figure}[htbp]
\begin{minipage}[c]{\textwidth}
 \centering
   \subfigure{\label{fig:4a}\includegraphics[width=6cm]{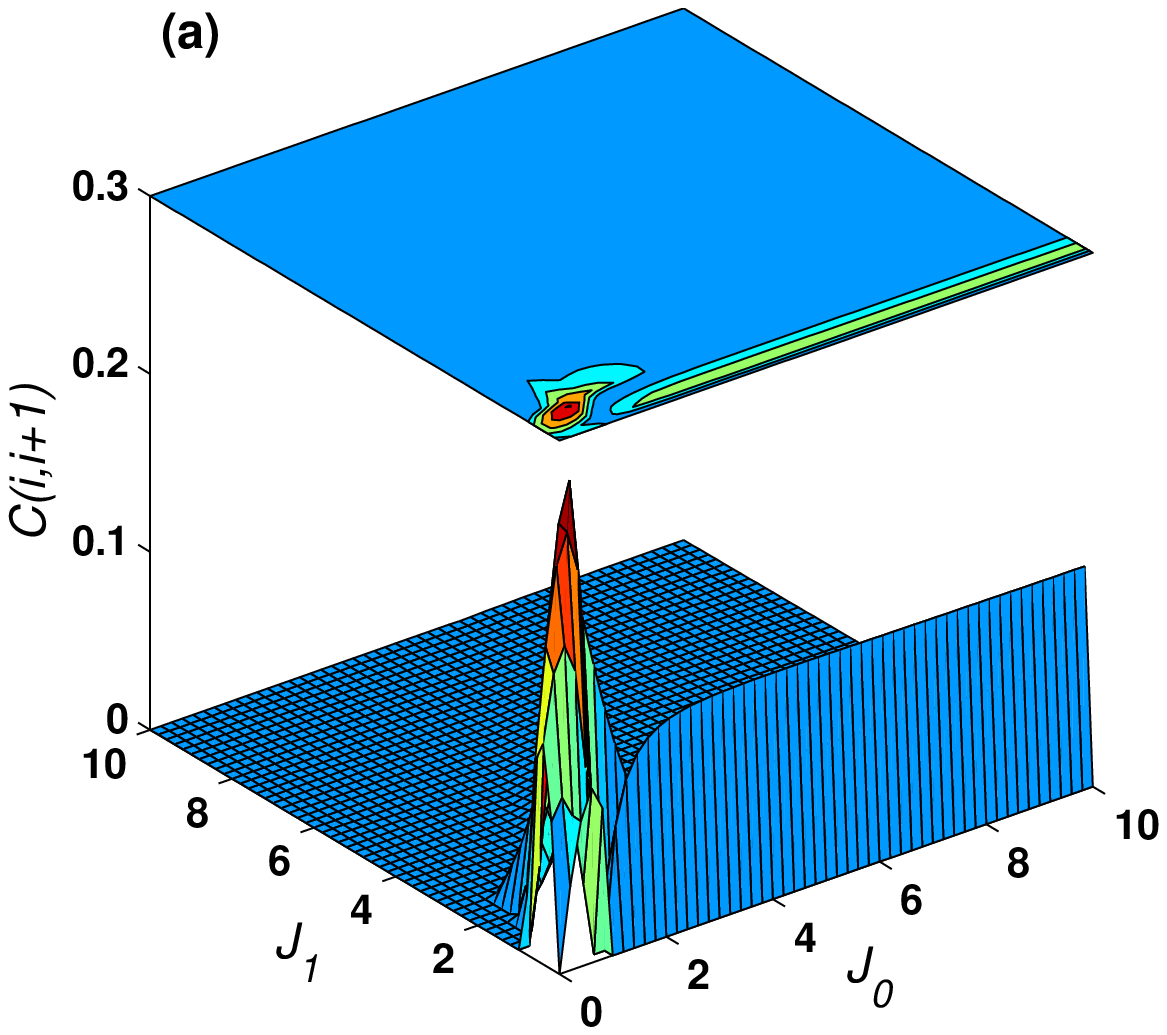}}\quad
   \subfigure{\label{fig:4b}\includegraphics[width=6cm]{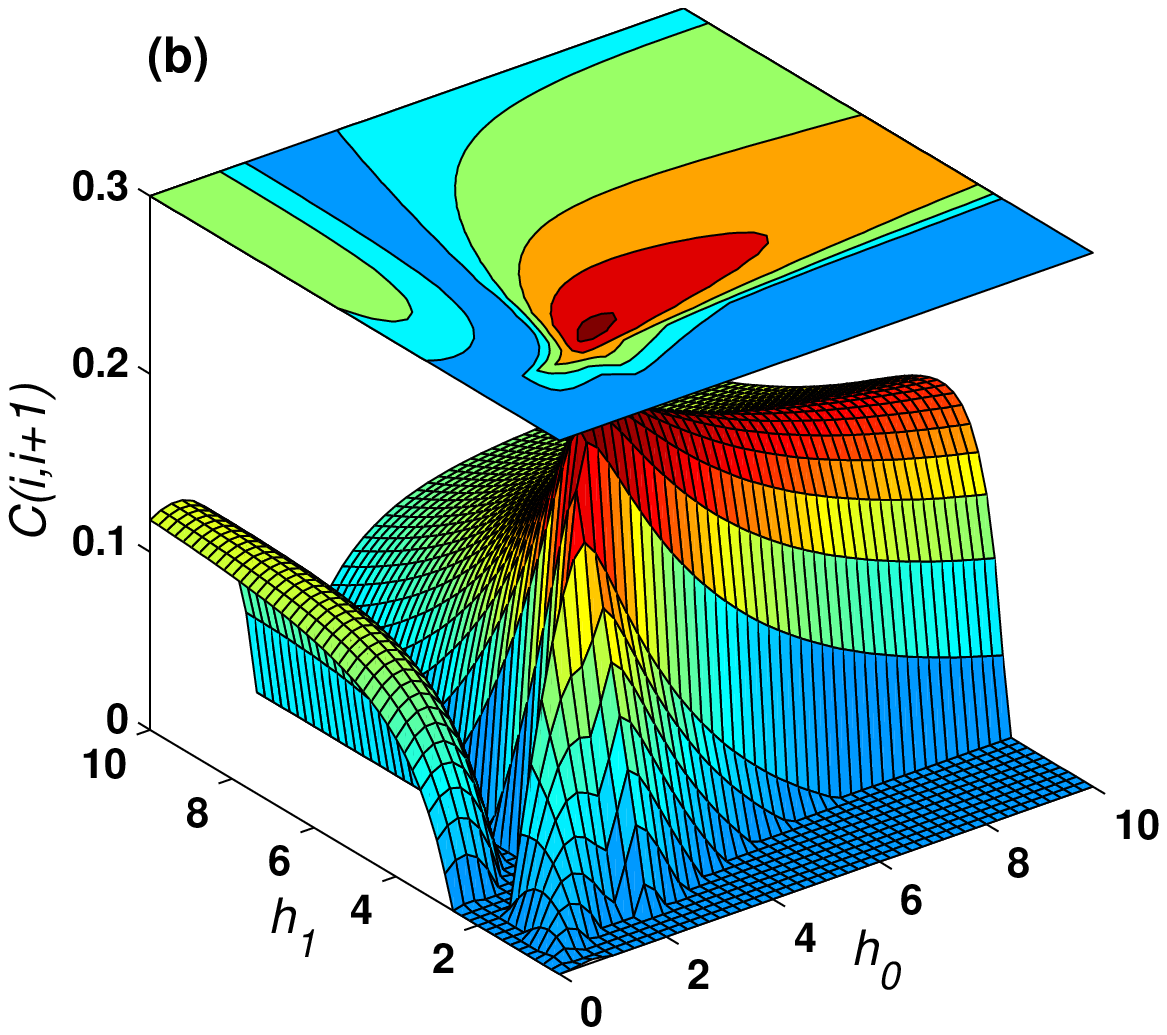}}
   \subfigure{\label{fig:4c}\includegraphics[width=6cm]{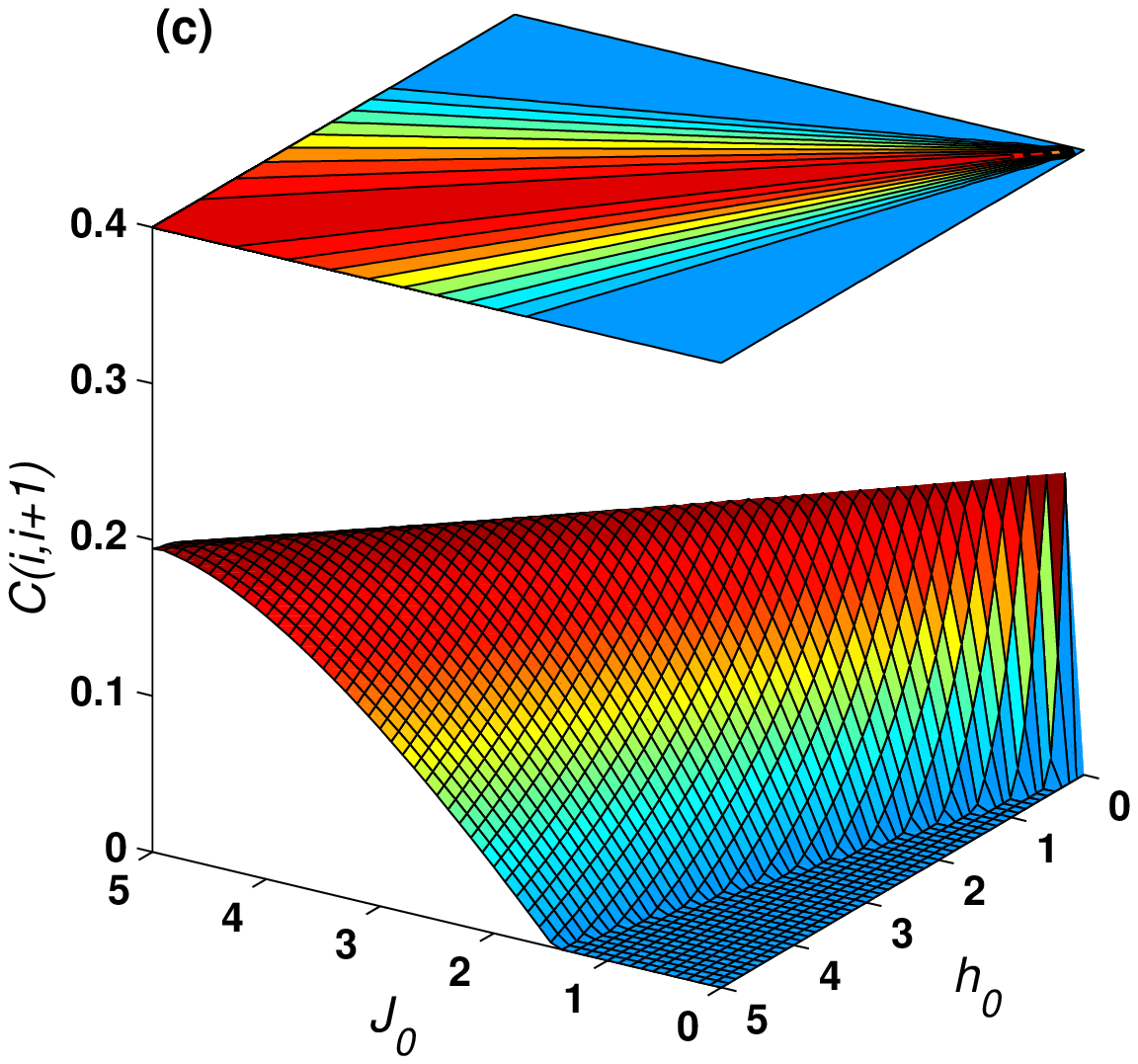}}\quad
		\subfigure{\label{fig:4d}\includegraphics[width=6cm]{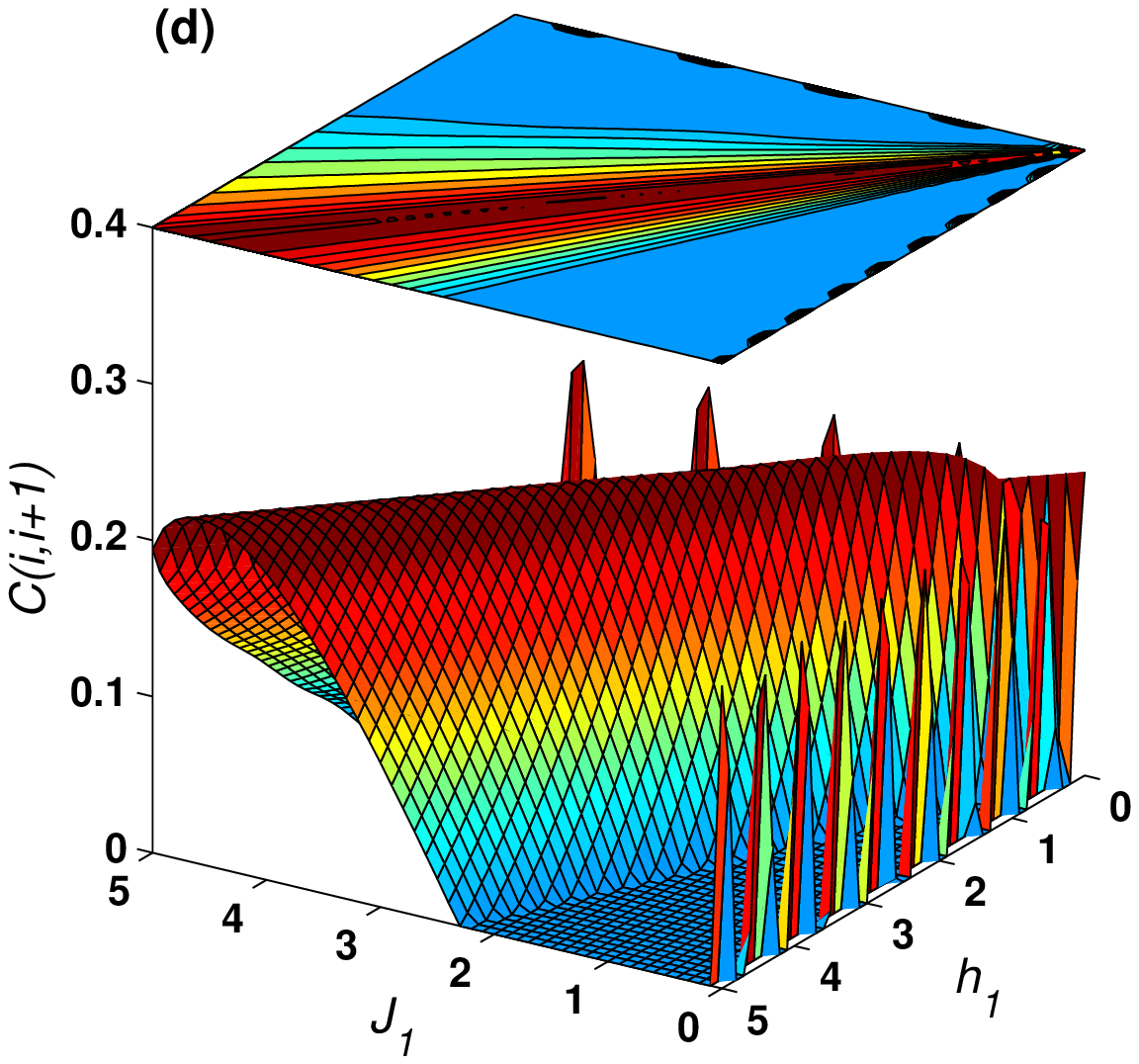}}
      \caption{{\protect\footnotesize (Color online) The asymptotic behavior of $C(i,i+1)$ as a function of (a) $J_{0}$ and $J_{1}$ with $\gamma=1$ and $h_{0}=h_{1}=1$; (b) $h_{0}$ and $h_{1}$ at $kT=0$ with $\gamma=1$ and $J_{0}=J_{1}=2$; (c) $h_{0}$ and $J_{0}$ at $kT=0$ with $\gamma=1$ and $h_{1}=J_{1}=1$; (d) $h_{1}$ and $J_{1}$ at $kT=0$ with $\gamma=1$ and $h_{0}=J_{0}=1$, where $h_{0}$, $h_{1}$, and $J_{0}$ are in units of $J_{1}$.}}
 \label{fig:4}
 \end{minipage}
\end{figure}

In Fig.~\ref{fig:4} we manifest the dependence of the asymptotic behavior (as $t \rightarrow \infty$) of the nearest-neighbor concurrence on the magnetic field and coupling parameters $h_0$, $h_1$, $J_0$, and $J_1$ at zero temperature. In Fig.~\ref{fig:4a} we present a three-dimensional plot for the concurrence versus $J_0$ and $J_1$ where we set the magnetic field at $h_0=h_1=1$. The concurrence starts with a zero value for $J_0=J_1=0$ and increases abruptly, reaching a maximum value of approximately $0.26$ at $J_{0}=J_{1}\approx 0.88$. For $J_{1} > 4$, $C(i,i+1)$ vanishes for all $J_{0}$ values. It is interesting to see that for all initial values $J_0 > 1$ and $J_1 < 1$, the asymptotic concurrence has a finite value which decays for higher values of $J_1$. This emphasis that starting with a finite coupling $J_0$ and reducing it to a very small value, $J_1$, leads to persisting entanglement in the system. Figure~\ref{fig:4b} shows the asymptotic behavior of the nearest-neighbor concurrence as a function of $h_{0}$ and $h_{1}$, while fixing the coupling parameter at $J_{0}=J_{1}=2$. The concurrence is zero at $h_1=h_0=0$ and increases as both increase until it reaches its maximum value at $h_0=h_{1}\approx 1.8$, i.e., close to the $\lambda_c$. If we start with a relatively large magnetic field (say $h_{0}>3$), $C(i,i+1)$ will vanish for $h_{1}<1$, reach a maximum value at $h_{1} \approx 2$, and then decrease gradually with further increase of $h_{1}$. However, if we start with a smaller magnetic field ($1<h_{0}<2$), $C(i,i+1)$ will have a maximum value if $h_{1}$ is kept within the same range and will vanish when $h_{1}$ is increased. Finally, if we start with a much smaller magnetic field ($h_{0}<<1$), $C(i,i+1)$ vanishes for $h_{1} <1.5$ but increases as $h_1$ increases and reaches a plateau, i.e., there will be a finite asymptotic concurrence left in the system for a very small initial magnetic field and large final one.
The asymptotic behavior of $C(i,i+1)$ as a function of $J_{0}$ and $h_{0}$ at $kT=0$ is illustrated in Fig.~\ref{fig:4c}. We studied this behavior for $J_{1}=h_{1}=0.5$, $J_{1}=h_{1}=1$, and $J_{1}=h_{1}=2$ and in all cases we got the same behavior. The largest entanglement is reached when $J_{0}=h_{0}$ but as $J_{0}$ differs from $h_{0}$, $C(i,i+1)$ decays in agreement with the physical interpretation discussed previously. We also study the asymptotic behavior of $C(i,i+1)$ as a function of $J_{1}$ and $h_{1}$ at $kT=0$ while setting $J_{0}=h_{0}=0.5, 1$, and $2$. Again, we get the same behavior in all three cases as shown in Fig.~\ref{fig:4d}. The largest entanglement is reached at $J_{1}=h_{1}$ but as $J_{1}$ diverges from $h_{1}$, $C(i,i+1)$ decreases. The oscillation appearing at $J_{1}=0$ indicates that the value of $h_{1}$ changes the phase of the oscillation.
\begin{figure}[htbp]
\begin{minipage}[c]{\textwidth}
 \centering 
   \subfigure{\label{fig:5a}\includegraphics[width=6cm]{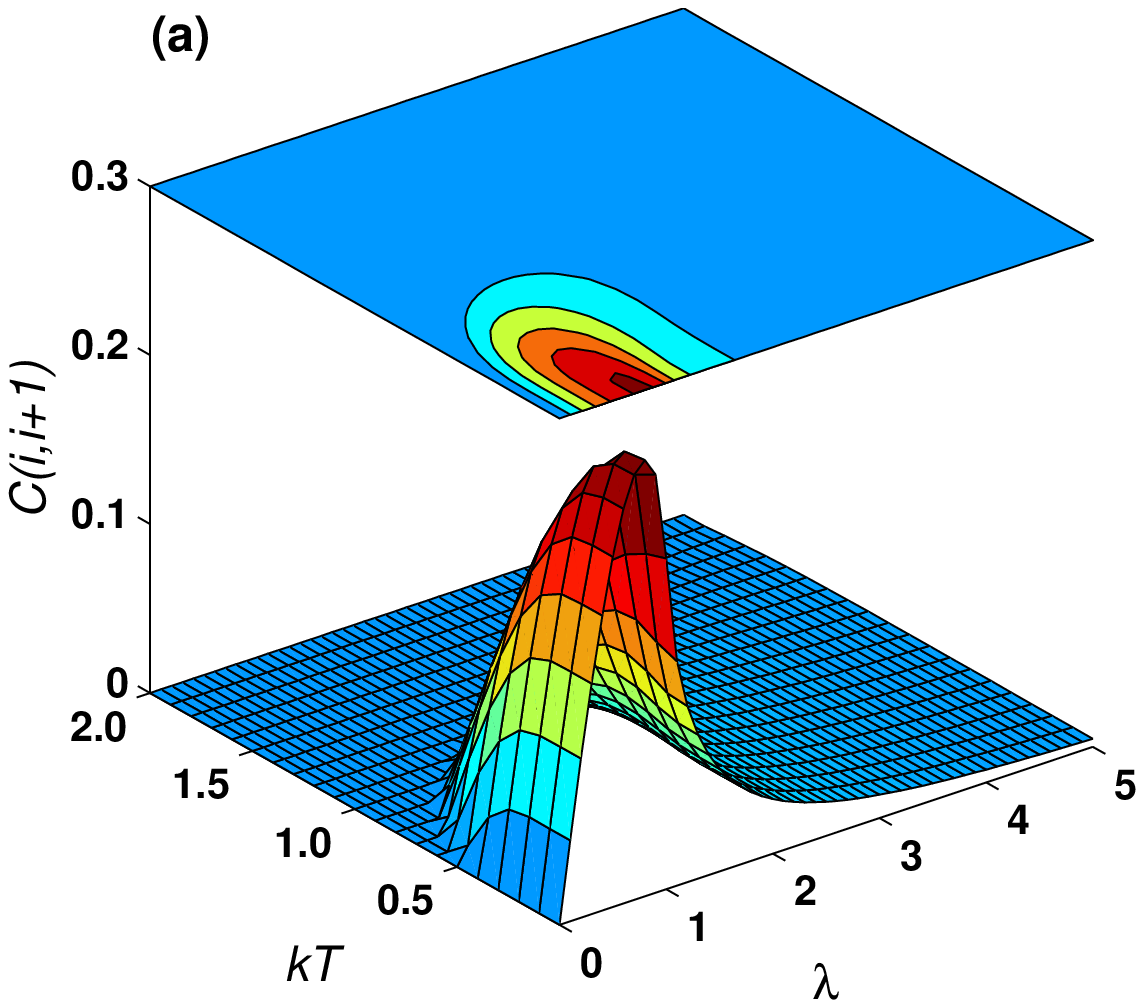}}\quad
   \subfigure{\label{fig:5b}\includegraphics[width=6cm]{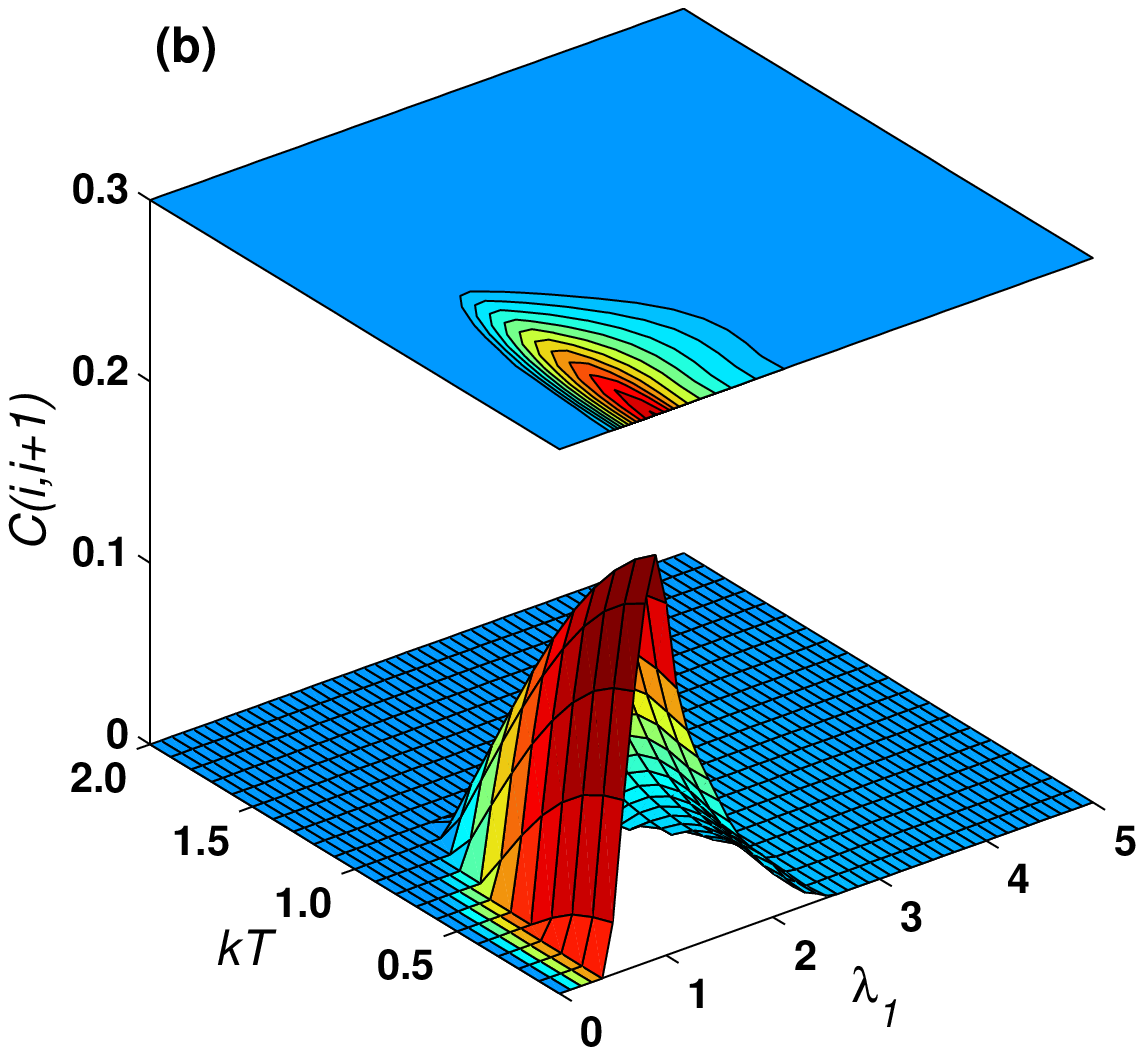}}
      \caption{{\protect\footnotesize (Color online) The asymptotic behavior of $C(i,i+1)$ as a function of (a) $\lambda$ and $kT$, in units of $J_1$, with $\gamma=1$, $h_{0}=h_{1}=1$, and $J_{0}=J_{1}$; (b) $\lambda_{1}$ and $kT$ with $\gamma=1$, $h_{0}=h_{1}=1$, and $J_{0}=1$.}}
 \label{fig:5}
 \end{minipage}
\end{figure}

There has been great interest in investigating the effect of temperature on the quantum entanglement and the critical behavior of many body systems and particularly spin systems \cite{Osborne-QPT},\cite{NielsenPhD}-\cite{Gunlycke}. Osborne and Nielsen have studied the persistence of quantum effects in the thermal state of the transverse Ising model as temperature increases \cite{Osborne-QPT}. They found that the largest amount of entanglement in the system takes place in the region of the parameter space close to the critical point $\lambda_c=1$. This means that at low temperatures the quantum effects are still very relevant to the system as manifested by entanglement. Here we investigate the persistence of quantum effects under both temperature and time evolution of the system in presence of the time-dependent coupling and magnetic field.

In Fig.~\ref{fig:5a} we reproduce, using our model, the behavior of $C(i,i+1)$ as a function of $\lambda$ and $kT$ for the static case where $\lambda=\lambda_{0}=\lambda_{1}$ and $h_{0}=h_{1}=1$. As one can see, the entanglement is maximum in the vicinity of the critical point $\lambda_c =1$ and the temperature $kT=0$. As the temperature increases or $\lambda$ diverges from the critical value, the entanglement decays rapidly as the thermal fluctuations destroy the quantum aspects of the system.
The asymptotic behavior, as $t \rightarrow \infty$, of the nearest-neighbor concurrence $C(i,i+1)$ as a function of $\lambda_{1}$ and $kT$, while fixing the parameters $J_{0}=1$ and $h_{0}=h_{1}=1$, is depicted in Fig.~\ref{fig:5b}. Interestingly, the entanglement shows a very similar profile to that it manifested in the static case, i.e., the system evolves in time preserving its quantum character in the vicinity of the critical point and $kT=0$ under the time varying coupling. Studying this behavior at different values of $J_0$ and shows that the threshold temperature, at which $C(i,i+1)$ vanishes, increases as $\lambda_{0}$ increases.
\begin{figure}[htbp]
\begin{minipage}[c]{\textwidth}
 \centering 
  \includegraphics[width=6cm]{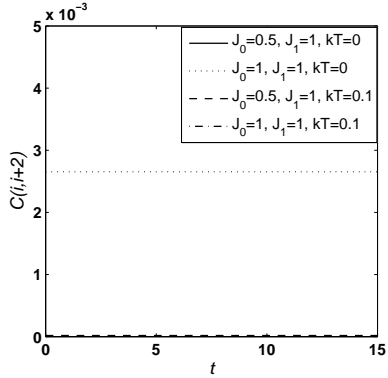}
   \caption{{\protect\footnotesize Dynamics of $C(i,i+2)$, where $t$ is in units of $J_{1}^{-1}$, with $\gamma=1$ and $h_{0}=h_{1}=1$ for various values of $J_{0}, J_{1}$ at $kT=0$ and $kT=0.1$.}}
 \label{fig:6}
 \end{minipage}
\end{figure}
Finally, we explore the evolution of next-to-nearest-neighbor concurrence $C(i,i+2)$, as shown in Fig.~\ref{fig:6}. As can be noted, $C(i,i+2)<<C(i,i+1)$ at the same circumstances and vanishes for nonzero temperature. Longer-range concurrence $C(i,i+r)$ for $r\geq 3$ was found to vanish even at zero temperature.
\section{Partially Anisotropic XY Model}
We now turn to the partially anisotropic system where $\gamma=0.5$. In this case the $x$-component of the coupling is triple its $y$ component (i.e., $J_{x} = 3 J_{y}$) and the Hamiltonian takes the form 
\begin{equation}
H=-\frac{3 J(t)}{4} \sum_{i=1}^{N} \sigma_{i}^{x} \sigma_{i+1}^{x}-\frac{J(t)}{4} \sum_{i=1}^{N} \sigma_{i}^{y} \sigma_{i+1}^{y}- \sum_{i=1}^{N} h(t) \sigma_{i}^{z}\; .
\label{eq:H_partially}
\end{equation}
\begin{figure}[htbp]
\begin{minipage}[c]{\textwidth}
 \centering 
   \subfigure{\label{fig:7a}\includegraphics[width=6cm]{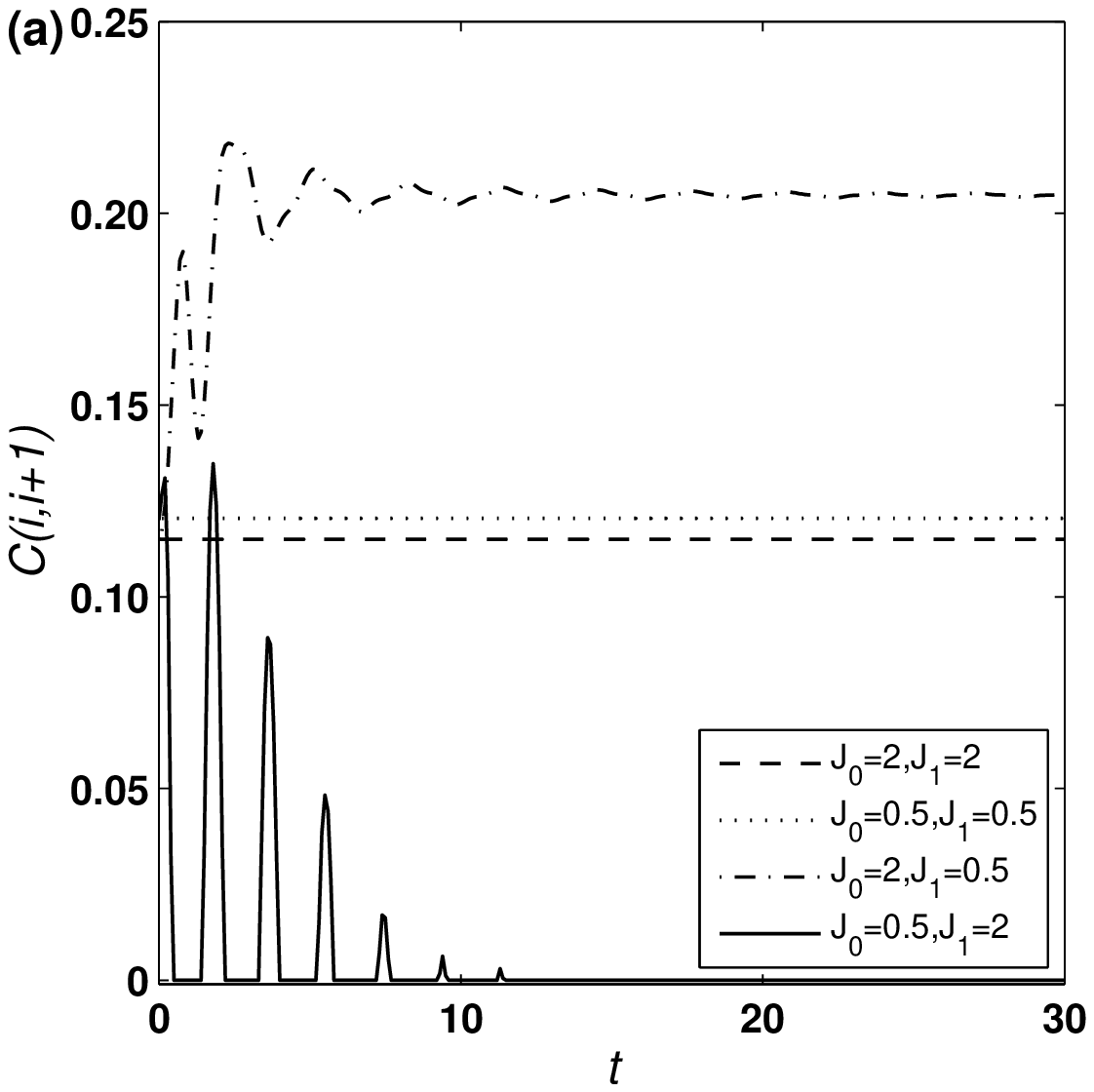}}\quad
   \subfigure{\label{fig:7b}\includegraphics[width=6cm]{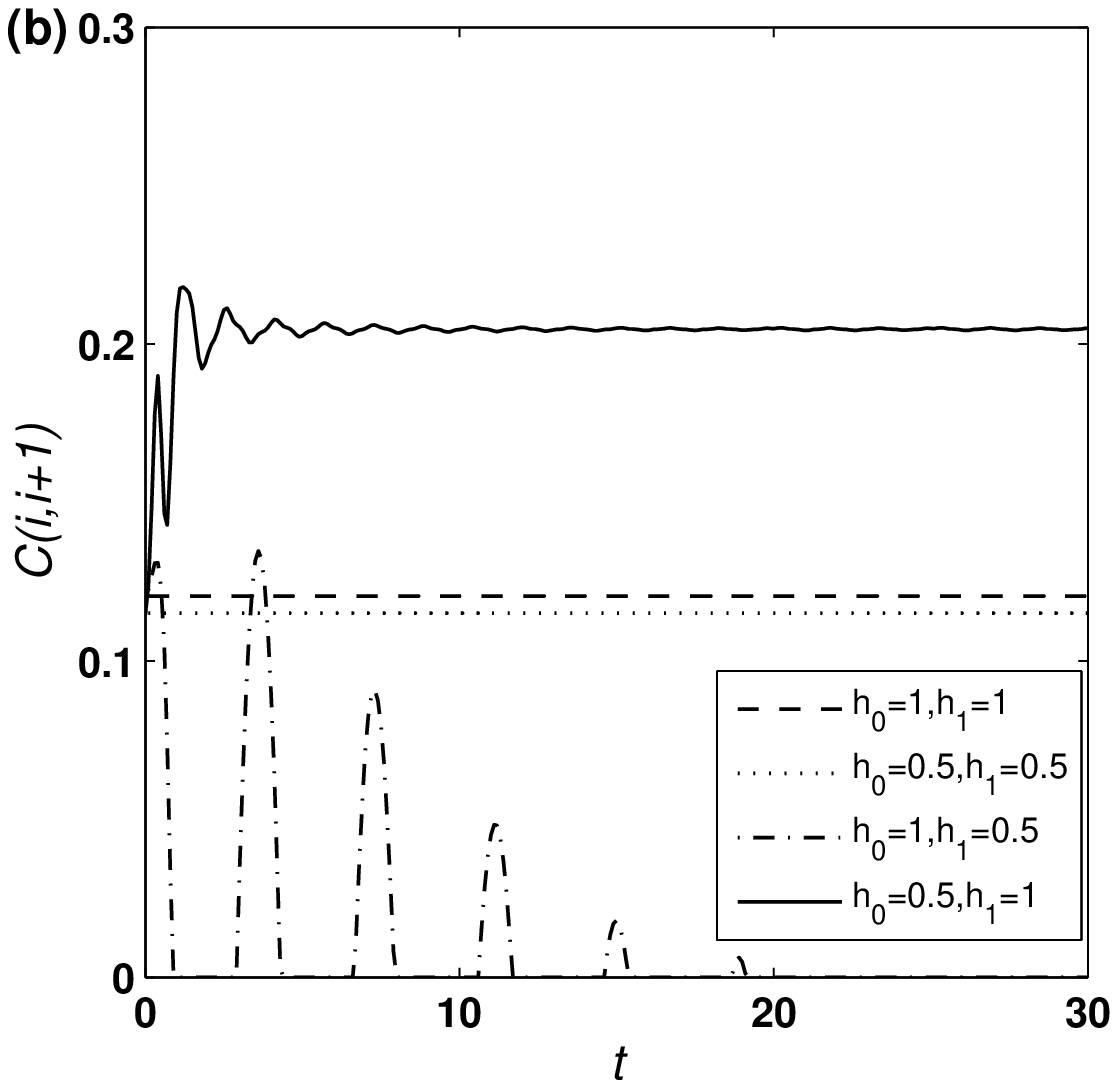}}\\
   \subfigure{\label{fig:7c}\includegraphics[width=6cm]{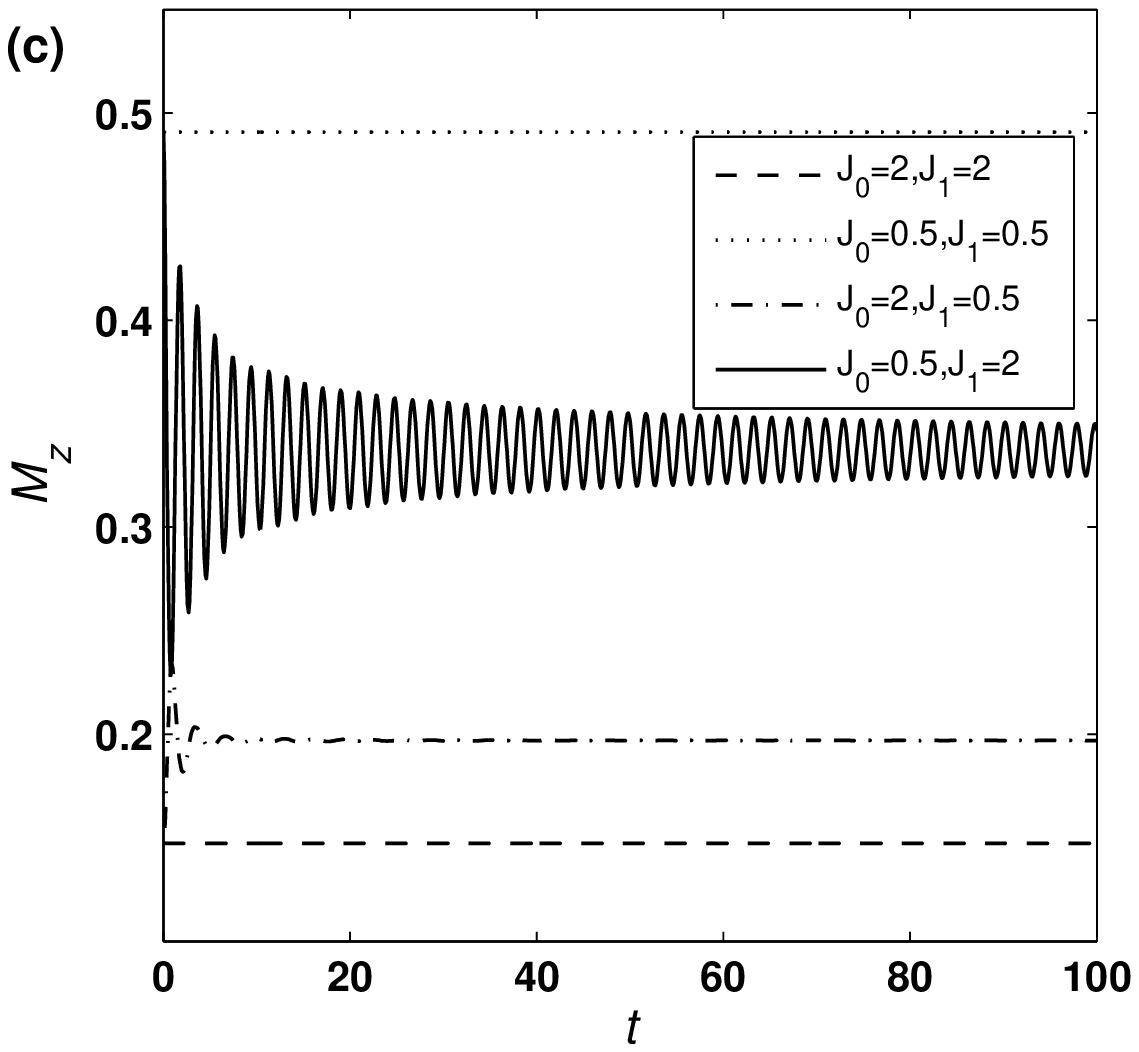}}\quad
   \subfigure{\label{fig:7d}\includegraphics[width=6cm]{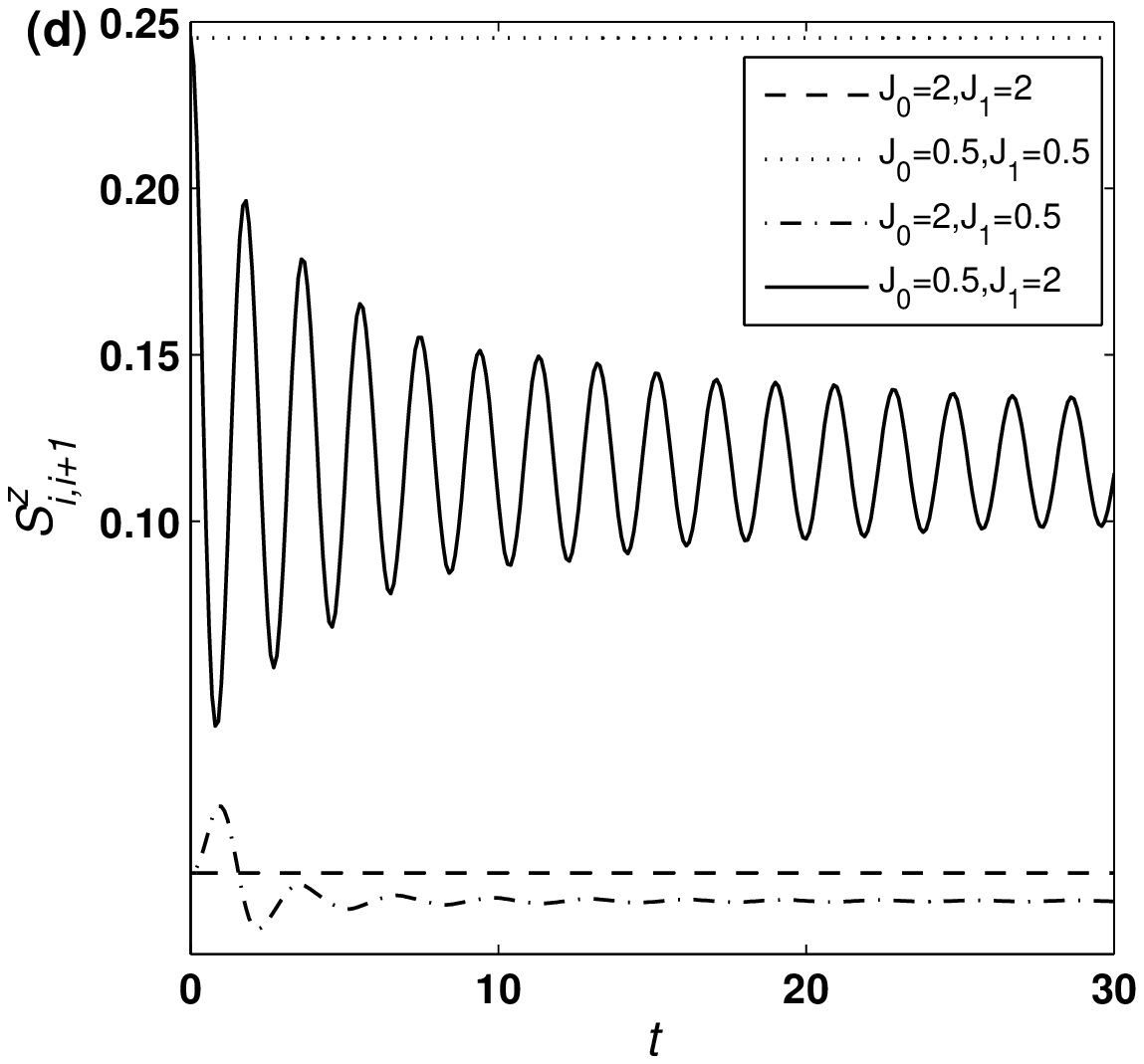}}
   \caption{{\protect\footnotesize Dynamics of the nearest-neighbor concurrence $C(i,i+1)$, where $t$ is in units of $J_{1}^{-1}$, with $\gamma=0.5$ and (a) $h_{0}=h_{1}=1$ for various values of $J_{0}$ and $J_{1}$ at $kT=0$; (b) $J_{0}=J_{1}=1$ for various values of $h_{0}$ and $h_{1}$ at $kT=0$. (c) Dynamics of the magnetization per spin and (d) the spin-spin correlation function in $z$ direction for fixed $h=h_{0}=h_{1}=1$ for various values of $J_{0}$ and $J_{1}$ at $kT=0$ with $\gamma=0.5$.}}
 \label{fig:7}
\end{minipage}
\end{figure}
First, we study the dynamics of nearest-neighbor concurrence for this model. In Fig.~\ref{fig:7a}, we choose the magnetic field to have a constant value of 1 while the coupling parameter is 0.5 or 2 or a step function changing from 0.5 to 2 (or 2 to 0.5). In Fig.~\ref{fig:7b}, we choose the coupling parameter to have a constant value of 1 while the magnetic field is 0.5 or 2 or a step function changing from 0.5 to 2 (or 2 to 0.5). As one can see, $C(i,i+1)$ shows a nonergodic behavior, similar to the isotropic case, which also follows from the nonergodic behavior of the spin correlation functions and magnetization shown in Figs.~\ref{fig:7c} and \ref{fig:7d}. Nevertheless, the equilibrium time in this case is much longer than the isotropic case as can be seen.
\begin{figure}[htbp]
\begin{minipage}[c]{\textwidth}
 \centering 
   \subfigure{\label{fig:8a}\includegraphics[width=6cm]{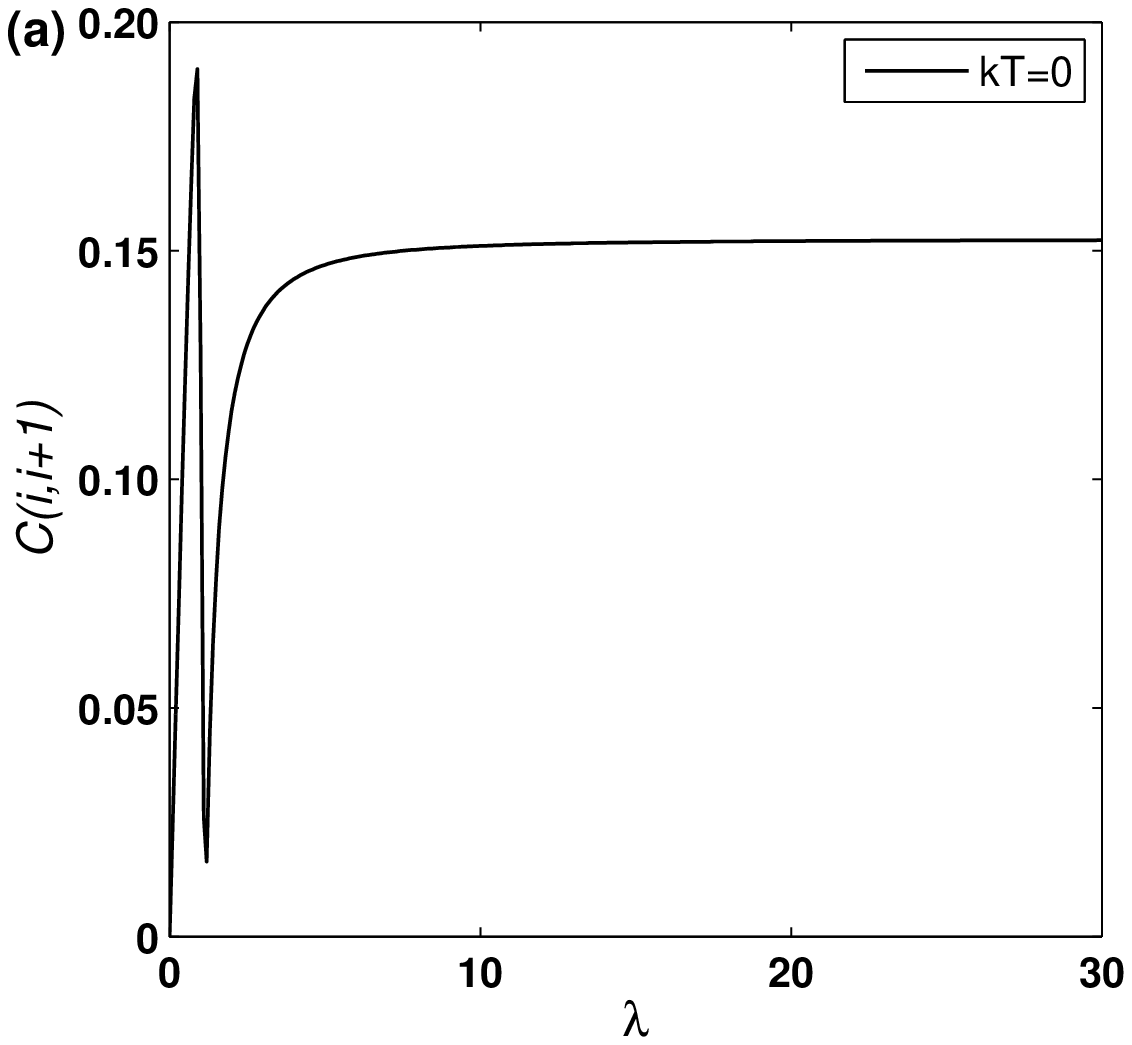}}\quad
   \subfigure{\label{fig:8b}\includegraphics[width=6cm]{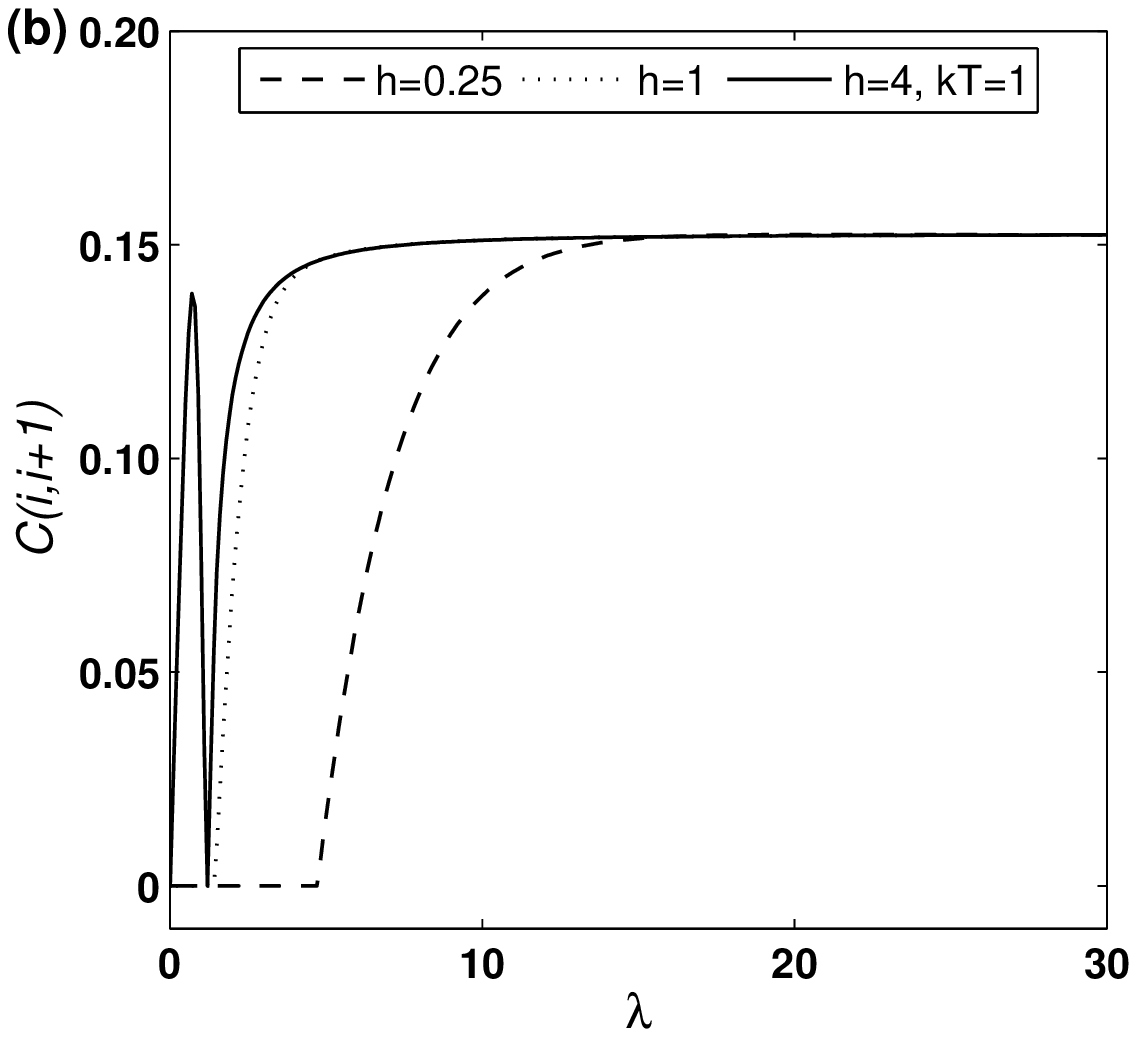}}\\
   \subfigure{\label{fig:8c}\includegraphics[width=6cm]{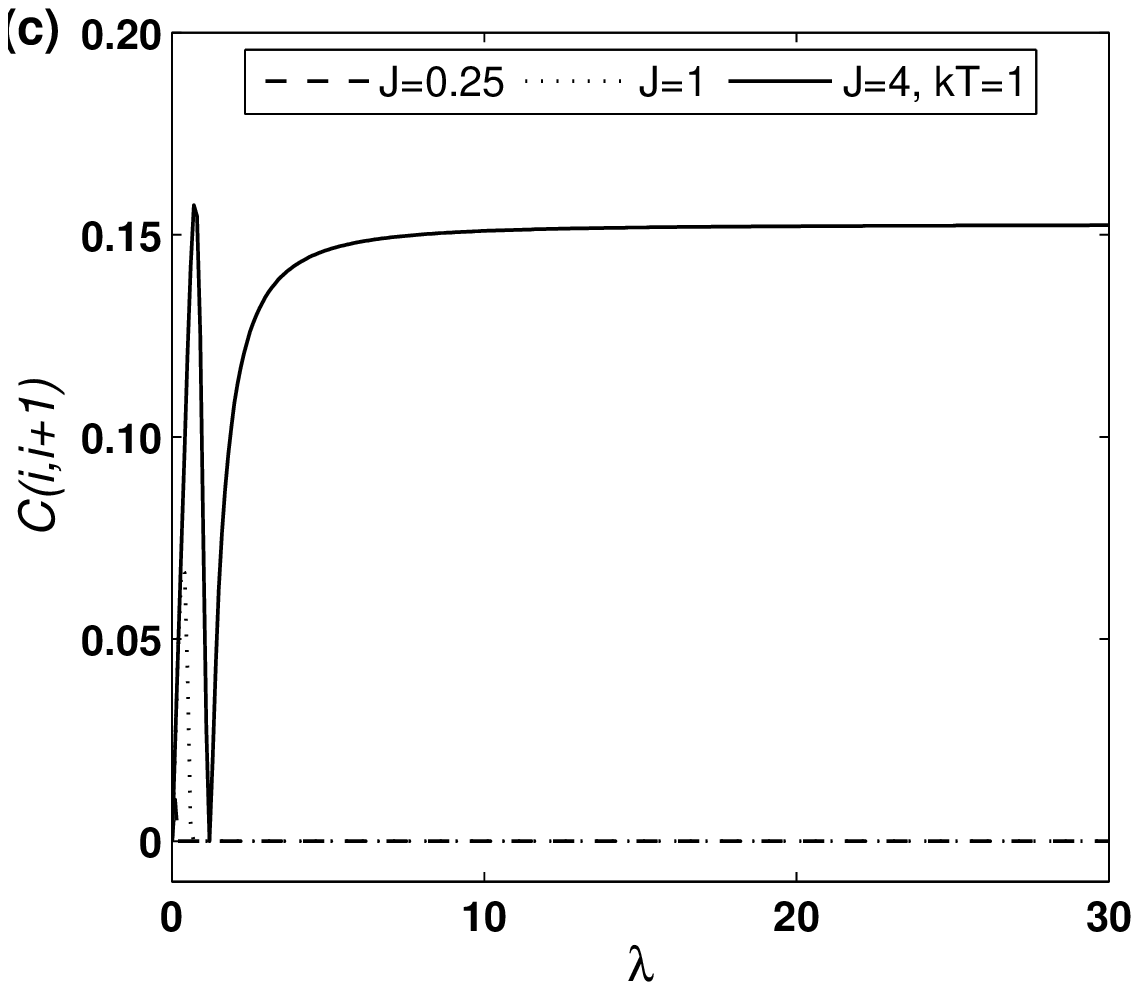}}\quad
   \subfigure{\label{fig:8d}\includegraphics[width=6cm]{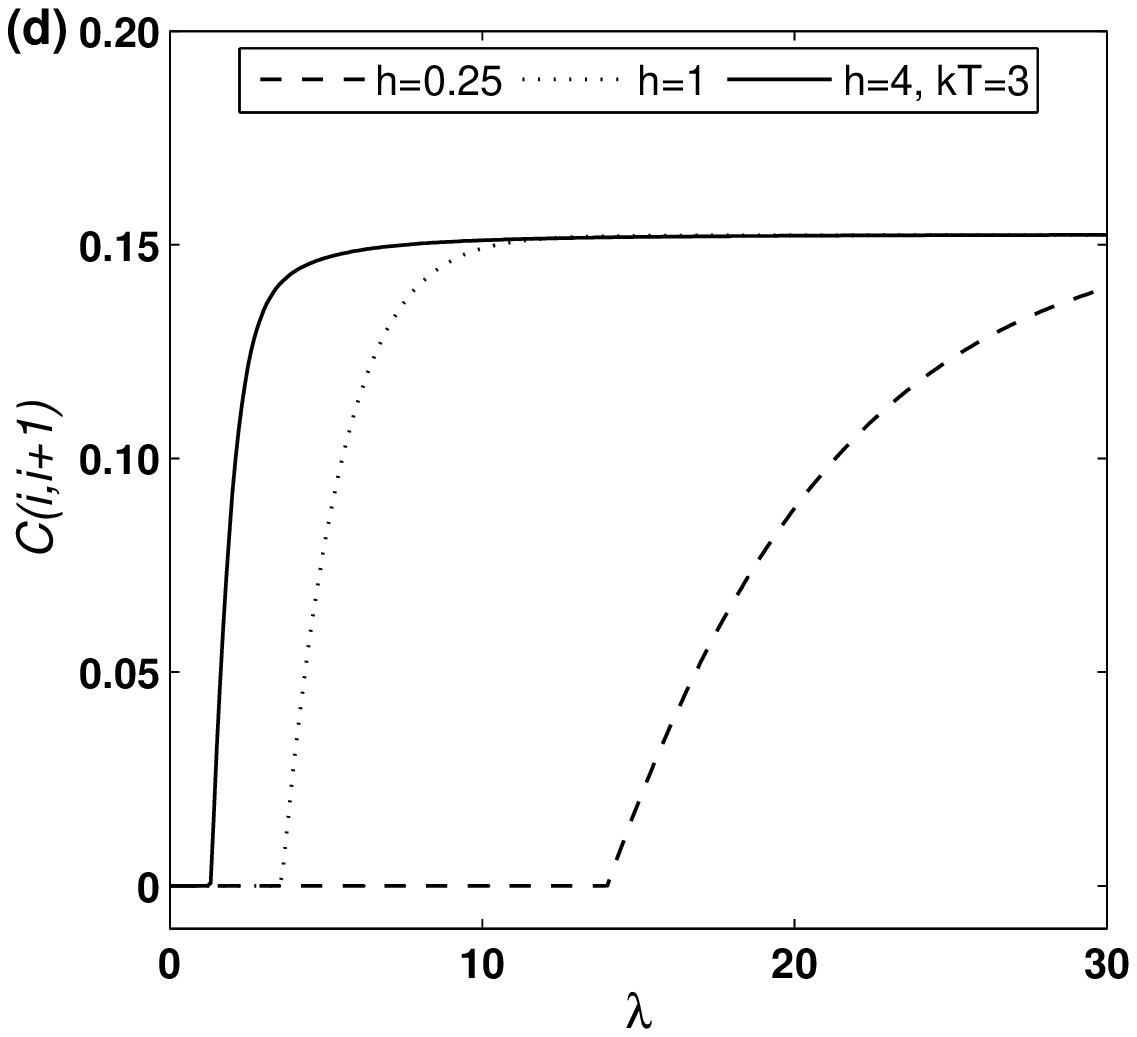}}
   \caption{{\protect\footnotesize The asymptotic behavior of $C(i,i+1)$ with $\gamma=0.5$ as a function of $\lambda$ when $h_{0}=h_{1}$ and $J_{0}=J_{1}$ at (a) $kT=0$ with any combination of constant $J$ and $h$; (b) $kT=1$ with $h_{0}=h_{1}=0.25, 1, 4$; (c) $kT=1$ with $J_{0}=J_{1}=0.25, 1, 4$; (d) $kT=3$ with $h_{0}=h_{1}=0.25, 1, 4$.}}
 \label{fig:8}
 \end{minipage}
\end{figure}

In Fig.~\ref{fig:8} we study $C(i,i+1)$ as a function of $\lambda$ for different values of $J$ and $h$ and at different temperatures. We first study the zero temperature case at different constant values of $J$ and $h$. For this particular case $C(i,i+1)$ depends only on the ratio of $J$ and $h$, similar to the isotropic case, rather than their individual values as shown in Fig.~\ref{fig:8a}. As can be noted, $C(i,i+1)$ starts from zero, reaches a maximum value at $\lambda\approx0.9$, drops to a very small value at $\lambda\approx 1.1$ and then increases rapidly, reaching a constant value for larger values of $\lambda$. The two extremes cases can be explained easily as for $h>>J$, i.e., $\lambda << 1$, the effect of the magnetic field is dominating and the spins are aligned into the $z$ direction and, as a result, $C(i,i+1)$ vanishes. On the other hand, when $h<<J$, i.e., $\lambda >> 1$, the effect of $J$ dominates. However, for this partial anisotropic case, increasing $J$ would increase both $J_{x}$ and $J_{y}$, which causes the spins not to be aligned in a particular direction and consequently $C(i,i+1)$ maintains an equilibrium finite value. Interestingly, the concurrence shows a complicated critical behavior in the vicinity of $\lambda=1$, where it reaches a maximum value first and immediately drops to a minimum (very small) value before raising again to its equilibrium value. The raising of the concurrence from zero as $J$ increases, for $\lambda < 1$, is expected as in that case part of the spins which were originally aligned in the $z$ direction change directions into the $x$ and $y$ directions. The sudden drop of the concurrence in the vicinity of $\lambda=1$, where $\lambda$ is slightly larger than 1, suggests that significant fluctuations is taken place and the effect of $J_x$ is dominating over both $J_y$ and $h$ which aligns most of the spins into the $x$ direction, leading to a reduced entanglement value. Studying the thermal concurrence in Figs.~\ref{fig:8b}, \ref{fig:8c}, and \ref{fig:8d} we note that the asymptotic value of $C(i,i+1)$ is not affected as the temperature increases. However, the critical behavior of the entanglement in the vicinity of $\lambda=1$ changes considerably as the temperature is raised and the other parameters are varied. As illustrated in Fig.~\ref{fig:8b}, the maximum entanglement value is reduced and the minimum value reaches zero at high magnetic fields at $kT=1$, but as the magnetic field is reduced the critical behavior disappears and the entanglement makes a direct transition from zero to the equilibrium value where the transition becomes sharper and takes place at smaller values of $\lambda$ as we increase the magnetic field. A similar behavior is shown in Fig.~\ref{fig:8c} where for small values of the coupling $J$ the critical behavior disappears as well. The effect of higher temperature is shown in Fig.~\ref{fig:8d} where the critical behavior of the entanglement disappears completely at all values of $h$ and $J$, which confirms that the thermal excitations destroy the critical behavior due to suppression of quantum effects. 
\begin{figure}[htbp]
\begin{minipage}[c]{\textwidth}
 \centering 
   \subfigure{\label{fig:9a}\includegraphics[width=6cm]{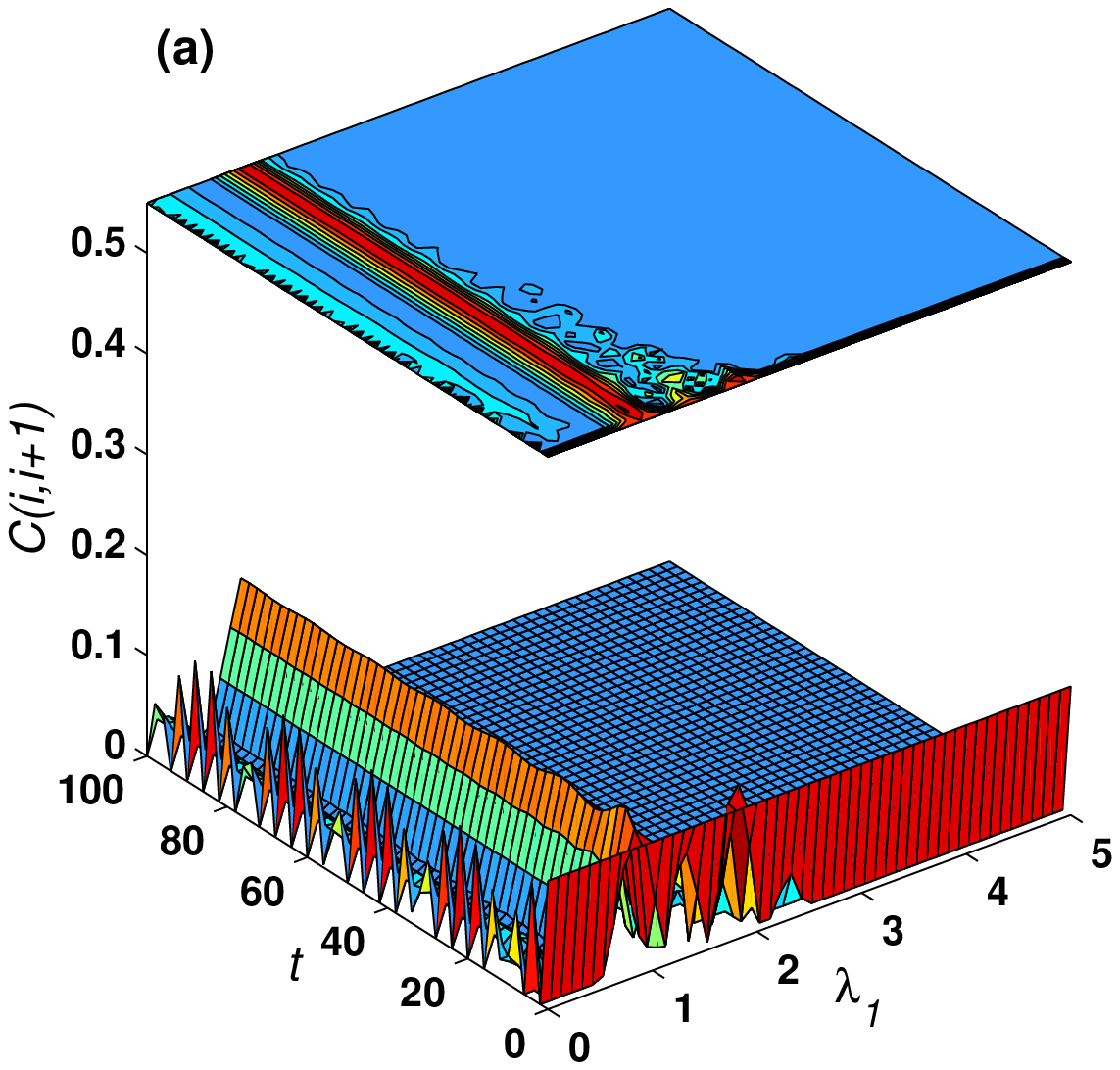}}\quad
   \subfigure{\label{fig:9b}\includegraphics[width=6cm]{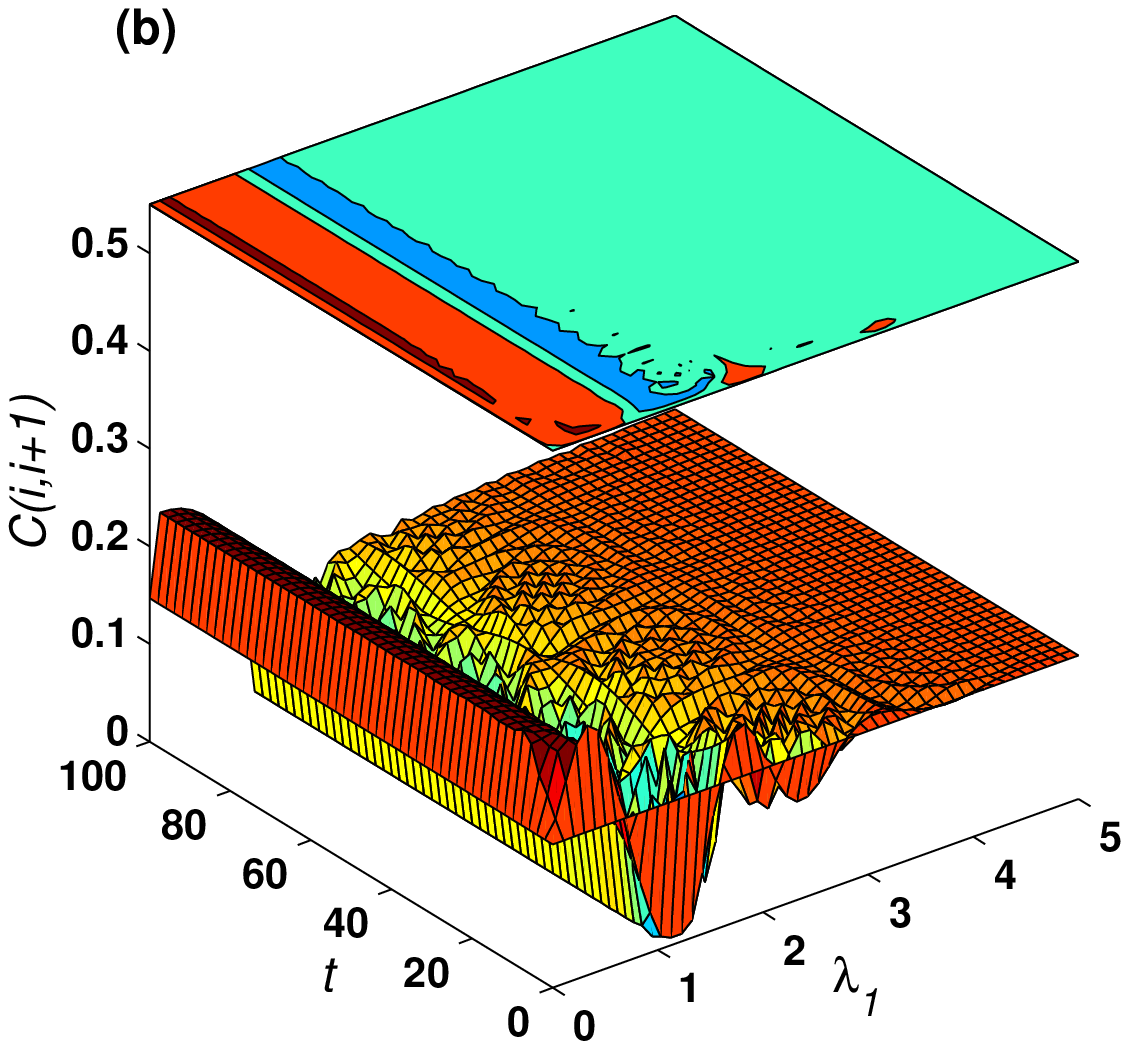}}\\
   \subfigure{\label{fig:9c}\includegraphics[width=6cm]{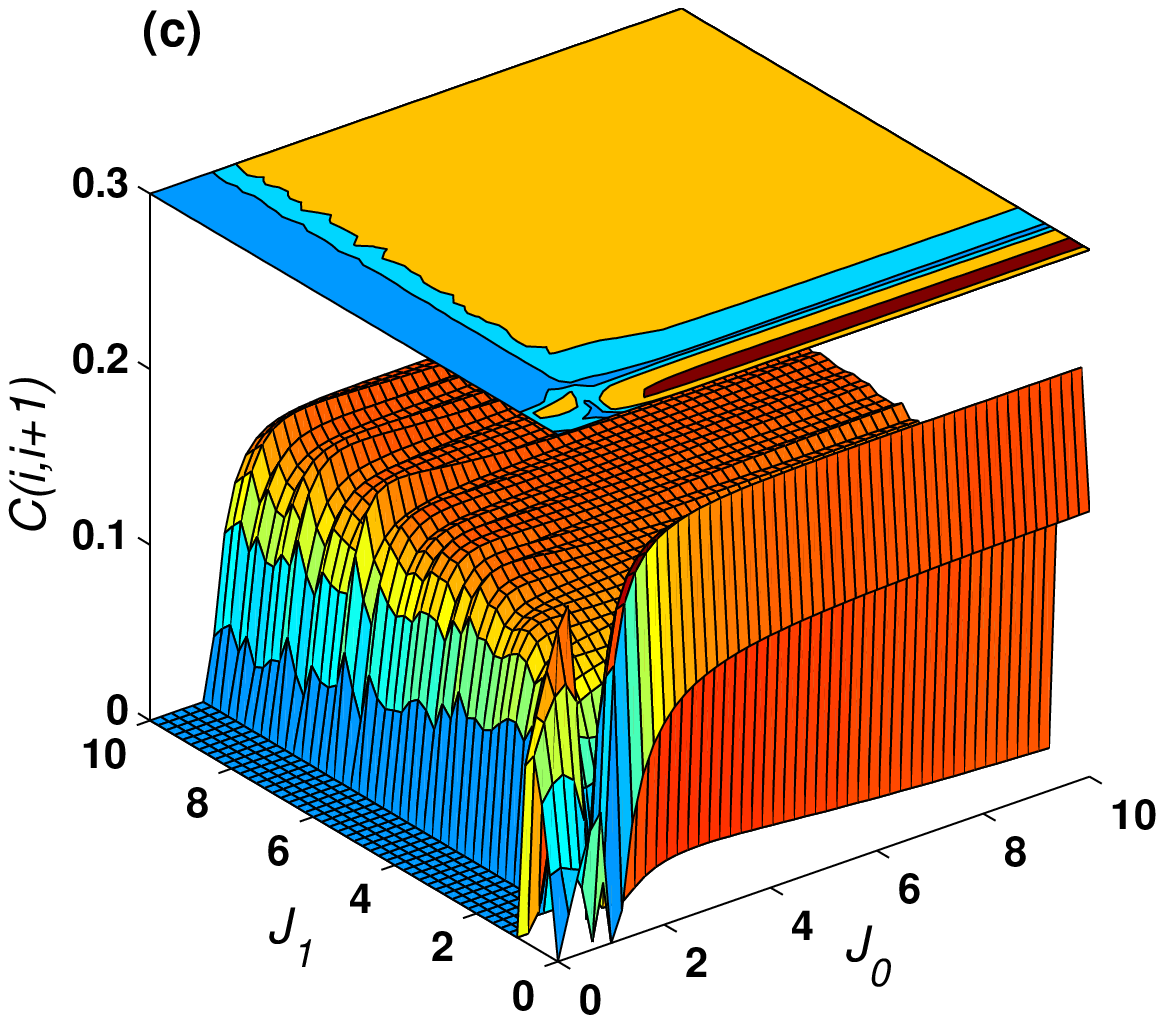}}\quad
   \subfigure{\label{fig:9d}\includegraphics[width=6cm]{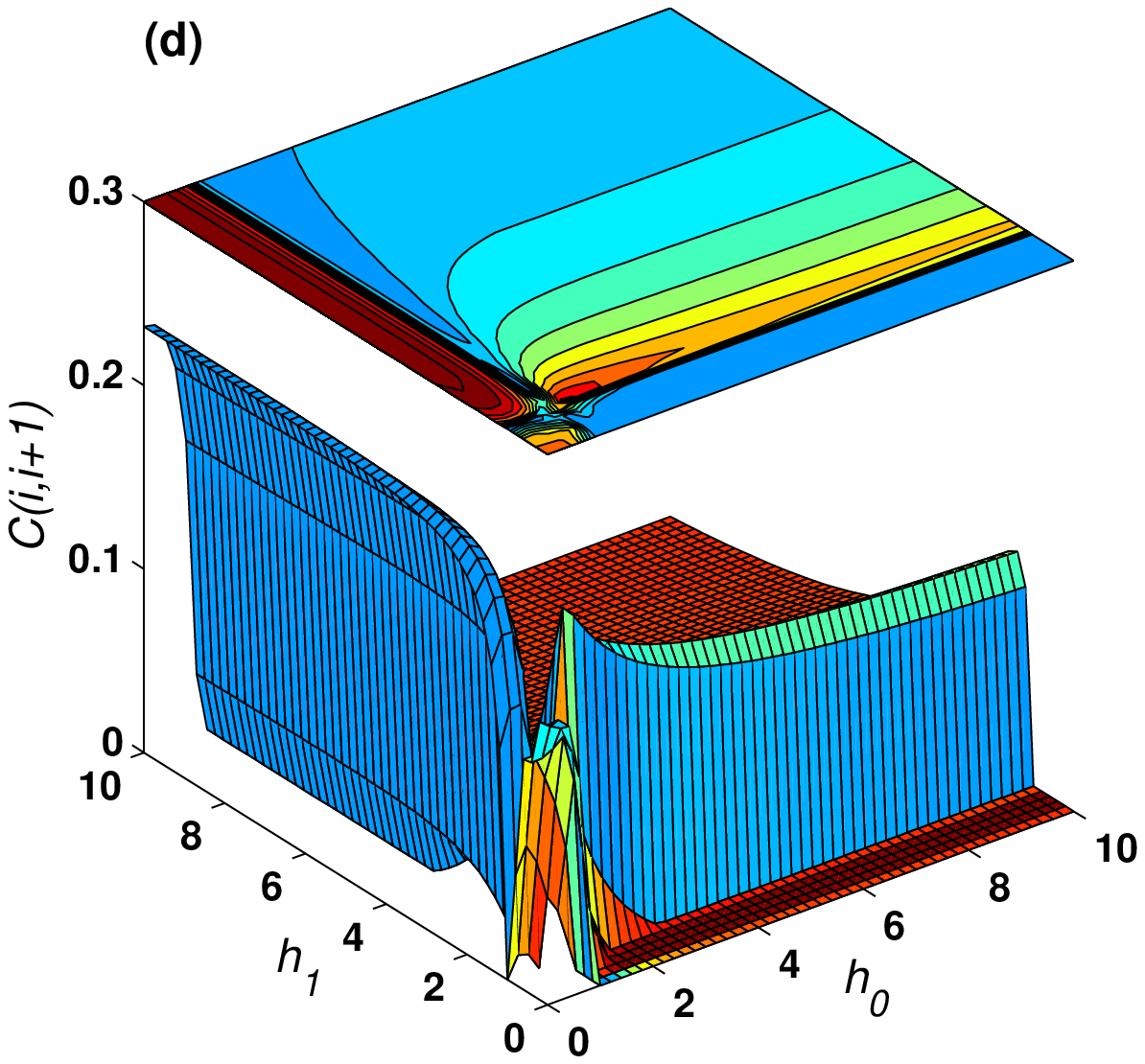}}
   \caption{{\protect\footnotesize (Color online) $C(i,i+1)$ as a function of $\lambda_{1}$ and $t$, in units of $J_{1}^{-1}$, at $kT=0$ with $\gamma=0.5$, $h_{0}=h_{1}=1$ and (a) $J_{0}=1$; (b) $J_{0}=5$. The asymptotic behavior of $C(i,i+1)$ as a function of (c) $J_{0}$ and $J_{1}$ at $kT=0$ with $\gamma=0.5$ and $h_{0}=h_{1}=1$; (d) $h_{0}$ and $h_{1}$ at $kT=0$ with $\gamma=0.5$ and $J_{0}=J_{1}=1$, where $h_{0}$, $h_{1}$, and $J_{0}$ are in units of $J_{1}$.}}
 \label{fig:9}
 \end{minipage}
\end{figure}

In Fig.~\ref{fig:9} we investigate the dependence of the time evolution and asymptotic behavior of concurrence on the different magnetic field and coupling parameters. We have studied $C(i,i+1)$ as a function of $\lambda_{1}$ and $t$ for many different selections of $h=h_0=h_1$ and $J_{0}$ and found that the concurrence behavior depends mainly on whether $\lambda_{0}>1$ or $\leq 1$. To test that behavior, Figs.~\ref{fig:9a} and ~\ref{fig:9b} show $C(i,i+1)$ as a function of $\lambda_{1}$ and $t$ with $J_{0}=1$ and $5$ respectively for fixed $h_{0}=h_{1}=1$ and $kT=0$. As can be seen in Fig.~\ref{fig:9a}, where $J_{0}=1$ ($\lambda_{0}=1$), the concurrence at any time $t > 0$ starts with a finite value (or zero) at $\lambda_{1}=\lambda_{0}$ and then decays to zero but increases again, reaching a maximum value in the vicinity of $\lambda_1 = 1$ and finally vanishes permanently as $\lambda_1$ increases. Interestingly at $\lambda_{1}=0$ the concurrence shows an oscillatory behavior in time, as was the case in the completely anisotropic model. For values of $\lambda_{1}$ around the critical value, the concurrence approximately maintains its initial value as time elapses. However, for larger values of $\lambda_{1}$, the concurrence starts initially with a finite value but decays sharply to zero in a very short period of time. On the other hand, in Fig.~\ref{fig:9b} where we set $J_{0}=5$ (i.e., $\lambda_{0}=5$), at any time $t > 0$ the concurrence starts with a finite value at $\lambda_{1}=0$ and increases rapidly as $\lambda_{1}$ increases because increasing $J_{1}$ reduces  the alignment in the $z$ direction. The concurrence $C(i,i+1)$ reaches a maximum value at $\lambda_{1}\approx0.2$ and vanishes at $\lambda_{1}\approx 1$. Finally, the concurrence increases, reaching a constant value for $\lambda_{1}\approx 2$ or larger. The variation of the concurrence in this case with time is very limited, as one can see; it approximately maintains its initial value especially for all $\lambda_{1} \leq 1$ and $\lambda_{1} \geq 3$. This critical dependence of the concurrence dynamics on the initial value of the coupling parameter is emphasized in Fig.~\ref{fig:9c}, where the asymptotic value of the concurrence is depicted as a function of both $J_0$ and $J_1$. In Fig.~\ref{fig:9d} the asymptotic behavior of the concurrence is explored as a function of $h_0$ and $h_1$ while fixing the coupling as $J_1=J_0=1$. The behavior of the concurrence is very close to the completely anisotropic case except for the region where $0 \leq h_0 \leq 1$ and $0 \leq h_1 \leq 1$ where the concurrence starts with a finite value at $h_0=h_1=0$ and decays gradually until it vanishes at $h_0 = h_1\approx 1$. For higher values of the coupling $J_1=J_0$ the rate of decay of the concurrence every where is smaller and the peaks are broadened. 
\begin{figure}[htbp]
\begin{minipage}[c]{\textwidth}
 \centering 
   \subfigure{\label{fig:10a}\includegraphics[width=6cm]{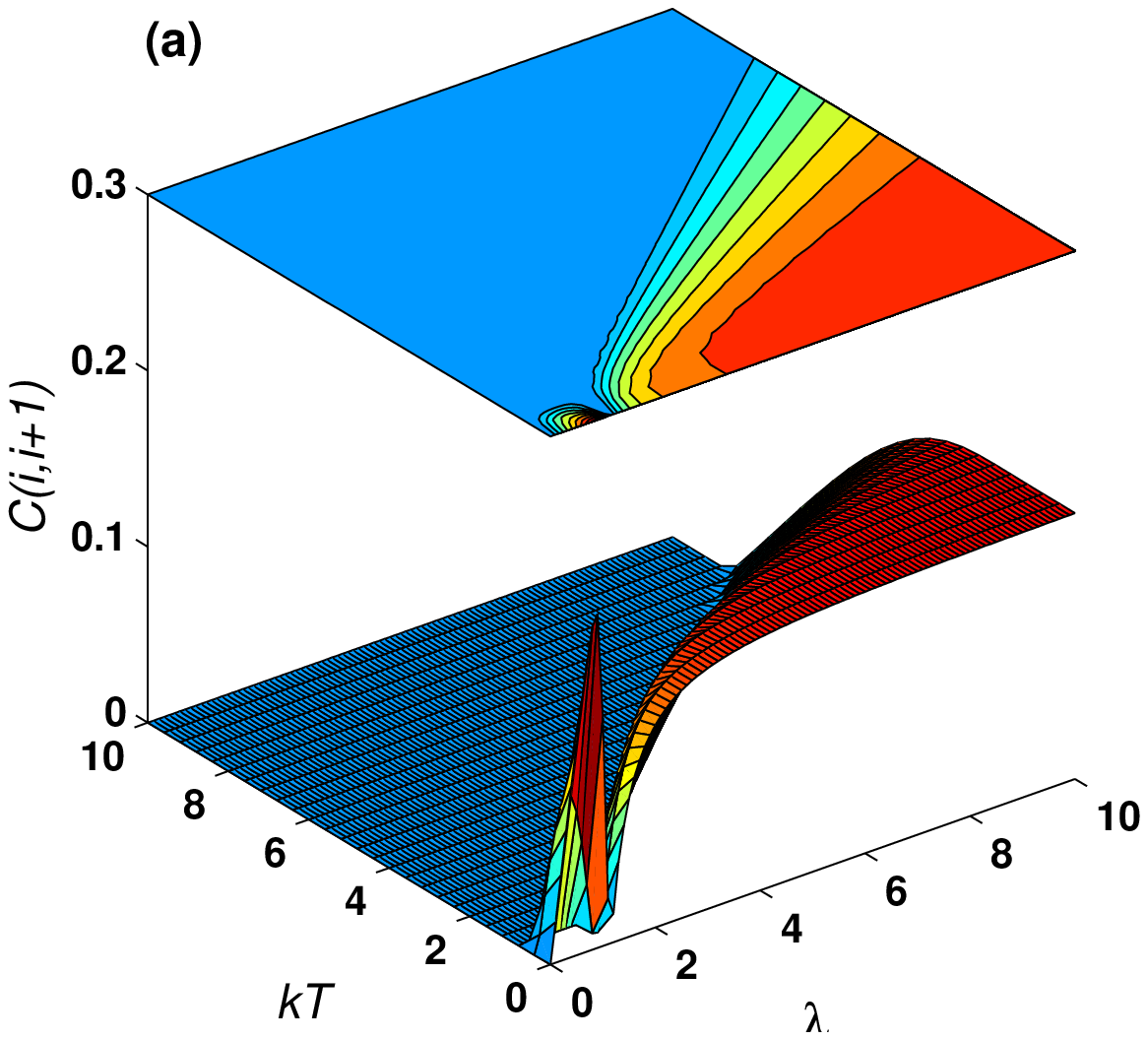}}\quad
   \subfigure{\label{fig:10b}\includegraphics[width=6cm]{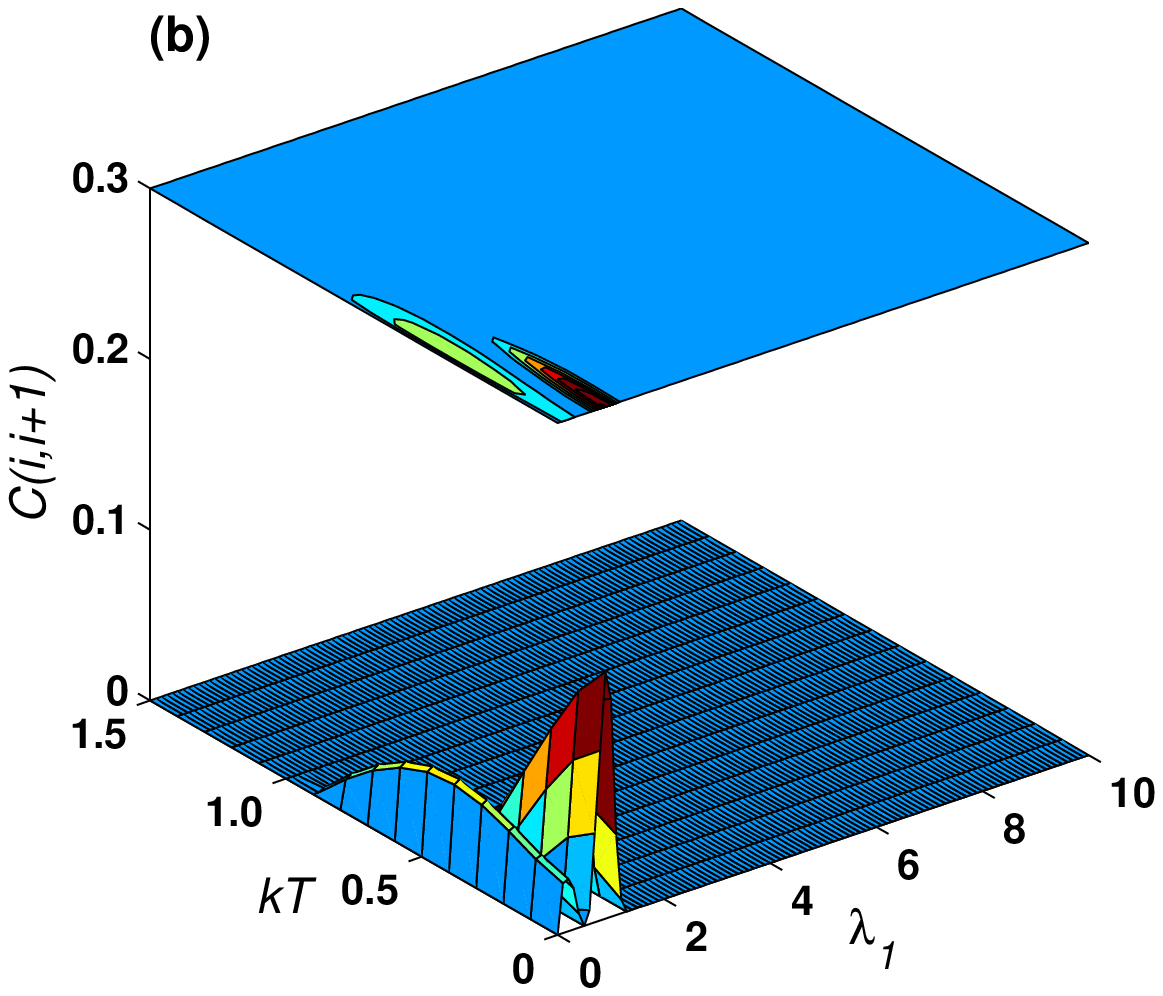}}
      \caption{{\protect\footnotesize (Color online) The asymptotic behavior of $C(i,i+1)$ as a function of (a) $\lambda$ and $kT$, in units of $J_{1}$, with $\gamma=0.5$, $h_{0}=h_{1}=1$, and $J_{0}=J_{1}$; (b) $\lambda_{1}$ and $kT$ with $\gamma=0.5$, $h_{0}=h_{1}=1$, and $J_{0}=1$.}}
 \label{fig:10}
 \end{minipage}
 \end{figure}

The persistence of quantum effects as temperature increases and time elapses in the partially anisotropic case is examined and presented in Fig.~\ref{fig:10}. In Fig.~\ref{fig:10a} the concurrence is plotted as a function of $\lambda \equiv \lambda_0=\lambda_1$ and $kT$ with $h_{0}=h_{1}=1$ and $J_{0}=J_{1}$. As one can see, the concurrence shows the expected behavior as a function of $\lambda$ and decays as the temperature increases. As one can see, the threshold temperature where the concurrence vanishes is determined by the value of $\lambda$, it increases as $\lambda$ increases. In Fig.~\ref{fig:10b} the asymptotic behavior of the concurrence as a function $\lambda_1$ and $kT$ is illustrated. Clearly the nonzero concurrence shows up at small values of $kT$ and $\lambda_1$. The concurrence has two peaks versus $\lambda_1$ but as the temperature increases, the second peak disappears. Very interestingly, the first peak raises as temperature increases then decays again and vanishes for $kT \approx 0.9$ or larger.
\begin{figure}[htbp]
\begin{minipage}[c]{\textwidth}
 \centering 
\includegraphics[width=6.5cm]{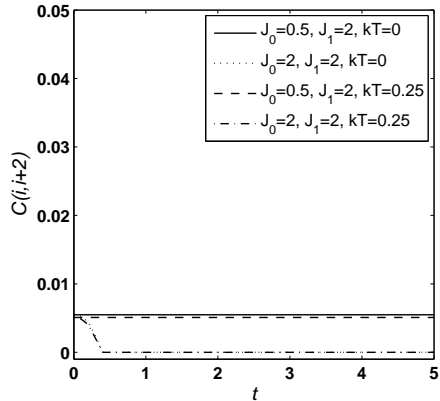}
   \caption{{\protect\footnotesize Dynamics of $C(i,i+2)$, where $t$ is in units of $J_{1}^{-1}$, with $\gamma=0.5$ and $h_{0}=h_{1}=1$ for various values of $J_{0}, J_{1}$ at $kT=0$ and $kT=0.25$.}}
 \label{fig:11}
 \end{minipage}
\end{figure}

The behavior of the next-to-nearest-neighbor concurrence $C(i,i+2)$, is shown in Fig.~\ref{fig:11}. As expected, $C(i,i+2)<<C(i,i+1)$ at the same circumstances. Studying longer-range concurrence $C(i,i+r)$ shows that it vanishes for $r\geq 3$.
\section{Isotropic XY Model}
In this section we consider the isotropic system with $\gamma=0$ (i.e., $J_{x}=J_{y}$); in this case the Hamiltonian assumes the form
\begin{equation}
H=-\frac{J(t)}{2} \sum_{i=1}^{N} \sigma_{i}^{x} \sigma_{i+1}^{x}-\frac{J(t)}{2} \sum_{i=1}^{N} \sigma_{i}^{y} \sigma_{i+1}^{y}- \sum_{i=1}^{N} h(t) \sigma_{i}^{z}
\label{eq:H_isotropic}
\end{equation}

\begin{figure}[htbp]
\begin{minipage}[c]{\textwidth}
 \centering
   \subfigure{\label{fig:12a}\includegraphics[width=6cm]{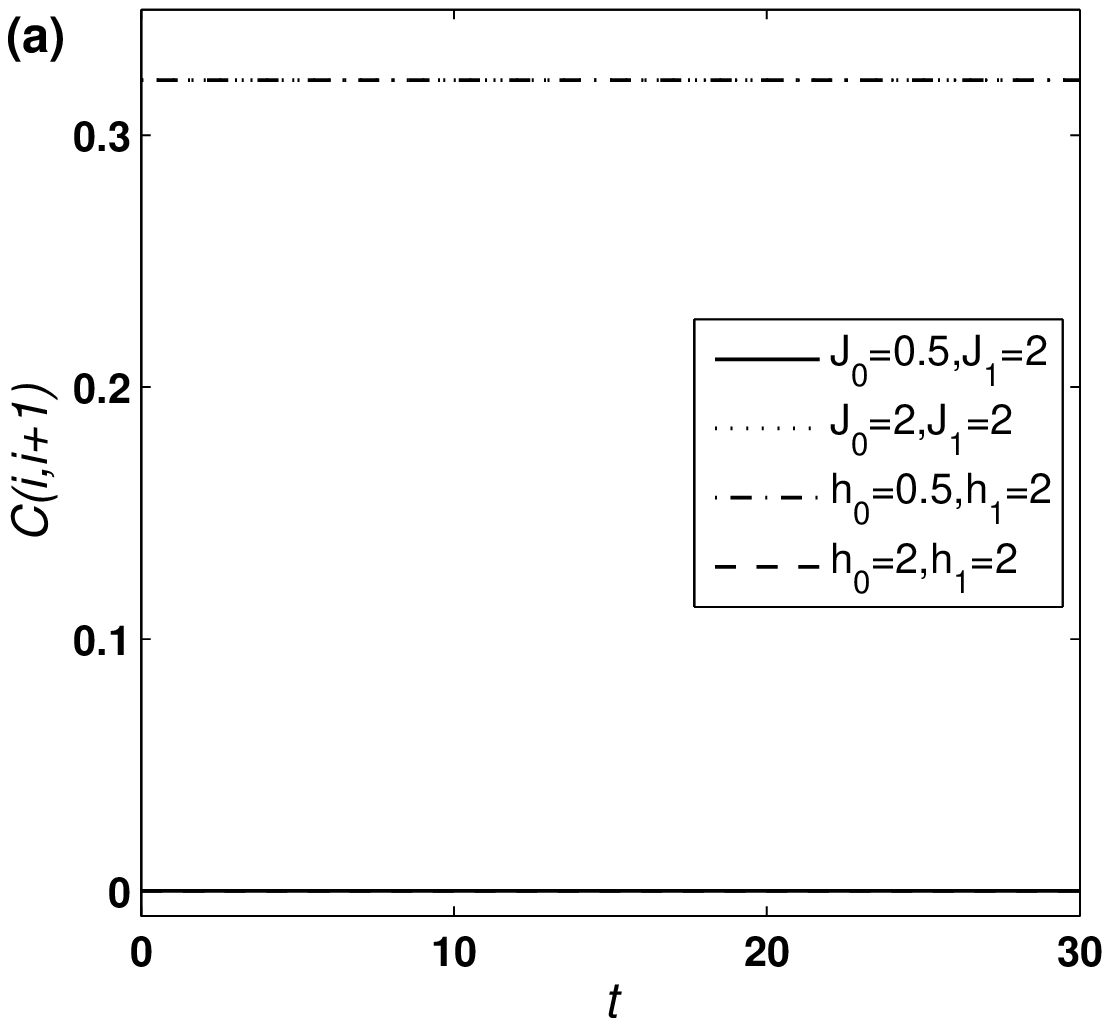}}\quad
   \subfigure{\label{fig:12b}\includegraphics[width=6cm]{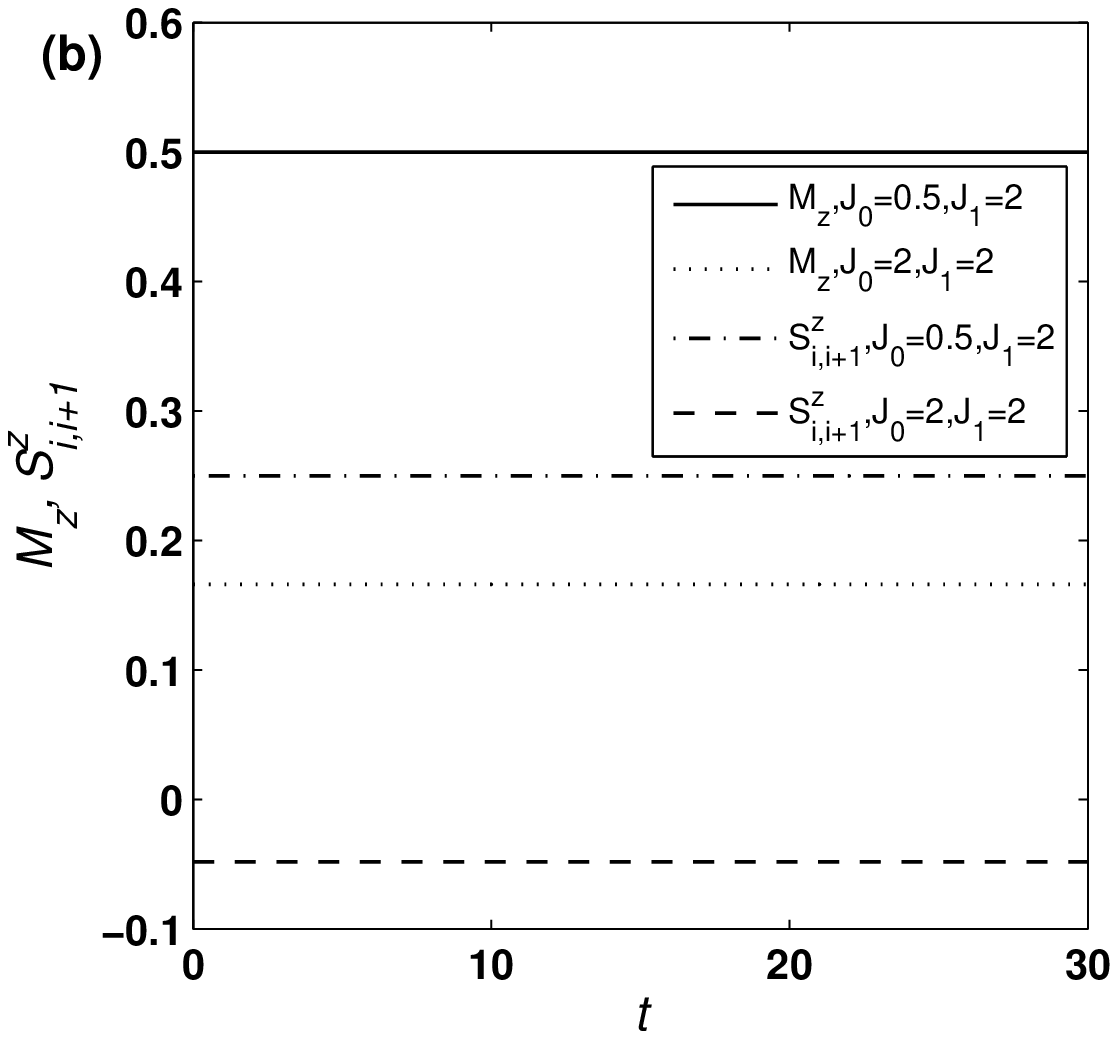}}\\
   \caption{{\protect\footnotesize Dynamics of the nearest-neighbor concurrence $C(i,i+1)$, where $t$ is in units of $J_{1}^{-1}$, with $\gamma=0$ for various values of $J_{0}, J_{1}, h_{0}$, and $h_{1}$ at $kT=0$. (b) Dynamics of the magnetization per spin and the spin-spin correlation function in the $z$ direction for fixed $h=h_{0}=h_{1}=1$ for various values of $J_{0}$ and $J_{1}$ at $kT=0$ with $\gamma=0$.}}
\label{fig:12}
\end{minipage}
\end{figure}

We start with the dynamics of the nearest-neighbor concurrence; in Fig.~\ref{fig:12}, we first choose the magnetic field to have a constant value of 1 while the coupling parameter is 2, or a step function changing between 0.5 and 2. We also study the case with a constant coupling parameter of 1 while the magnetic field is 2, or a step function changing between 0.5 and 2. Testing the concurrence for several different values of the magnetic field and coupling parameter, we note that $C(i,i+1)$ takes a constant value that does not depend on the final value of the coupling $J_{1}$ and magnetic field $h_{1}$. This follows from the dependence of the spin correlation functions and the magnetization on the initial state only as shown in Fig.~\ref{fig:12b}.
This is because the initial coupling parameters $J_x$ and $J_y$, which are equal, force the spins to be equally aligned into the $x$ and $y$ directions, apart from those in the $z$-direction, causing a finite concurrence. Increasing the coupling parameters strength would not change that distribution or the associated concurrence at constant magnetic field. 
\begin{figure}[htbp]
\begin{minipage}[c]{\textwidth}
 \centering 
   \subfigure{\label{fig:13a}\includegraphics[width=6cm]{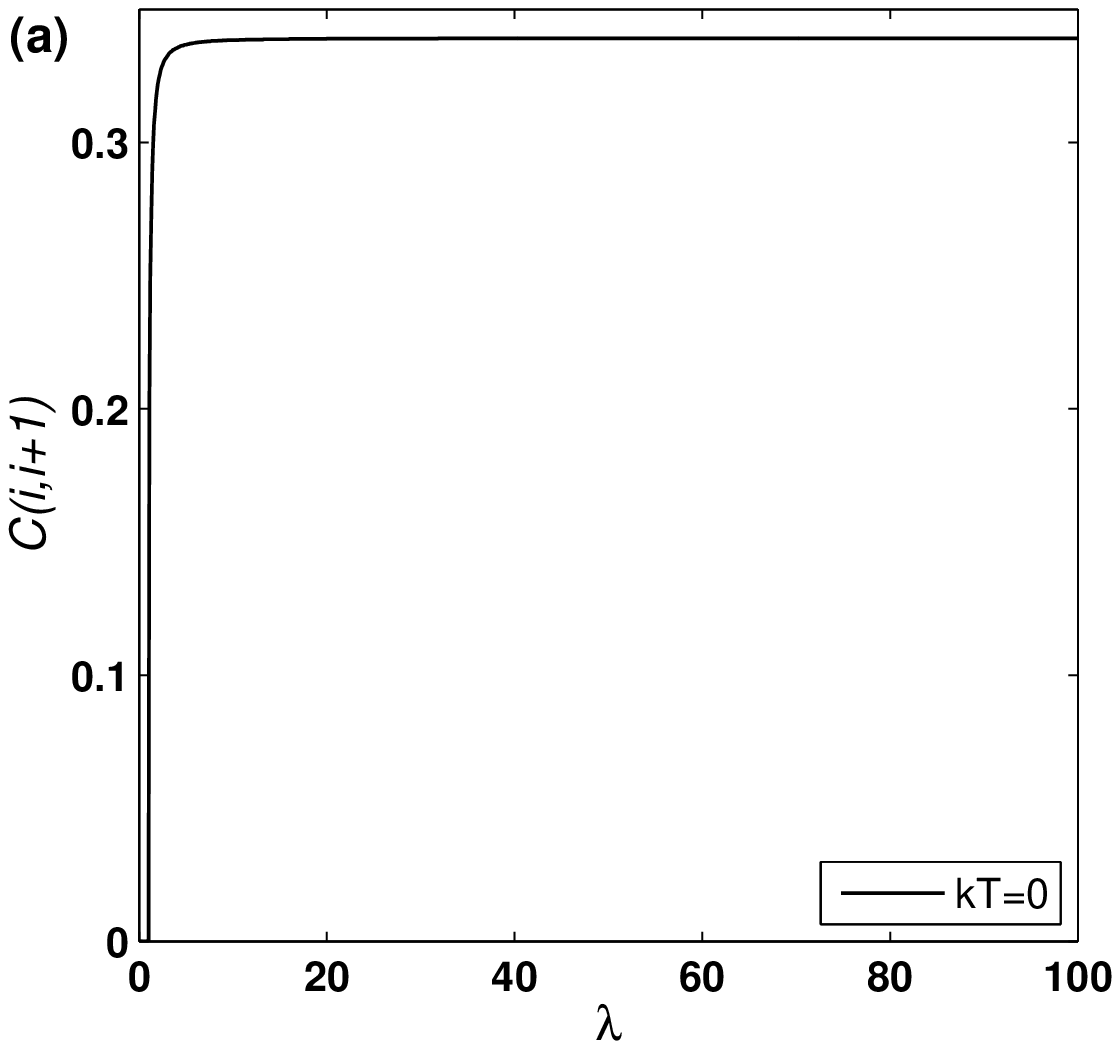}}\quad
   \subfigure{\label{fig:13b}\includegraphics[width=6cm]{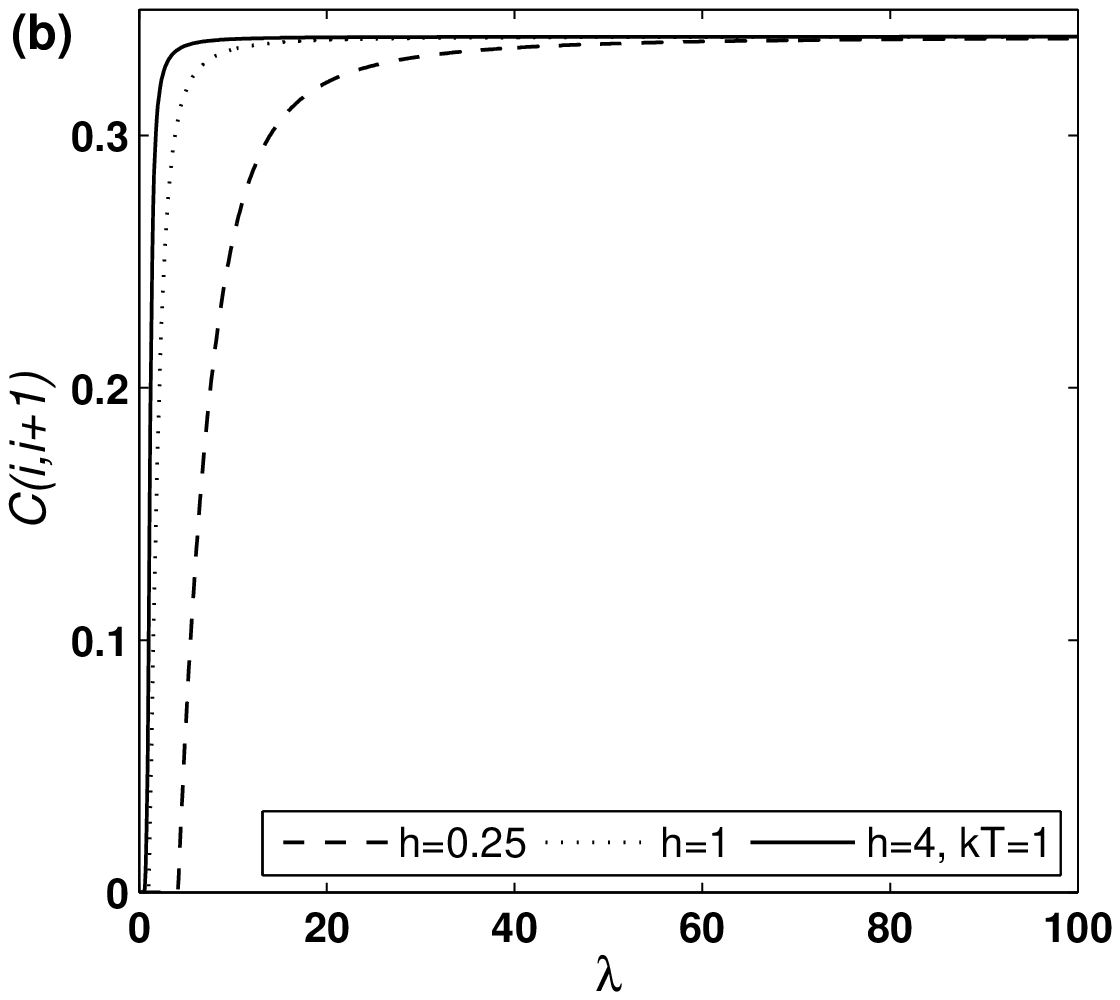}}\\
   \subfigure{\label{fig:13c}\includegraphics[width=6cm]{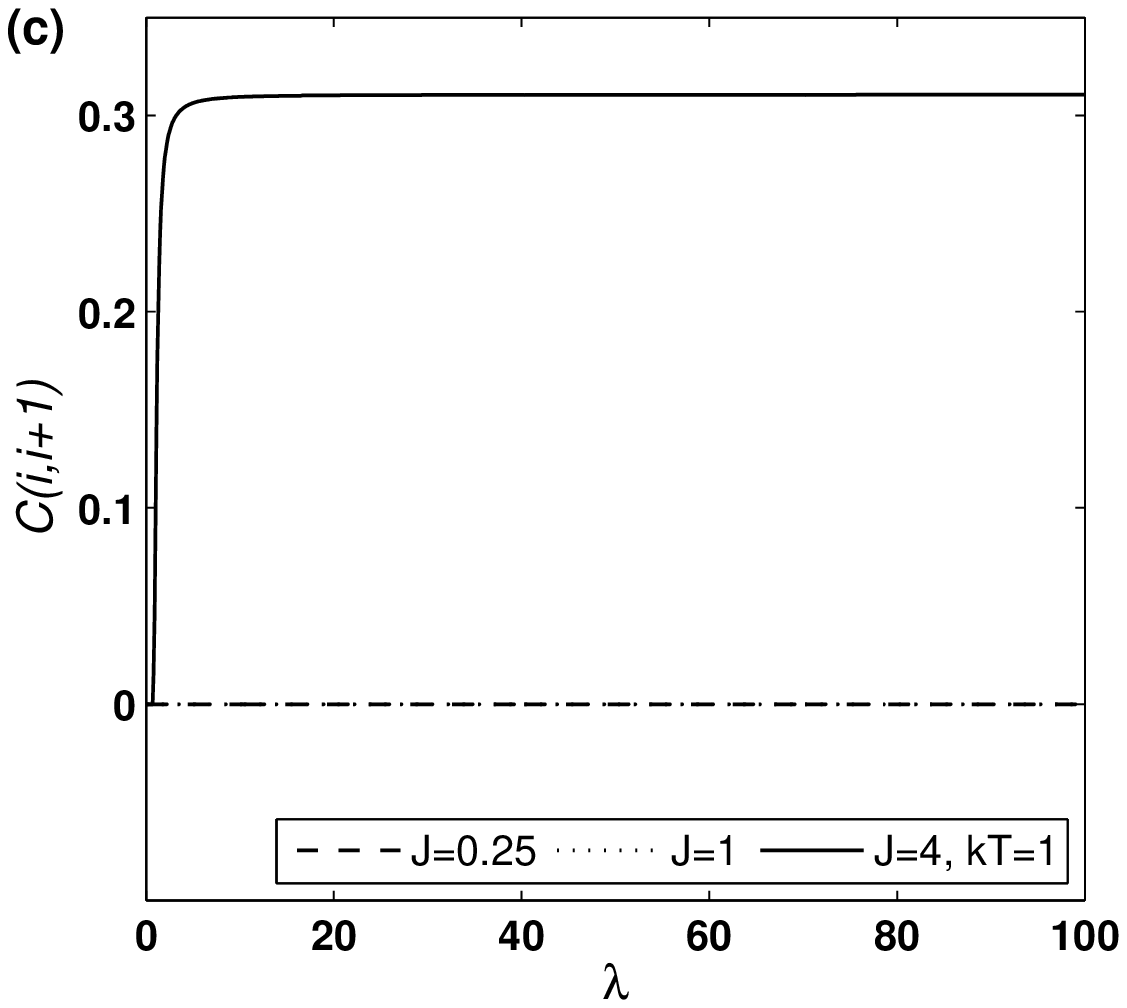}}\quad
   \subfigure{\label{fig:13d}\includegraphics[width=6cm]{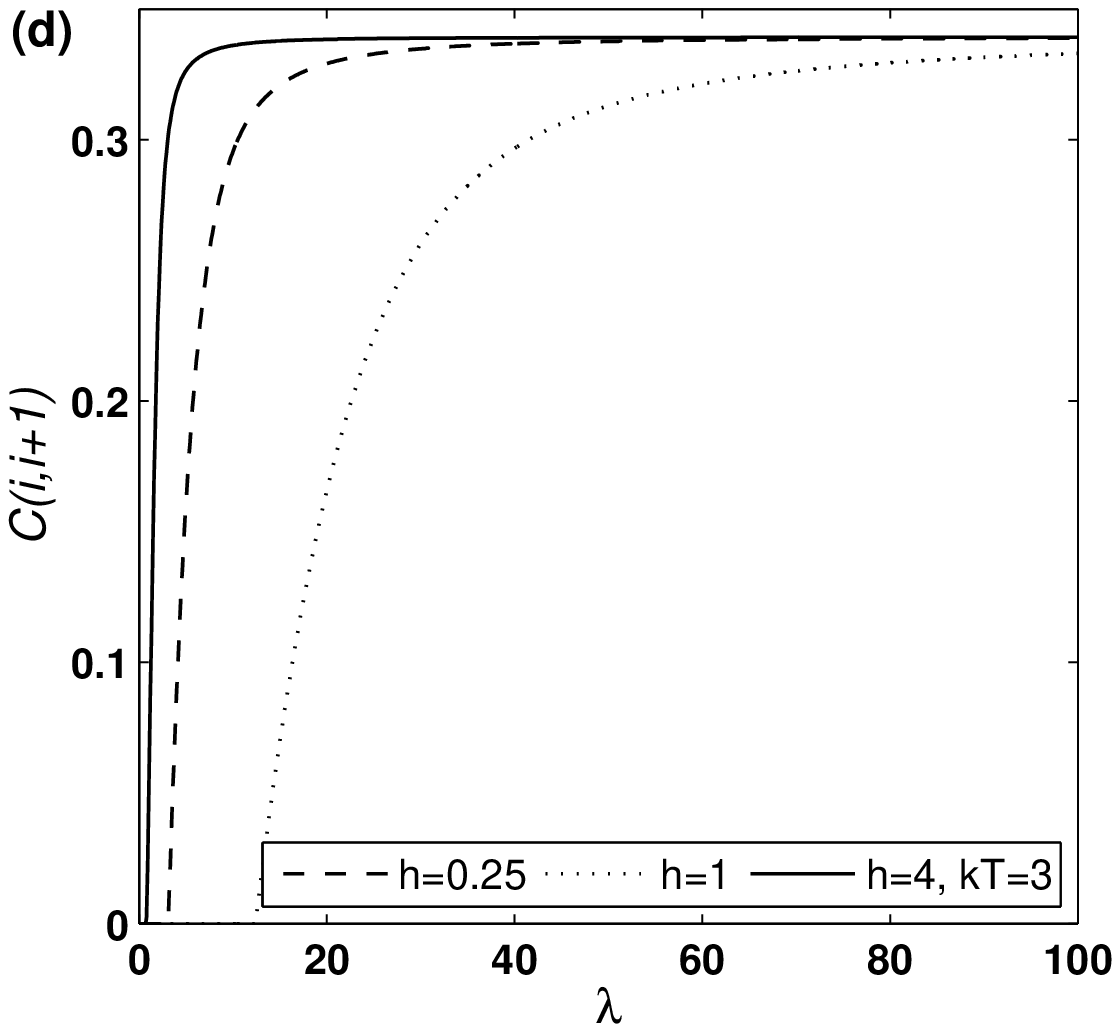}}
   \caption{{\protect\footnotesize The asymptotic behavior of $C(i,i+1)$ with $\gamma=0$ as a function of $\lambda$ when $h_{0}=h_{1}$ and $J_{0}=J_{1}$ at (a) $kT=0$ with any combination of constant $J$ and $h$; (b) $kT=1$ with $h_{0}=h_{1}=0.25, 1, 4$; (c) $kT=1$ with $J_{0}=J_{1}=0.25, 1, 4$; (d) $kT=3$ with $h_{0}=h_{1}=0.25, 1, 4$.}}
 \label{fig:13}
 \end{minipage}
\end{figure}

In Fig.~\ref{fig:13} we study $C(i,i+1)$ as a function of time-independent $\lambda$ for different values of $J=J_0=J_1$ and $h=h_0=h_1$ and at different temperatures. Again $C(i,i+1)$ depends only on the ratio of $J$ and $h$ rather than their individual values at $kT=0$ as shown in Fig.~\ref{fig:13a}. As can be seen, $C(i,i+1)$ starts from zero, increases rapidly at $\lambda=1$, and then maintains a constant value as $\lambda$ increases. In this case, when $h>>J$, i.e., $\lambda << 1$, the effect of magnetic field dominates, causing the spins to be aligned to the $z$ direction and as a result $C(i,i+1)$ vanishes. On the other hand, when $h<<J$, i.e., $\lambda >> 1$, the effect of the coupling dominates and the spins are equally aligned in both $x$ and $y$ directions and the concurrence maintains a constant finite value.
Interestingly, raising the temperature as shown in Fig.~\ref{fig:13b} does not reduce the concurrence but causes the values of the coupling and the magnetic field to affect the concurrence independently at nonzero temperature. Also in Fig.~\ref{fig:13b}, one can notice that decreasing $h$ at $kT=1$ causes the change in $C(i,i+1)$ to be less rapid and to reach equilibrium value at larger values of $\lambda$. In Fig.~\ref{fig:13c}, we fix the coupling at different values and study the concurrence as the magnetic field changes. The concurrence $C(i,i+1)$ vanishes when $J < 1$, i.e., when the magnetic field dominates, while for $J \geq 1$ it manifests the same behavior as before. The effect of higher temperatures is shown in Fig.~\ref{fig:13d} where it causes smoother change in the concurrence.
\begin{figure}[htbp]
\begin{minipage}[c]{\textwidth}
 \centering 
   \subfigure{\label{fig:14a}\includegraphics[width=6cm]{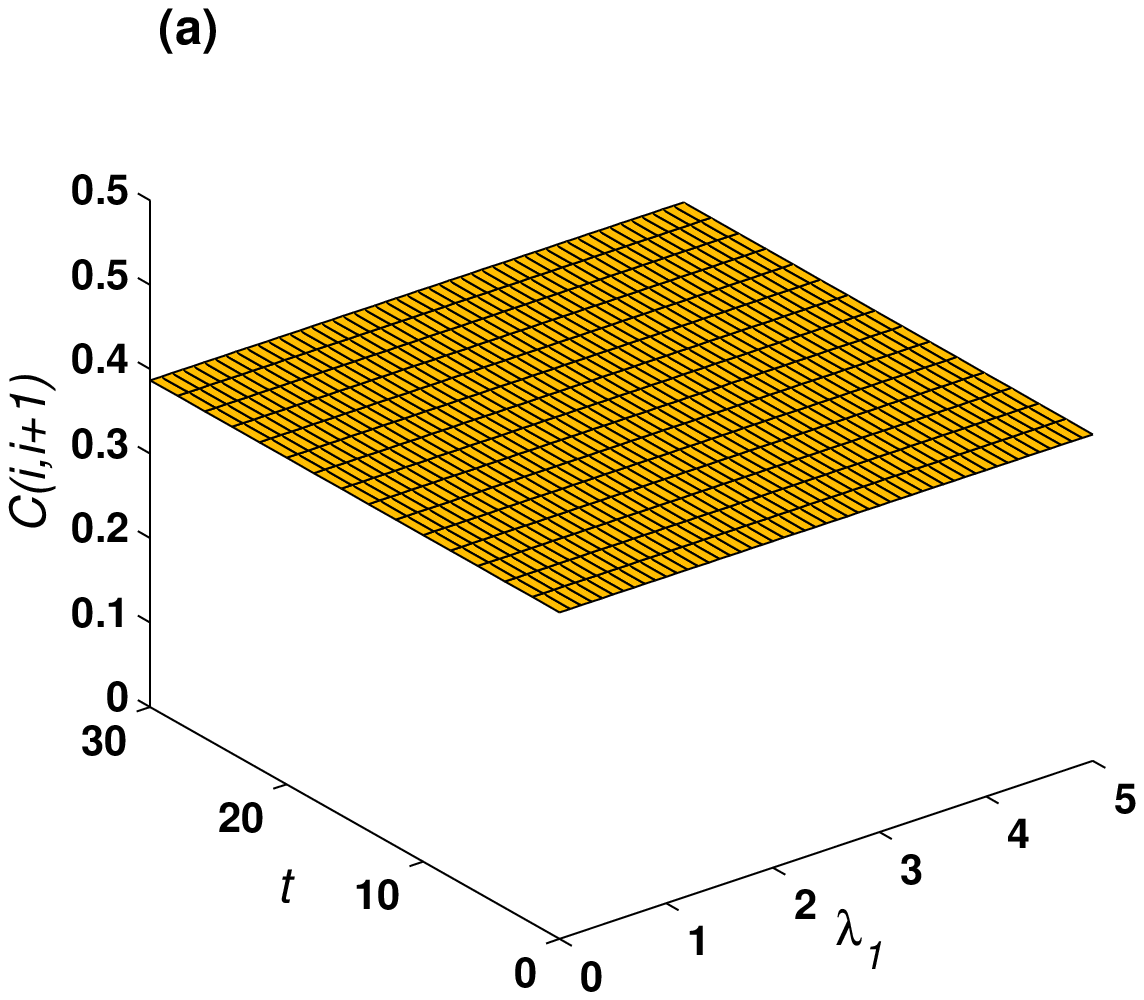}}\quad
   \subfigure{\label{fig:14b}\includegraphics[width=6cm]{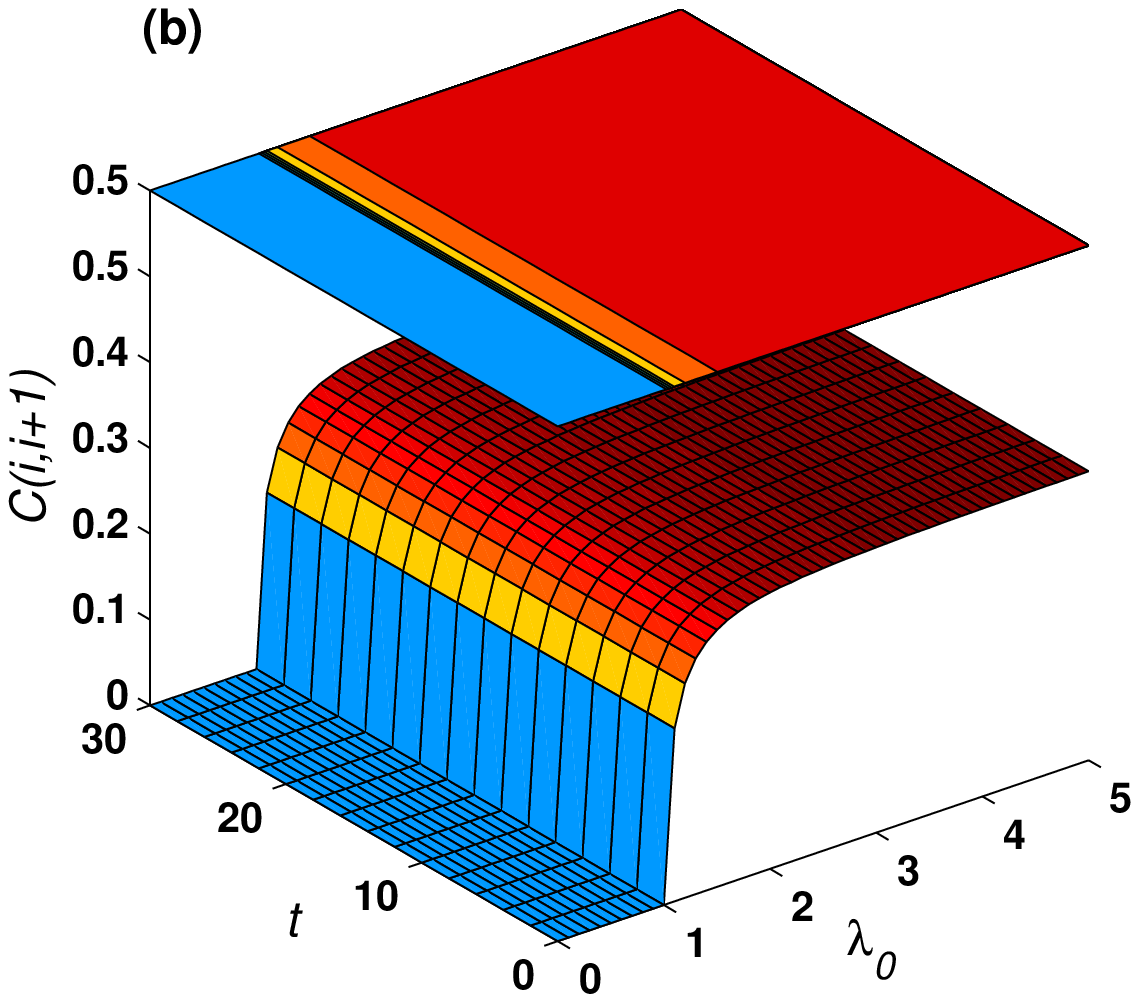}}\\
   \caption{{\protect\footnotesize (Color online) (a) $C(i,i+1)$ as a function of $\lambda_{1}$ and $t$, in units of $J_{1}^{-1}$, at $kT=0$ with $\gamma=0$, $h_{0}=h_{1}=1$ and $J_{0}=5$; (b) $C(i,i+1)$ as a function of $\lambda_{0}$ and $t$ at $kT=0$ with $\gamma=0$, $h_{0}=h_{1}=1$ and $J_{1}=5$.}}
 \label{fig:14}
 \end{minipage}
\end{figure}

The time evolution of nearest-neighbor concurrence as a function of the time-dependent coupling is explored in Fig.~\ref{fig:14}, where we fix the magnetic field. Figure~\ref{fig:14a} shows $C(i,i+1)$ as a function of $\lambda_{1}$ and $t$ where $h_{0}=h_{1}=1$ and $J_{0}=5$ at $kT=0$. Clearly, $C(i,i+1)$ is independent of $\lambda_{1}$. Studying $C(i,i+1)$ as a function of $\lambda_{0}$ and $t$ with $h_{0}=h_{1}=1$ at $kT=0$ for various values of $J_{1}$, we note that the results are independent of $J_{1}$. Figure~\ref{fig:14b} represents the case where $J_{1}=5$. Again, as can be noticed when $J_{0}<h_{0}$, the magnetic field dominates and $C(i,i+1)$ vanishes. While for $J_{0} \geq h_{0}$, $C(i,i+1)$ has a finite value, as discussed above.
\begin{figure}[htbp]
\begin{minipage}[c]{\textwidth}
 \centering
   \subfigure{\label{fig:15a}\includegraphics[width=6cm]{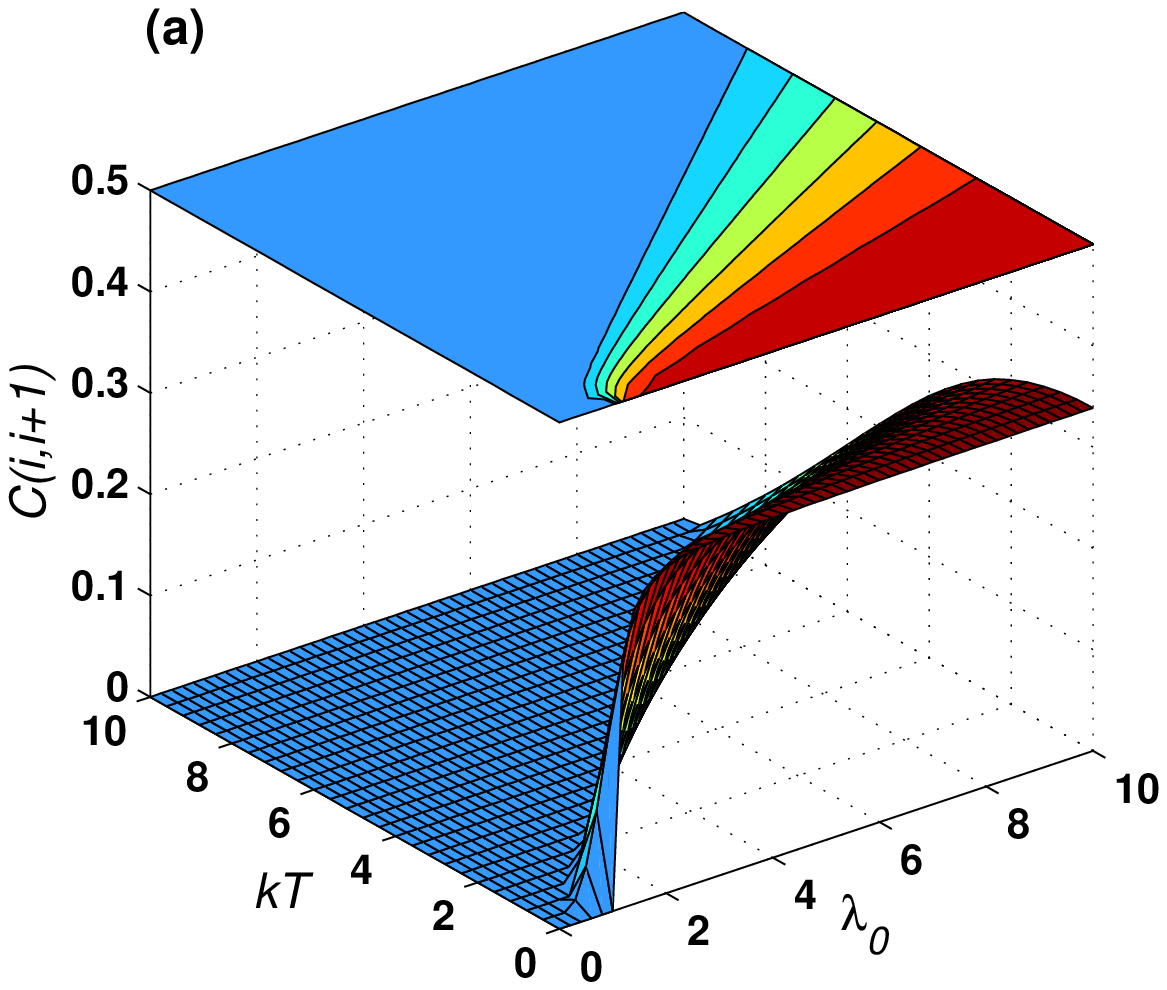}}\quad
   \subfigure{\label{fig:15b}\includegraphics[width=6cm]{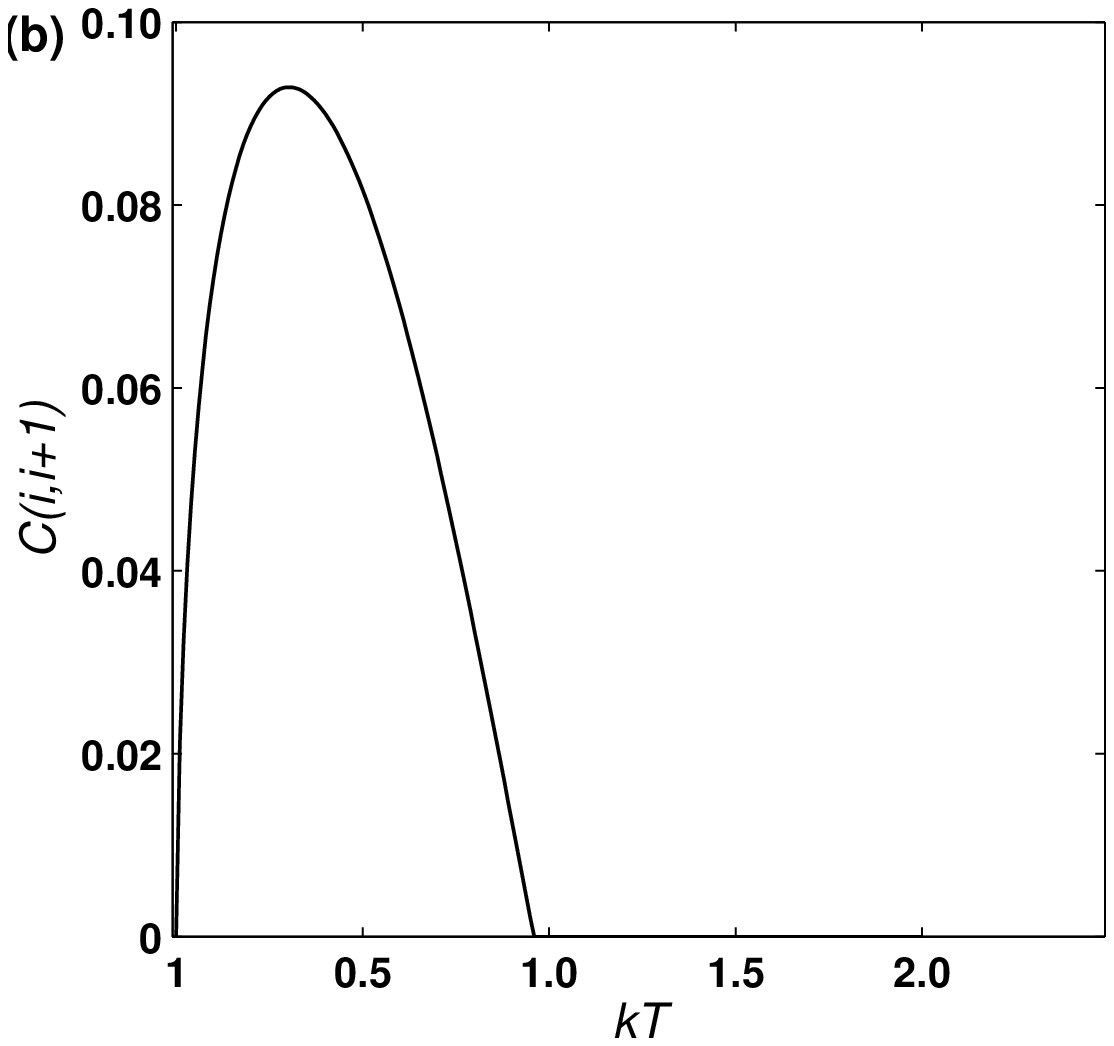}}
      \caption{{\protect\footnotesize (Color online) The asymptotic behavior of $C(i,i+1)$ as a function of (a) $\lambda_{0}$ and $kT$, in units of $J_{1}$, with $\gamma=0$, $h_{0}=h_{1}=1$, and $J_{1}=1$; (b) $kT$ with $\gamma=0$, $h_{0}=h_{1}=1$, and $J_{0}=J_{1}=1$.}}
 \label{fig:15}
 \end{minipage}
\end{figure}

The effect of temperature on concurrence is investigated in Fig.~\ref{fig:15}. In Fig.~\ref{fig:15a} we plot the asymptotic concurrence $C(i,i+1)$ as a function of $\lambda_{0}$ and $kT$. Clearly, as $kT$ increases the threshold $\lambda_{0}$, at which $C(i,i+1)$ starts to have a finite value, increases. An interesting behavior of the concurrence is featured here, studying the asymptotic concurrence $C(i,i+1)$ as a function of $\lambda_{1}$ and $kT$, one observes that for $\lambda\leq 1$, $C(i,i+1)$ starts from $0$ at $kT=0$, grows up as $kT$ increases, reaching a maximum value at $kT \approx 0.3$, and then vanishes again for $kT \approx 0.9$ as shown in Fig.~\ref{fig:15b}.
\begin{figure}[htbp]
\begin{minipage}[c]{\textwidth}
 \centering 
   \includegraphics[width=6cm]{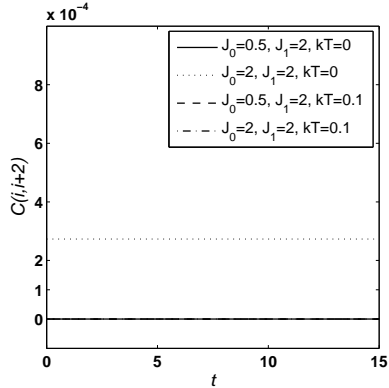}
   \caption{{\protect\footnotesize Dynamics of $C(i,i+2)$, where $t$ is in units of $J_{1}^{-1}$, with $\gamma=0$ and $h_{0}=h_{1}=1$ for various values of $J_{0}, J_{1}$ at $kT=0$ and $kT=0.1$.}}
 \label{fig:16}
 \end{minipage}
\end{figure}
In Fig.~\ref{fig:16}, we depict the time evolution of next-to-nearest-neighbor concurrence $C(i,i+2)$. As expected, $C(i,i+2) << C(i,i+1)$ at the same circumstances. The longer-range concurrence $C(i,i+r)$ vanishes for $r\geq 3$.

\begin{figure}[htbp]
\begin{minipage}[c]{\textwidth}
 \centering 
   \subfigure{\label{fig:17a}\includegraphics[width=6.5cm]{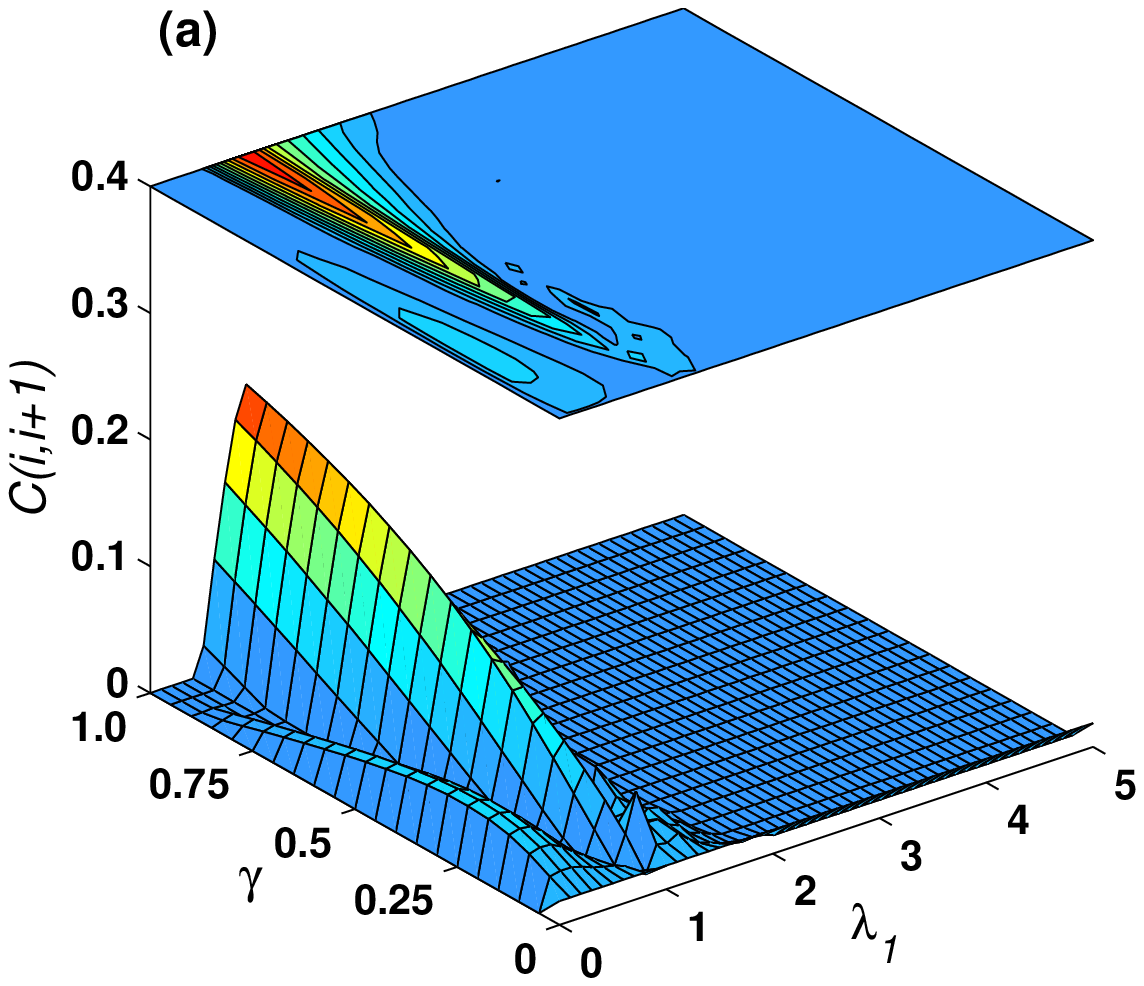}}\quad
   \subfigure{\label{fig:17b}\includegraphics[width=6.5cm]{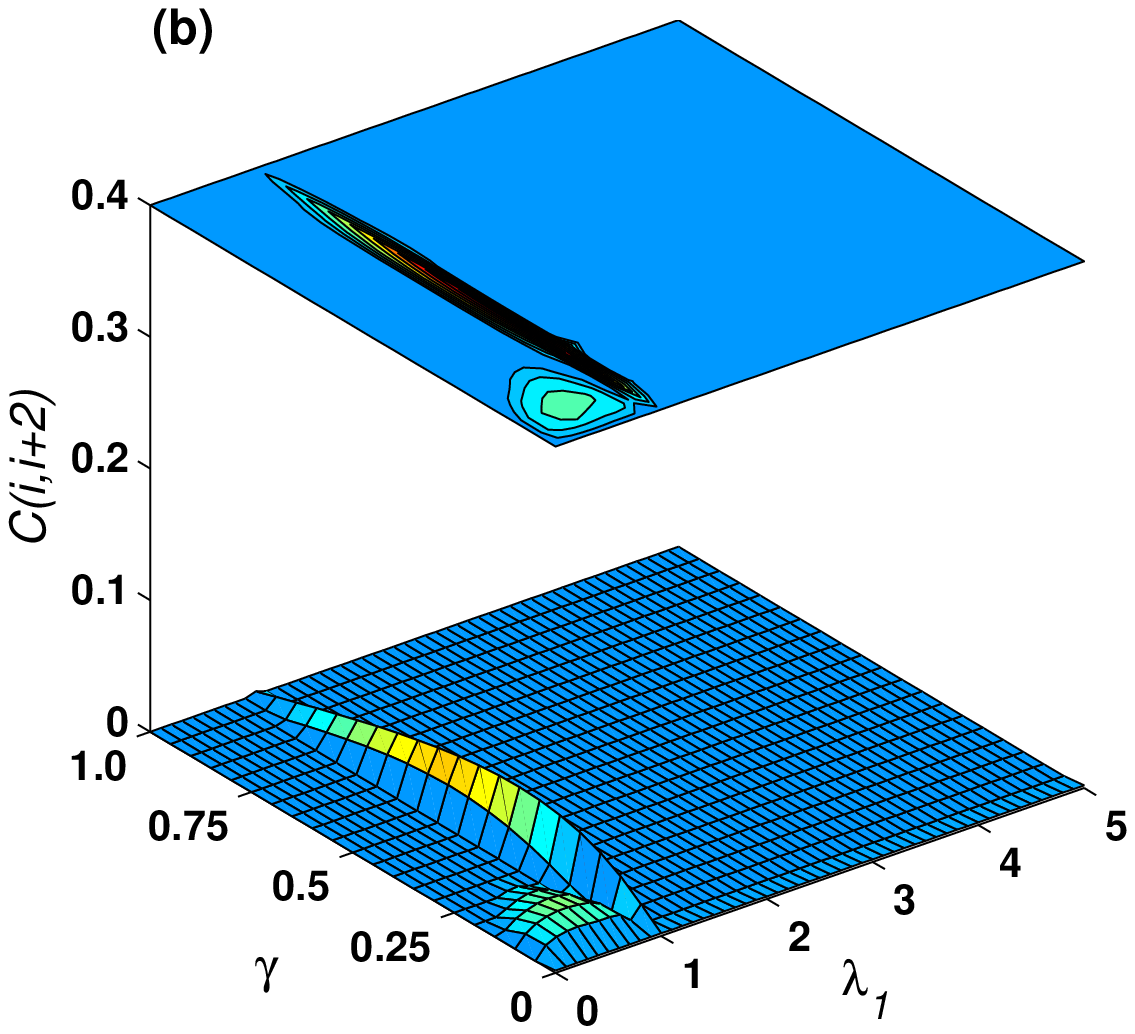}}\\
   \caption{{\protect\footnotesize (a) (Color online) The asymptotic behavior of $C(i,i+1)$ as a function of $\lambda_{1}$ and $\gamma$ with $h_{0}=h_{1}=1$ and $J_{0}= 1$ at $kT=0$. (b) The asymptotic behavior of $C(i,i+2)$ as a function of $\lambda_{1}$ and $\gamma$ with $h_{0}=h_{1}=1$ and $J_{0}= 1$ at $kT=0$.}}
 \label{fig:17}
\end{minipage}
\end{figure}

Finally, we explore the asymptotic behavior of the nearest-neighbor and next-to-nearest-neighbor concurrence in the $\lambda$-$\gamma$ phase space of the one-dimensional $XY$ spin system under the effect of a time-dependent coupling $J(t)$. In Figs.~\ref{fig:17a} and \ref{fig:17b} we plot $C(i,i+1)$ and $C(i,i+2)$ respectively as a function of the parameter $\lambda_1$ and the degree of anisotropy $\gamma$ for constant magnetic field $h_{0}=h_{1}=1$ and $J_0=1$ at $kT=0$. As one can notice, the nonvanishing concurrences appear in the vicinity of $\lambda=1$ or lower and vanishes for higher values. One interesting feature is that the maximum achievable nearest-neighbor concurrence takes place at $\gamma=1$, i.e., in a completely anisotropic system, while the maximum next-to-nearest-neighbor concurrence is achievable in a partially anisotropic system, where $\gamma \approx 0.3$.
\section{Conclusions and Future Directions}
We have investigated the entanglement evolution in an infinite one-dimensional $XY$ model in an external time-dependent magnetic field at zero and finite temperature. The nearest-neighbor interaction between the spins were considered time dependent. An exact solution was presented for a step function form of both the time-dependent coupling and magnetic field. The system showed nonergodic and critical behavior at all degrees of anisotropy and for all different choices of coupling and magnetic field. At zero temperature and constant magnetic field and coupling, the asymptotic behavior of the system at the infinite time limit depends only on the ratio of the coupling to the magnetic field, not their individual values but changes for nonzero temperature. For many system setups the initial values of the coupling and magnetic field dictate the asymptotic behavior of the entanglement regardless of their final values at different degrees of anisotropy and may lead to asymptotic residual entanglement in the system. Interestingly, studying the dynamics of entanglement at zero and finite temperature showed that the quantum properties of the system are preserved within certain regions of the coupling, magnetic field, and temperature space that vary significantly depending on the degree of anisotropy of the system. Particularly, the quantum effects in the transverse Ising model persist in the vicinity of both its critical phase transition point and zero temperature as it evolves in time. In future work, it would be interesting to study the $XY$ model under the effect of time-dependent coupling and magnetic field where the function form for each one could differ from the other and take other forms of practical interest, such as the sinusoidal and exponential. This would need in that case the use of numerical methods along with the analytical ones to treat the system.
\section*{Acknowledgments}
This work was supported in part by the deanship of scientific research and the research center, College of Science, King Saud University.

\end{document}